\documentclass[aps,pra,reprint,groupedaddress,longbibliography]{revtex4-2}
\usepackage[T1]{fontenc}

\usepackage{graphicx}
\usepackage{bxtexlogo}

\usepackage{amsmath,amssymb}
\usepackage{mathtools}

\usepackage{amsthm}
\usepackage{amsfonts}
\usepackage{subcaption}

\usepackage{lmodern}

\usepackage{enumitem}
\usepackage{mathrsfs}

\theoremstyle{definition}
\newtheorem{definition}{Definition}[section] %[] が x.x なら x.x.1 から始まる
\newtheorem{theorem}[definition]{Theorem} %[] が x.x.x なら x.x.(x+1) というように続く

\newtheorem{lemma}[definition]{Lemma}
\newtheorem{corollary}[definition]{Corollary}
\newtheorem{example}[definition]{Example}

\newtheorem{innerproblem}{Problem}

\newenvironment{problem}[1]{
    \renewcommand\theinnerproblem{#1}\innerproblem
}{\endinnerproblem}

\newcommand{\bra}{\langle}% 
\newcommand{\ket}{\rangle}% 

\newcommand{\bs}[1]{\boldsymbol{#1}}
\newcommand{\mbb}[1]{\mathbb{#1}}
\newcommand{\mcal}[1]{\mathcal{#1}}
\newcommand{\mrm}[1]{\mathrm{#1}}
\newcommand{\msf}[1]{\mathsf{#1}}
\newcommand{\mscr}[1]{\mathscr{#1}}
\newcommand{\mfrak}[1]{\mathfrak{#1}}

\newcommand{\ran}{\operatorname{ran}}

\newcommand{\Tr}{\operatorname{Tr}}

\newcommand{\im}{\mathrm{i}}

\newcommand{\hash}{\#}

\newcommand{\epower}[1]{\mathrm{e}^{#1}}
\newcommand{\hatmcal}[1]{\hat{\mcal{#1}}}

\newcommand{\Ad}[1]{\operatorname{Ad}_{#1}}
\newcommand{\ie}{{i.e.}}
\usepackage{comment}
\newcommand{\defeq}{\coloneqq}
\newcommand{\vect}[1]{\boldsymbol{#1}}

\usepackage[luatex,pdfencoding=auto]{hyperref}
\hypersetup{% hyperrefオプションリスト
setpagesize=false, 
 bookmarksnumbered=true, %
 bookmarksopen=true, %
 colorlinks=true, %
 linkcolor=blue, 
 citecolor=blue, 
 urlcolor=blue,
}

\usepackage{tikz}
\usepackage{tikz-3dplot}
\usetikzlibrary {arrows.meta}
\usetikzlibrary{intersections, calc, arrows}
\usetikzlibrary{positioning}
\usetikzlibrary{shapes.geometric}
\usetikzlibrary{shapes.misc}
\usetikzlibrary{decorations.pathreplacing,calligraphy}
\usetikzlibrary{shadings} 
\newcommand{\numberthis}{\addtocounter{equation}{1}\tag{\theequation}} %数式番号をつける

\allowdisplaybreaks

\newcommand{\delete}[1]{\relax}

\begin{document}

%\preprint{}

\title{{ A Necessary and Sufficient Condition for Quantum Realizability of Correlations for Arbitrary Normalized Observables in the Clauser--Horne--Shimony--Holt Setup}}


\author{Ryosuke Nogami}
\email[]{nogami.ryosuke.v3@s.mail.nagoya-u.ac.jp}
%\thanks{}
\affiliation{Graduate School of Informatics, Nagoya University, Furo-cho, Chikusa-Ku, Nagoya, 466-8601, Japan}

\author{Jaeha Lee}
\email[]{lee@iis.u-tokyo.ac.jp}
\affiliation{Institute of Industrial Science, the University of Tokyo, 5-1-5 Kashiwanoha, Kashiwa, Chiba, 277-8574, Japan}

\date{\today}

% \begin{abstract}
%     We establish a necessary and sufficient condition for the existence of a quantum state that reproduces given correlation values in the Clauser–Horne–Shimony–Holt (CHSH) setup for arbitrarily fixed binary observables. The condition takes the form of a nonlinear matrix inequality depending explicitly on the angles between the observables and fully characterizes the quantum realizability of CHSH-type correlation data under fixed measurement settings. Our result reveals that the set of quantum realizable correlations can be strictly smaller than the set of classically realizable ones, even in the simplest (2,2,2) scenario. The formulation also recovers Tsirelson's bound as a special case. Furthermore, we employ symmetry reduction techniques to simplify the realizability problem and derive a closed-form inequality for anti-commuting observables. These findings clarify the statistical constraints imposed by quantum mechanics when measurement settings are not allowed to vary.
%     \end{abstract}

\begin{abstract}
    
    We establish a necessary and sufficient condition for the existence of a quantum state that reproduces given correlation values in the Clauser--Horne--Shimony--Holt (CHSH) setup for any fixed normalized observables.
    This result addresses a fundamental question shared by both local realism and quantum mechanics: under what conditions a given set of observed data can be reproduced by a physical model.
    While previous studies have mainly addressed conditions for correlations achievable without specifying the measurement settings, our result gives a finer characterization by treating the observables as fixed in advance.
    The resulting quantum condition strengthens previously known constraints, such as Tsirel'son's inequalities and the Tsirel'son--Landau inequality, by characterizing statistical constraints explicitly for each specified set of observables. 
    In particular, we show that our condition applies to Bell's original scenario and reveals that whether Bell's original inequality is violated depends sensitively on the chosen observables.
    More broadly, this perspective offers new insights into how quantum violations of local realism depend on the measurement settings.

\end{abstract}

%\keywords{}

\maketitle

% References should be done using the \cite,~\ref, and \label commands

%%%%%%%%%%%%%%%%%%%%%%%%%%%%%%%%%%%%%%%%%%%%%%%%%%%%%%%%%%%%%%%
%%%%%%%%%%%%%%%%%%%%%%%%%%%%%%%%%%%%%%%%%%%%%%%%%%%%%%%%%%%%%%%
%%%%%%%%%%%%%%%%%%%%%%%%%%%%%%%%%%%%%%%%%%%%%%%%%%%%%%%%%%%%%%%
\section{Introduction\label{sec:introduction}}

% \subsection{}
% \subsubsection{}

The celebrated work~\cite{einstein1935can} by Einstein, Podolsky, and Rosen (EPR) was made public in 1935 with the intent of questioning the completeness of quantum theory. While quantum mechanics had enjoyed great success in explaining various physical phenomena beyond the reach of classical theory,
%{such as atomic spectra and black-body radiation}
the authors yet believed that the world should be described by a theory grounded in what is now called \textit{local realism}, to which the quantum theory did not appear to adhere.

%It was not until almost 30 years later that the possibility of the predictions of quantum theory admitting a local realistic description was given a viable solution;  
It was not until almost 30 years later that the possibility of a local realistic description of the predictions of quantum theory was given a viable solution;
Bell offered a negative answer to this problem by demonstrating that quantum theory is capable of violating a set of statistical constraints~\cite{bell1964einstein} that every local realistic theory must necessarily comply with.  In honour to his pioneering work, such constraints are now collectively referred to as the \textit{Bell inequalities}.  Clauser, Horne, Shimony, and Holt (CHSH) later derived an alternative set of constraints~\cite{clauser1969proposed,clauser1978bell}, but now under less restrictive and therefore more practical assumptions than those of Bell's original formulation.

In contemporary language, a local realistic theory is understood to satisfy two major principles: \textit{locality}, which forbids superluminal interaction to ensure compatibility with the theory of relativity, and \textit{realism}, which assumes the objective existence of the values of physical observables independent of the act of measurement.  As for the latter, realism is commonly equated with the existence of (without loss of generality~\cite{fine1982joint}, deterministic) \textit{hidden variables}, which dictates that the value $X(\lambda)$ of an observable $X$ should be uniquely determined by a hidden variable $\lambda \in \Lambda$ specifying the state of the system.  Realism thus entails the existence of a joint probability distribution
\begin{equation}\label{eq:JPD_in_Realism}
%= (\vect{X}_{\ast} w)(\vect{x}) \defeq
P_{\vect{X}}(\vect{x}) \defeq \int_{\vect{X}^{-1}(\vect{x})} \,dw(\lambda)
\end{equation}
for any tuple $\vect{X}(\lambda) \defeq (X_{1}(\lambda), \ldots, X_{N}(\lambda))$ of observables given the probability distribution $w(\lambda)$ on $\Lambda$.  Note that by construction, the existence of such joint probability distributions conversely implies realism~\cite{fine1982joint}, as one may specifically choose $\Lambda$ to be the set of every possible tuples of the values of all observables under consideration, and accordingly, $w$ to be their joint probability distribution.

\delete{Clauser, Horne, Shimony, and Holt (CHSH) \cite{clauser1969proposed,clauser1978bell} derived a similar set of statistical constraints for local realism under less restrictive and therefore more practical assumptions than those of Bell's original formulation. }
In the CHSH setup, two spatially separated experimenters, Alice and Bob, each measure two binary observables (Alice: $A_1,A_2$; Bob: $B_1,B_2$).
It is assumed that Alice and Bob can freely choose which observable to measure independently of the choice made by the other. 
The correlation between $A_i$ and $B_j$ is defined as
\begin{equation}
    C_{ij} := \sum_{a,b} ab \,P_{A_i B_j}(a,b),
    \label{eq:definition of correlation}
\end{equation}
where $P_{A_i B_j}(a,b)$ is the joint probability distribution for $A_i$ and $B_j$.
If the system can be described by local realism, the correlation $C_{ij}$ is given by
\begin{equation}
    C_{ij} = \int_{\Lambda} d w(\lambda) A_i(\lambda) B_j(\lambda).
    \label{eq:correlation in LR}
\end{equation}
% \delete{where $P_{A_i B_j}$ is the joint probability distribution for $A_i$ and $B_j$, and $w(\lambda)$ denotes the probability distribution of the hidden variable $\lambda$. }
The left-hand side of Eq.~\eqref{eq:correlation in LR}, as defined by \eqref{eq:definition of correlation}, represents a measurable quantity in experiments, while the equality connecting it with the right-hand side expresses the assumption that there exists a hidden variable $\lambda$ that explains the experimental data.
As stated above, alternatively, and equivalently \cite{fine1982joint}, one may assume the existence of a joint probability distribution $P(A_1,A_2,B_1,B_2)$ for all four variables $A_1,A_2,B_1,B_2$.
Under these conditions, if $A_i$ and $B_j$ take values in $\{\pm 1\}$, the following CHSH inequalities hold \cite{clauser1969proposed,clauser1978bell}:
\begin{equation}
    \left|\sum_{i,j \in \{1,2\} } C_{ij}  -2C_{kl}\right| \leq 2, 
    \quad  \forall k,l \in \{1,2\}.
    \label{eq:CHSH ineq}
\end{equation}

On the other hand, in quantum mechanics, each observable is specified by a self-adjoint operator, and the correlation between commuting observables $A_i$ and $B_j$ under the quantum state described by a density operator $\rho$ is calculated by
\begin{equation}
    C_{ij} = \Tr[\rho A_i B_j].
    \label{eq:quantum correlation}
\end{equation}
If $A_i$ and $B_j$ take values in $\{\pm 1\}$, we can assume that the spectra of the corresponding operators are $\{\pm 1\}$. Then, quantum correlations must satisfy the following Tsirel'son's inequalities \cite{cirel1980quantum}:
\begin{equation}
    \left|\sum_{i,j \in \{1,2\} } C_{ij} -2 C_{kl} \right| \leq 2\sqrt{2}, 
    \quad \forall k,l \in \{ 1,2\},
    \label{eq:Tsirel'son ineq}
\end{equation}
The equality can be attained in quantum mechanics for each $k,l \in \{1,2\}$, indicating that in such cases, the corresponding CHSH inequality must be violated. Violations of the CHSH inequalities have been experimentally verified \cite{aspect1981experimental,aspect1982experimentalEPR,aspect1982time,freedman1972experimental,weihs1998violation}.

Bell inequalities can be considered necessary conditions for a system to be described by local realism. Conversely, one can also consider sufficient conditions for a system to be described by local realism. 
Fine \cite{fine1982joint,fine1982hidden} demonstrated that the validity of the CHSH inequalities is a necessary and sufficient condition for a system to be described by local realism in the CHSH setting (see the next section for the precise statement of the theorem). 
Furthermore, Peres \cite{peres1999all} proposed a more generalized problem: under the $(n,m,d)$ setup, where $n$ spatially separated observers can each measure one of $m$ observables with $d$ possible values, what is the necessary and sufficient condition for a system to be described by local realism? 
The inequalities that provide the answer to this problem are collectively referred to as \textit{`all the Bell inequalities'}. The CHSH setting corresponds to the $(2,2,2)$ setup, and by Fine's theorem, the CHSH inequalities~\eqref{eq:CHSH ineq} constitute `all the Bell inequalities' for the $(2,2,2)$ setup. 

There have been numerous studies on the Bell inequalities.  
One direction of research involved methods based on convex analysis \cite{froissart1981constructive,garg1984farkas,peres1999all}, recognizing that the set of locally realistic probability distributions forms a convex set known as the correlation polytope \cite{pitowsky1991correlation}.
Building on this approach or through alternative methods, a number of studies derived explicit inequalities.
For small specific values of $n,m$, and $d$, see \cite{pitowsky2001optimal,Kaszlikowski2002clauser,bacon2003bell,collins2004relevant,sliwa2003symmetries,ito2006bell,wiesniak2007explicit,brunner2008partial,pal2009quantum}; for generalizations to $n$-partite systems, see \cite{werner2001all,zukowski2002bell,paterek2006series,zukowski2006tight,wu2008extending}; and for generalizations to $m$-measurement settings, $d$-valued observables, or both, see \cite{collins2004relevant,collins2002bell,massar2002resistant,nagata2006bell,masanes2003tight,ji2008multisetting,liang2009reexamination}.
For further references, refer to \cite{open_quantum_problems}.
The tightness (\ie, whether the inequalities represent boundaries of the correlation polytope) and completeness (\ie, whether the inequalities constitute the complete set of `all the Bell inequalities') of the inequalities derived in these prior studies have not been fully clarified. 
A barrier to determining `all the Bell inequalities' is the exponential growth in computational complexity with respect to $n,m$, and $d$ poses, as the problem is known to be NP-complete \cite{pitowsky1989quantum}.

Some of the aforementioned studies also examine the violations of these inequalities in quantum mechanics, and several studies \cite{dur2001multipartite,acin2001distillability,acin2002bell,acin2002violation,wehner2006tsirelson,zohren2008maximal,doherty2008quantum,li2020exact} specifically focus on such quantum violations.
While the local realistic statistical restrictions can be expressed using a finite number of linear inequalities, more complex inequalities are needed to completely characterize the restrictions on quantum distributions
\cite{tsirelson1987quantum,landau1988empirical,avis2006relationship,navascues2007bounding,Ishizaka2017cryptographic,Ishizaka2018necessary}.
However, a complete understanding of the statistical constraints imposed by quantum mechanics remains an open problem. In particular, the detailed structure of the constraints under fixed measurement settings has not yet been fully explored.

% This paper aims to systematically formulate the realizability problem of experimental data under local realism and quantum mechanics, and derive the constraints on quantum correlations for arbitrarily given observables within the CHSH setting.
% This paper investigates the realizability problem in quantum mechanics by distinguishing between two settings: one in which the observables are freely chosen, and another in which they are fixed in advance.
% Focusing on the CHSH setup, we derive a necessary and sufficient condition for the quantum realizability of given correlation values under the assumption that the measurement observables are fixed arbitrarily beforehand.
% This result provides an exact characterization of which correlation values can be reproduced by quantum mechanics when measurement settings are predetermined rather than subject to free choice.
% Our approach employs characteristic functions, which convert marginal constraints on observed correlations into decoupled algebraic conditions.
% To treat both quantum and local realistic scenarios within a unified framework, we introduce quasi-joint probability distributions.
% In addition, we develop a symmetry-based reduction technique that significantly simplifies the analysis of the realizability problem without loss of generality.
% Compared to previously known results, which consider the case where observables are allowed to vary freely, our findings offer a finer characterization of the statistical constraints imposed by quantum mechanics.
We investigate the problem of characterizing which correlation values can be realized by quantum mechanics in the CHSH setting. In particular, we distinguish between two variants of the problem depending on whether the measurement observables are fixed or not.
In the unfixed observables setting, we ask whether a given set of correlation values can be realized by some choice of quantum observables and state, without prior constraints on the observables.
In contrast, the fixed observables setting considers the case where the measurement observables are predetermined, and the question becomes whether there exists a quantum state that realizes the given correlations under this fixed measurement setting.
The fixed setting provides a more fine-grained characterization of quantum realizability, as the feasibility condition in the unfixed setting can be obtained as a logical disjunction of those in the fixed setting over all possible choices of observables.

In this paper, we focus on the fixed observable setting in the CHSH scenario and derive a necessary and sufficient condition for the quantum realizability of correlations. This contributes to a more precise understanding of the limitations and capabilities of quantum correlations under realistic constraints on experimental settings.

This paper is organized as follows. 
% First, we describe the problems under consideration in Section~\ref{sec:problem}. 
% In Section~\ref{sec:preliminaries}, we introduce the necessary mathematical background.
In Section~\ref{sec:quantum restrictions in the CHSH setting}, we derive a set of exact quantum constraints on correlations for arbitrarily given normalized observables in the CHSH setting. In Section~\ref{sec:comparison}, we compare our results with the CHSH inequalities and quantum constraints derived in prior studies. In Section~\ref{sec:summary and discussion}, we provide a summary of the paper and discuss its findings.

\section{Quantum restrictions on correlations for given observables in the CHSH setting \label{sec:quantum restrictions in the CHSH setting}}

\subsection{Statement of the Main Theorem}
We consider a quantum problem for fixed observables in the CHSH setting, \ie, the $(2,2,2)$ setup. 
The following theorem, corresponding to the quantum problem provides a quantum statistical restriction on the correlations.

\begin{figure}
    \centering
    \includegraphics[width=80mm]{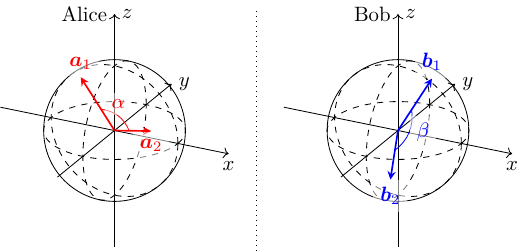}
    \caption{The two vectors $\bs{a}_1$ and $\bs{a}_2$, which are associated with the observables $A_1$ and $A_2$, respectively, form the angle $\alpha \in [0,\pi]$. Similarly, the two vectors $\bs{b}_1$ and $\bs{b}_2$, which are associated with the observables $B_1$ and $B_2$, respectively, form the angle $\beta \in [0,\pi]$.}
    \label{fig:bloch sphere}
\end{figure}

To this end, let $A_i := a_{i0} I_2 + \bs{a}_i\cdot \bs{\sigma}$ and $B_j := b_{j0} I_2 + \bs{b}_j \cdot \bs{\sigma}$ be binary self-adjoint operators on $\mathcal{H} = \mbb{C}^2$ with $a_{i0},b_{j0} \in \mbb{R}$ and $\bs{a}_i,\bs{b}_j \in \mbb{R}^3 \setminus \{\bs{0}\}$ for $i,j \in \{1,2\}$, where $\bs{\sigma} := (\sigma_x, \sigma_y, \sigma_z)$ denotes the Pauli matrices. We then introduce their normalizations
\begin{equation}
\tilde{A}_i := \frac{A_i - a_{i0} I_2}{\lVert\bs{a}_i\rVert}, \quad \tilde{B}_j := \frac{B_j - b_{j0}I_2}{\lVert\bs{b}_j\rVert}, 
\end{equation}
where $\lVert\bs{x}\rVert \defeq (\sum_{i} x_{i}^{2})^{1/2}$ denotes the standard Euclidean norm.  Note that $\tilde{A}_i$ and $\tilde{B}_i$ are binary operators with the spectrum $\{\pm 1\}$.  We also introduce the angles $\alpha \in [0,\pi]$ and $\beta \in [0,\pi]$ denote the angles between $\bs{a}_1$ and $\bs{a}_2$, and between $\bs{b}_1$ and $\bs{b}_2$, respectively, namely $\cos{\alpha} = \vect{a}_{1} \cdot \vect{a}_{2}/(\lVert \bs{a}_{1} \rVert \lVert \bs{a}_{2} \rVert)$ and $\cos{\beta} = \vect{b}_{1} \cdot \vect{b}_{2}/(\lVert \bs{b}_{1} \rVert \lVert \bs{b}_{2} \rVert)$ (see Fig.~\ref{fig:bloch sphere}). 
%%%%%

%%%%%
\begin{theorem}\label{thm:main theorem}
	Let $A_{1}$, $A_{2}$, $B_{1}$, and $B_{2}$ be binary observables on a two-level system, and let $\gamma_{ij} \in \mathbb{R}$, $i, j = 1, 2$.
    Then, there exists a density operator $\rho$ on the Hilbert space $\mbb{C}^2 \otimes \mbb{C}^2$ of the composite system that reproduces the correlations
    \begin{equation}
        \gamma_{ij} = \Tr[\rho \tilde{A}_i \otimes \tilde{B}_j], \quad \forall i,j ,\in \{1,2\}
        \label{eq:gamma_ij}
    \end{equation}
    of the normalized observables
        %for a given real vector $\bs{\gamma} := (\gamma_{11}, \gamma_{21}, \gamma_{12}, \gamma_{22})^\top \in \mbb{R}^4$ 
    if and only if the inequality
    \begin{equation}
        \sqrt{\bs{\gamma}^\top F_{\alpha,\beta}\, \bs{\gamma}} + \sqrt{\bs{\gamma}^\top F_{\alpha,-\beta}\, \bs{\gamma}} \leq 2
        \label{eq:main inequality}
    \end{equation}
    holds, where $\bs{\gamma} := (\gamma_{11}, \gamma_{21}, \gamma_{12}, \gamma_{22})^{\top}$ and
    \begin{multline}
        F_{\alpha,\beta} :=  \frac{1}{\sin^2\alpha \sin^2\beta} \\
        \times \begin{pmatrix}
            1 & -\cos\alpha & - \cos \beta & \cos(\alpha + \beta) \\
            -\cos\alpha & 1 & \cos(\alpha-\beta) &  -\cos \beta \\
            -\cos\beta & \cos(\alpha-\beta) & 1 & -\cos\alpha \\
            \cos(\alpha+\beta) & -\cos\beta & -\cos\alpha & 1
        \end{pmatrix}.
    \end{multline}
In particular, $\rho$ can be chosen so that $\bra \tilde{A}_i \ket_{\rho} = \bra \tilde{B}_j \ket_{\rho} = 0$ for all $i,j \in \{1,2\}$.
\end{theorem}
Here, the functions $\sqrt{\bs{\gamma}^\top F_{\alpha,\pm \beta}\, \bs{\gamma}}$ appearing in the left-hand side of~\eqref{eq:main inequality} are defined as the unique continuous extensions of those defined on the dense subset $(0,\pi)\times (0,\pi)$.  In general, the extensions are allowed to take values in the extended real numbers $\mbb{R}\cup\{\pm \infty\}$, whereas the restrictions are real valued.  

We provide a proof of Theorem~\ref{thm:main theorem} in Appendix~\ref{sec:proof of main thm}.
In the proof, we introduce a quasi-joint probability distribution for the observables $A_i$ and $B_j$ to treat the quantum problem in parallel with the local realism case.
We then utilize its characteristic function to translate the marginal constraints into decoupled algebraic conditions.
A symmetry reduction technique is also applied to reduce the degrees of freedom, which plays a crucial role in the proof.
% For traceless $A_{i}$ and $B_{j}$, the correlations of the original observables $C_{ij} = \Tr[\rho A_i \otimes B_j]$ and their normalized counterparts are related to one another by
% %\begin{equation}
% $\gamma_{ij} = 4 C_{ij} / (\lVert A \rVert \lVert B \rVert)$,
% %\end{equation}
% where $\lVert X \rVert$ here denotes the Hilbert--Schmidt norm of an operator $X$.

When $\alpha$ or $\beta$ equals $0$ or $\pi$, \ie, $\bs{a}_1 \parallel \bs{a}_2$ or $\bs{b}_1 \parallel \bs{b}_2$, the left-hand side of~\eqref{eq:main inequality} can diverge to $\infty$. By considering different cases separately, we derive more concrete expressions of the condition in Theorem~\ref{thm:main theorem} as follows:
\begin{enumerate}[label={(\roman{*})}]
    \item When $\bs{a}_1 \nparallel \bs{a}_2$ and $\bs{b}_1 \nparallel \bs{b}_2$: 
    \begin{equation}
        \sqrt{\bs{\gamma}^\top F_{\alpha,\beta} \bs{\gamma}} + \sqrt{\bs{\gamma}^\top F_{\alpha,-\beta} \bs{\gamma}} \leq 2.
        \label{eq:main inequality (i)}
    \end{equation}
    \item When $\bs{a}_1 \nparallel \bs{a}_2$ and $\bs{b}_1 \parallel \bs{b}_2$: 
    \begin{gather}
        \sqrt{\gamma_{11}^2 + \gamma_{21}^2 - 2\cos(\alpha) \gamma_{11} \gamma_{21}} \leq \sin(\alpha) , \\
        \gamma_{11} = s_b \gamma_{12}, \quad
        \gamma_{21} = s_b \gamma_{22},
        \label{eq:main inequalitiy (B_1 parallel B_2)}
    \end{gather}
    where $s_b := \operatorname{sgn}(\bs{b}_1 \cdot \bs{b}_2)$. 
    The case where $\bs{a}_1 \parallel \bs{a}_2$ and $\bs{b}_1 \nparallel \bs{b}_2$ yields a similar result.
    \item When $\bs{a}_1 \parallel \bs{a}_1$ and $\bs{b}_1 \parallel \bs{b}_2$: 
    \begin{equation}
        -1 \leq \gamma_{11} = s_a \gamma_{21} = s_b \gamma_{12} = s_a s_b \gamma_{22} \le 1,
    \end{equation}
    where $s_a := \operatorname{sgn}(\bs{a}_1 \cdot \bs{a}_2)$.
\end{enumerate}

To prove these conditions are equivalent to the condition in Theorem~\ref{thm:main theorem}, it is sufficient to show that taking the limit of $\alpha$ or $\beta$ approaching $0$ or $\pi$ in the case (i) reduces the result to the case (ii) or the case (iii).
Let $G_{\alpha,\beta}$ be the numerator of $F_{\alpha,\beta}$. If we take $\beta \to 0$, the left-hand side of~\eqref{eq:main inequality} diverges when $\bs{\gamma} \not\in \ker G_{\alpha,0}$.
Therefore, the limit of the region represented by~\eqref{eq:main inequality (i)} is contained in $\ker G_{\alpha,0}$. If the condition
\begin{equation}
    \bs{\gamma} \in \ker G_{\alpha,0} = \mathrm{span} \left\{ \begin{pmatrix}
        1 \\ 0 \\ 1 \\ 0
    \end{pmatrix}, \begin{pmatrix}
        0 \\ 1 \\ 0 \\ 1
    \end{pmatrix}\right\}
\end{equation}
holds, then we have
\begin{equation}
    \gamma_{11} = \gamma_{12}, \quad \gamma_{21} = \gamma_{22}.
    \label{eq:gamma_11=gamma_12,gamma_21=gamma_22}
\end{equation}
Substituting~\eqref{eq:gamma_11=gamma_12,gamma_21=gamma_22} into~\eqref{eq:main inequality}, we obtain
\begin{equation}
    \sqrt{
        \frac{2}{1 + \cos \beta} (\gamma_{11}^2 + \gamma_{21}^2 - 2 \gamma_{11} \gamma_{21}\cos \alpha)
    }
    \leq \sin \alpha.
    \label{eq:inequality in ker G_alpha0}
\end{equation}
Therefore, taking the limit $\beta \to 0$ in~\eqref{eq:inequality in ker G_alpha0} recovers~\eqref{eq:main inequalitiy (B_1 parallel B_2)} with $s_b=1$.
Similarly, taking the limit of $\beta \to \pi$ in~\eqref{eq:inequality in ker G_alpha0} yields~\eqref{eq:main inequalitiy (B_1 parallel B_2)} with $s_b=-1$.
the case (iii) can be addressed in a similar manner. 

\subsection{Remarks}
Our results depend on the angle $\alpha$ and $\beta$ between the pairs of the observables $(\tilde{A}_1,\tilde{A}_2)$ and $(\tilde{B}_1,\tilde{B}_2)$, respectively; however, they are independent of the relative orientation between these two sets. 
This can be understood as follows.
Consider a unitary transformation $S$ that rotates the observables $\tilde{B}_1$ and $\tilde{B}_2$ while preserving the angle between them. The transformed observables are then given by
\begin{equation}
    \tilde{B}_j' = S \tilde{B}_i S^\dagger , \quad \forall j \in \{1,2\}.
\end{equation}
The correlations realizable with the observables $\tilde{B}_1,\tilde{B}_2$ and a density operator $\rho$ are equally realizable with the transformed observables $\tilde{B}_1',\tilde{B}_2'$ and the conjugated density operator $(I\otimes S) \rho (I\otimes S^\dagger)$:
\begin{equation}
    \Tr[\rho \tilde{A}_i \otimes \tilde{B}_{j}] = \Tr[(I\otimes S )\rho (I\otimes S^\dagger) \tilde{A}_i \otimes S \tilde{B}_j S^\dagger].
\end{equation} 
It follows that the set of realizable correlations is invariant under such transformations, and consequently, does not depend on the absolute orientation of $\tilde{B}_1$ relative to $\tilde{A}_1$.

% We also remark that the inequality~\eqref{eq:main inequality} constrains the correlations for observables defined on a Hilbert space of arbitrary dimension. It follows from the equivalence between conditions (ii) and (iii) in Theorem~\ref{thm:QI in (2,m,2) setup} (\cite[Theorem 2.1]{tsirelson1987quantum}) that, for any given correlations realized by observables on a Hilbert space of arbitrary dimension, there exists a set of observables and a quantum state on the Hilbert space $\mbb{C}^2 \otimes \mbb{C}^2$ that reproduces the same correlations.

The correlations $\gamma_{ij}$ of the normalized observables are related to the correlations $C_{ij} = \Tr[\rho A_i \otimes B_j]$ of the original observables $A_i$ and $B_j$ by
\begin{align*}
    \gamma_{ij}
    &= \frac{C_{ij} - a_{i0} \bra B_j \ket - b_{j0} \bra A_i \ket + a_{i0}b_{j0}}{\lVert \bs{a}_i \rVert \lVert \bs{b}_j \rVert} \\
    &= \frac{4  C_{ij} - 2\Tr A_i \bra B_j \ket  - 2\Tr B_j \bra A_i \ket + \Tr A_i \Tr B_j }{\lVert A \rVert \lVert B \rVert},
     \numberthis
    \label{eq:gamma_ij in terms of C_ij}
\end{align*}
where $\lVert X \rVert$ denotes the Hilbert--Schmidt norm of an operator $X$.
When the observables take values in $\{ \pm 1\}$, $a_{i0} = b_{j0}=0$ and $\lVert \bs{a}_i \rVert = \lVert \bs{b}_j \rVert = 1$ hold, which leads to the relation $\gamma_{ij} = C_{ij}$.
Therefore, we obtain the following corollary of Theorem~\ref{thm:main theorem}:
\begin{corollary}\label{cor:corollary of main theorem}
    Let $A_{1}$, $A_{2}$, $B_{1}$, and $B_{2}$ be observables on a two-level system taking values in $\{ \pm 1\}$, and let $C_{ij} \in \mbb{R}$, $i,j \in \{1,2\}$.
    Then, there exists a density operator $\rho$ on the Hilbert space $\mbb{C}^2 \otimes \mbb{C}^2$ of the composite system that reproduces the correlations
    \begin{equation}
        C_{ij} = \Tr[\rho A_i \otimes B_j], \quad \forall i,j \in \{1,2\}
        \label{eq:C_ij=Tr[rho A_i B_j]}
    \end{equation}
    if and only if the inequality
    \begin{equation}
        \sqrt{\bs{C}^\top F_{\alpha,\beta} \bs{C}} + \sqrt{\bs{C}^\top F_{\alpha,-\beta} \bs{C}} \leq 2
        \label{eq:main inequality (C)}
    \end{equation}
    holds, where $\bs{C} := (C_{11}, C_{21}, C_{12}, C_{22})^{\top}$.
    In particular, $\rho$ can be chosen so that $\bra A_i \ket_{\rho} = \bra B_j \ket_{\rho} = 0$ for all $i,j \in \{1,2\}$.
\end{corollary}

In addition to the assumption that all observables $A_1$, $A_2$, $B_1$, and $B_2$ take values in ${\pm 1}$, we are often interested in the case where $A_1$ and $A_2$ anti-commute, and so do $B_1$ and $B_2$:
\begin{equation}
A_1 A_2 = - A_2 A_1, \quad B_1 B_2 = - B_2 B_1.
\end{equation}
A typical example of such observables is given by $A_1 = \sigma_x$, $A_2 = \sigma_y$, $B_1 = \sigma_x$, and $B_2 = \sigma_y$.
In this case, the parameter values $\alpha = \beta = \frac{\pi}{2}$ are realized.
This leads to the following corollary.
\begin{corollary}
    Let $A_1$, $A_2$, $B_1$, and $B_2$ be observables on a two-level system taking values in $\{\pm 1\}$, and assume that $A_1$ and $A_2$, as well as $B_1$ and $B_2$, are respectively anti-commuting. Then, there exists a density operator $\rho$ on the Hilbert space $\mbb{C}^2 \otimes \mbb{C}^2$ of the composite system that satisfies~\eqref{eq:C_ij=Tr[rho A_i B_j]} for a given set of $C_{ij} \in \mbb{R} \, (i,j \in \{1,2\})$ if and only if the following inequality holds:
    \begin{multline}
        \sqrt{
            (C_{11}+C_{22})^2 + (C_{12}-C_{21})^2
        } \\
        + \sqrt{
            (C_{11}-C_{22})^2 + (C_{12}+C_{21})^2
        }
        \leq 2.
        \label{eq:main inequality for anti-commuting observables}
    \end{multline}
    In particular, $\rho$ can be chosen so that $\bra A_i \ket_{\rho} = \bra B_j \ket_{\rho} = 0$ for all $i,j \in \{1,2\}$.
\end{corollary}

%%%%%%%%%%%%%%%%%%%%%%%%%%%%%%%%%%%%%%%%%%%%%%%%%%%%%%%%%%%%%%%
%%%%%%%%%%%%%%%%%%%%%%%%%%%%%%%%%%%%%%%%%%%%%%%%%%%%%%%%%%%%%%%
%%%%%%%%%%%%%%%%%%%%%%%%%%%%%%%%%%%%%%%%%%%%%%%%%%%%%%%%%%%%%%%
\section{Comparison to related works \label{sec:comparison}}
We compare the results obtained in Section~\ref{sec:quantum restrictions in the CHSH setting} with relevant previous works. In this section, we assume that $A_i$ and $B_j$ take values from $\{\pm 1\}$, and thus $\gamma_{ij}=C_{ij}=\Tr[\rho A_i B_j]$ for $i,j \in \{1,2\}$. In particular, under this setting, we can apply Corollary~\ref{cor:corollary of main theorem} directly.

\subsection{CHSH Inequalities}
The CHSH inequalities~\eqref{eq:CHSH ineq} are linear in the correlation coefficients. In contrast, for a fixed set of observables $\bs{X} = (A_1, A_2, B_1, B_2)$, our inequality~\eqref{eq:main inequality (C)} cannot be represented as a finite collection of linear inequalities.

Correlations satisfying~\eqref{eq:main inequality (C)} can violate the CHSH inequalities. 
For example, if $C_{11}=C_{12}=C_{21}=-C_{22}=\frac{\sqrt{2}}{2}$, then the equality in~\eqref{eq:main inequality for anti-commuting observables}, \ie,~\eqref{eq:main inequality (C)} with $\alpha=\beta=\frac{\pi}{2}$, is satisfied, while one of the CHSH inequalities is violated: $C_{11}+C_{12}+C_{21}-C_{22}=2\sqrt{2}>2$.
Therefore, quantum correlations that cannot be described by local realism exist.
% Furthermore, for arbitrary angles $\alpha \in (0,\pi)$ and $\beta \in (0,\pi)$, there exist correlations satisfying~\eqref{eq:main inequality (C)} that violate the CHSH inequalities~\eqref{eq:CHSH ineq C}, since the joint-spectral measure

On the other hand, correlations satisfying the CHSH inequalities~\eqref{eq:CHSH ineq} can also violate the quantum restriction~\eqref{eq:main inequality (C)} for a given set $\bs{X}$ of observables.
For example, if $C_{11}=C_{12}=1$ and $C_{21}=C_{22}=0$, the CHSH inequalities~\eqref{eq:CHSH ineq} are satisfied, whereas~\eqref{eq:main inequality for anti-commuting observables} is violated. Therefore, the correlations cannot be reproduced by a quantum state and observables with $\alpha=\beta=\frac{\pi}{2}$. 
This implies that if we fix a set $\bs{X}$ of observables, local realistic correlations that cannot be described by quantum mechanics exist.
A similar observation was made in a more complicated setting \cite{isobe2010method}, whereas our result demonstrates it in a simpler setup.

\subsection{Tsirel'son's Inequalities}
The validity of Tsirel'son's inequalities~\eqref{eq:Tsirel'son ineq} is a necessary condition on correlations for the existence of a quantum state that reproduces given correlations, while our inequality~\eqref{eq:main inequality (C)} provides a necessary and sufficient condition for the existence of such a state, given an arbitrary set of measured observables. 
Hence, correlations satisfying~\eqref{eq:main inequality (C)} must also satisfy Tsirel'son's inequalities~\eqref{eq:Tsirel'son ineq}.

Although we have not yet succeeded in deriving Tsirel'son's inequalities~\eqref{eq:Tsirel'son ineq} directly from~\eqref{eq:main inequality (C)} in the general setting, Tsirel'son's inequalities~\eqref{eq:Tsirel'son ineq} follow from~\eqref{eq:main inequality for anti-commuting observables}, \ie,~\eqref{eq:main inequality (C)} with $\alpha=\beta=\frac{\pi}{2}$. For example, four of Tsirel'son's inequalities~\eqref{eq:Tsirel'son ineq} are derived as follows:
\begin{align*}
    2&\geq   (\text{l.h.s. of~\eqref{eq:main inequality for anti-commuting observables}}) \\
    &\geq \sqrt{(C_{11}+C_{22})^2 + (C_{12}-C_{21})^2} \\
    &\geq \sqrt{2} \cdot \frac{|C_{11} + C_{22}| + |C_{12}-C_{21}|}{2}\\
    &\geq \frac{| C_{11}+ C_{22} \pm (C_{12} - C_{21})|}{\sqrt{2}}, \numberthis
\end{align*}
where the third inequality follows from the concavity of the square root function.

\subsection{Tsirel'son--Landau Inequality}

Tsirel'son \cite{tsirelson1987quantum} addressed the quantum problem in a setting that includes the $(2,m,2)$ setup, and provided a concrete expression for the solution in the CHSH setup.
\begin{theorem}[{\cite[Theorem 2.2]{tsirelson1987quantum}}]\label{thm:QI in CHSH setup}
    Given real numbers $C_{ij}$ satisfying $|C_{ij}| \leq 1$, $i,j \in \{1,2\}$, a density operator $\rho \in \mcal{D}(\mbb{C}^2 \otimes \mbb{C}^2)$ and a set $(A_1\otimes I_2, A_2\otimes I_2, I_2 \otimes B_1, I_2 \otimes B_2)$ of binary observables taking values in $\{\pm 1\}$ that satisfies $C_{ij} = \Tr[\rho A_i \otimes B_j]$ exist if and only if at least one of the following two inequalities holds:
    \begin{gather}
            \begin{aligned}
                 0 \leq 
                (C_{12}C_{21}- C_{11} C_{22}) (C_{11}C_{12} &- C_{21}C_{22}) \\
                &\times (C_{11}C_{21}- C_{12}C_{22}) 
            \end{aligned}
            \notag \\
            \le \frac{1}{4}\left(\sum_{i,j }  {C_{ij}^2}\right)^2 - \frac{1}{2} \sum_{i,j} {C_{ij}^4} - 2  \prod_{i,j }C_{ij}, \\
            0  \leq  2\max_{i,j } C_{ij}^4 - \left(\max_{i,j }C_{ij}^2\right)\sum_{i,j }  C_{ij}^2  +  2\prod_{i,j }C_{ij}.
        \end{gather}
\end{theorem}

In another context, Landau \cite{landau1988empirical} derived a necessary condition on a quantum distribution in the $(2,2,d)$ setup:
\begin{multline}
    \big|D_{11} D_{12} - D_{21} D_{22} \big| \\
    \leq \sqrt{1-D_{11}^2} \sqrt{1-D_{12}^2} + \sqrt{1-D_{21}^2} \sqrt{1-D_{22}^2},
    \label{eq:Landau's inequality (original)}
\end{multline}
where
\begin{equation}
    D_{ij} := \frac{\bra A_i B_j \ket_{\rho}}{\bra A_i^2 \ket_\rho \bra B_j^2 \ket_\rho}, \quad \forall i,j \in \{1,2\}.
\end{equation}
{Note that $A_1,A_2,B_1,B_2$ are not necessarily binary observables.}
Landau's inequality~\eqref{eq:Landau's inequality (original)} can be equivalently expressed as follows:
\begin{equation}
    \left|\sum_{i,j \in \{1,2\}} \arcsin D_{ij}  -2 \arcsin D_{kl}\right| \leq \pi, \quad k,l \in \{1,2\}.
    \label{eq:Landau's inequality (symmetric)}
\end{equation}
Although the symmetry under the exchange of Alice and Bob is not apparent in~\eqref{eq:Landau's inequality (original)}, it becomes evident in~\eqref{eq:Landau's inequality (symmetric)}.

If the observables $A_1,A_2,B_1,B_2$ are binary, taking values in $\{\pm 1\}$, then $D_{ij}=C_{ij}$ for all $i,j \in \{1,2\}$.
In this case,~\eqref{eq:Landau's inequality (original)} reduces to
\begin{multline}
    \big| C_{11}C_{12} - C_{21} C_{22} \big| \\
    \leq \sqrt{1-C_{11}^2} \sqrt{1-C_{12}^2} + \sqrt{1-C_{21}^2} \sqrt{1-C_{22}^2},
    \label{eq:Landau's inequality for pm1 observables (original)}
\end{multline}
or equivalently
\begin{equation}
    \left| \sum_{i,j \in \{1,2\}}\arcsin C_{ij} -2\arcsin C_{kl} \right| \le \pi, \quad \forall k,l \in \{1,2\}.
    \label{eq:Landau's inequality for pm1 observables (symmetric)}
\end{equation}
Although~\eqref{eq:Landau's inequality for pm1 observables (original)} and~\eqref{eq:Landau's inequality for pm1 observables (symmetric)} are derived as necessary conditions on a quantum distribution, they are actually equivalent to the conditions stated in Theorem~\ref{thm:QI in CHSH setup} \cite{tsirelson1993some} (see Appendix~\ref{sec:equivalence of Tsirel'son--Landau}), which provides a necessary and sufficient condition for the existence of a quantum state and observables.
In this sense, we refer to~\eqref{eq:Landau's inequality for pm1 observables (original)} (or~\eqref{eq:Landau's inequality for pm1 observables (symmetric)}) as the Tsirel'son--Landau inequality (or inequalities).

% \subsection{Unfixed Observables Settings}

In contrast, Corollary~\ref{cor:corollary of main theorem} provides a solution to the quantum problem for each fixed set of measured observables. Therefore, the union of the regions of correlations represented by~\eqref{eq:main inequality (C)}, taken over all possible sets of observables, must coincide with the region described by the Tsirel'son--Landau inequality (Theorem~\ref{thm:QI in CHSH setup}).

Unlike the region of correlations given by~\eqref{eq:main inequality (C)} for each set of observables, the region described the Tsirel'son--Landau inequality encompasses the region represented by the CHSH inequalities~\eqref{eq:CHSH ineq}.
This is not a trivial fact, since the dimension of the Hilbert space of the composite system is restricted to $4$ in Theorem~\ref{thm:QI in CHSH setup}.
If a Hilbert space of arbitrary dimension were allowed, it would be easy to see that any local realistic correlation could also be realized in quantum theory. As a trivial example, given a joint probability distribution $P$ for a set of four observables $\bs{X}=(A_1,A_2,B_1,B_2)$, where each $X_i$ takes values in $V_i=\{v_{i,k}\}_{k\in[d]}$, one can construct a density operator $\rho$ and a set of operators on a Hilbert space of dimension $d^4$ as follows:
\begin{align}
    \rho &:= \sum_{\bs{k} \in [d]^4} P(\bs{k})|\bs{k}\ket \bra \bs{k}|, \label{eq:density operator that reproduces LR}\\
    X_i &:= \sum_{\bs{k} \in [d]^4} v_{i,k_i} |\bs{k}\ket \bra \bs{k} |, \quad i \in \{1,2,3,4\},\label{eq:observables that reproduce LR}
\end{align}
where $|\bs{k}\ket := |k_1\ket \otimes |k_2 \ket\otimes |k_3\ket \otimes |k_4 \ket$, and $\{|k\ket\}_{k \in [d]}$ is an orthonormal basis of $\mbb{C}^{d}$.
However, the Hilbert space $\mbb{C}^2 \otimes \mbb{C}^2$, whose dimension is $2^2=4$, is sufficient to reproduce any local realistic correlations of binary observables. 
This follows from Theorem 2.1 in \cite{tsirelson1987quantum}. See Appendix~\ref{sec:more on type I} for details.

\begin{figure*}[htbp]
    \begin{minipage}[b]{\linewidth}
        \centering
        
        \includegraphics[width=180mm]{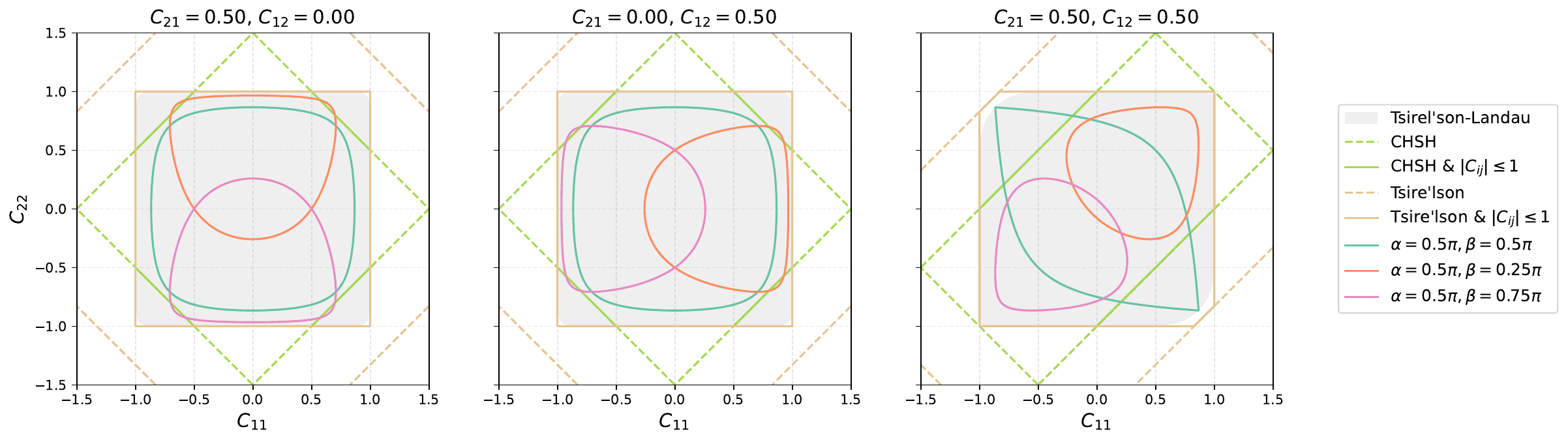}
        \subcaption{$C_{11}$--$C_{22}$ cross sections.}
        \label{fig:C11-C22}
    \end{minipage}
    \begin{minipage}[b]{\linewidth}
        \centering
        \includegraphics[width=180mm]{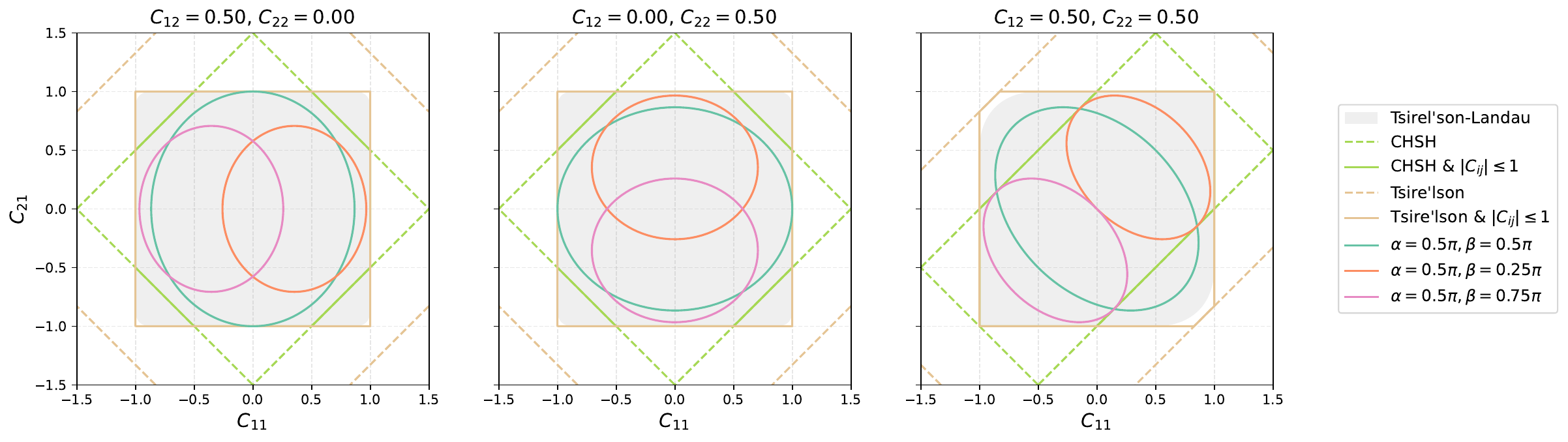}
        \subcaption{$C_{11}$--$C_{21}$ cross sections.}
        \label{fig:C11-C21}
    \end{minipage}

    \begin{minipage}[b]{\linewidth}
        \centering
        \includegraphics[width=180mm]{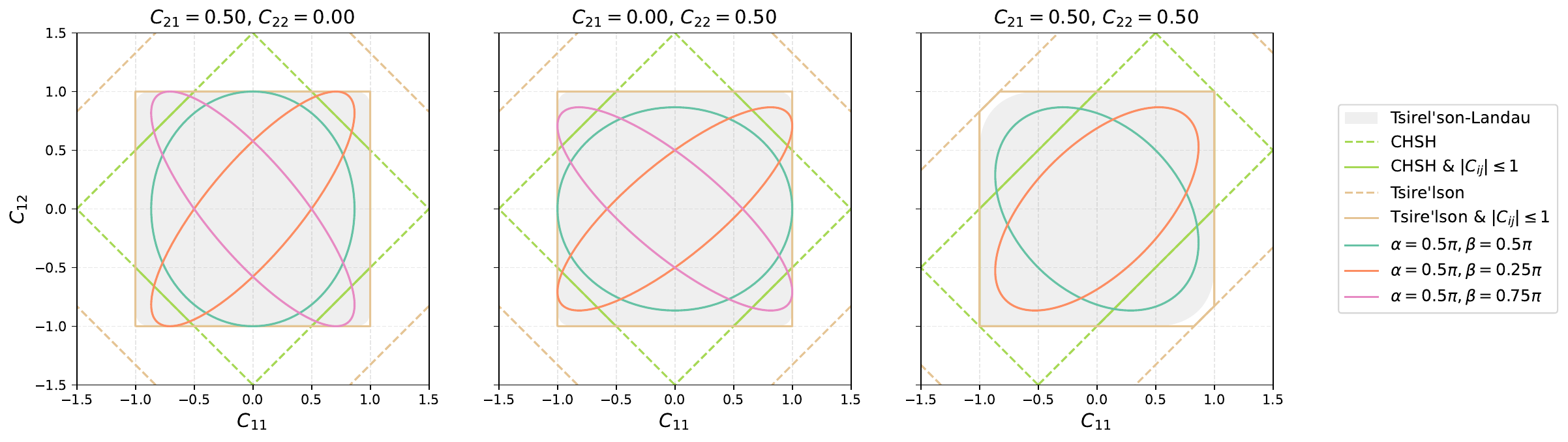}
        \subcaption{$C_{11}$--$C_{12}$ cross sections. The region represented by our inequality for $\alpha = \frac{\pi}{2}, \beta = \frac{3\pi}{4}$ has no intersection with the subspace $C_{21}=C_{22}=\frac{1}{2}$.}
        \label{fig:C11-C12}
    \end{minipage}
    \caption{Two-dimensional cross sections of the regions defined by our inequality~\eqref{eq:main inequality (C)} for $(\alpha,\beta)=\big(\frac{\pi}{2},\frac{\pi}{2}\big),\big(\frac{\pi}{2},\frac{\pi}{4}\big),\big(\frac{\pi}{2},\frac{3\pi}{4}\big)$, the Tsirel'son--Landau inequality~\eqref{eq:Landau's inequality for pm1 observables (original)}, the CHSH inequalities~\eqref{eq:CHSH ineq}, and Tsirel'son's inequalities~\eqref{eq:Tsirel'son ineq}. The regions defined by our inequality for  $(\alpha,\beta)=\big(\frac{\pi}{2},\frac{\pi}{2}\big),\big(\frac{\pi}{2},\frac{\pi}{4}\big),\big(\frac{\pi}{2},\frac{3\pi}{4}\big)$ are represented by the region enclosed by the teal, orange, and magenta curves, respectively. The Tsirel'son--Landau inequality is represented by the light gray region, the CHSH inequalities by the region enclosed by green dashed line segments, and the Tsirel'son inequalities by the region enclosed by the beige dashed line segments. The boundary of the intersection of CHSH and $|C_{ij}|\leq 1$ constraints is shown by the solid green line segments, while that of Tsirel'son and $|C_{ij}|\leq 1$ constraints is depicted by the solid beige line segments.}
    \label{fig:cross sections}
\end{figure*}

Two-dimensional cross-sections of the regions defined by our inequality~\eqref{eq:main inequality (C)} for $(\alpha,\beta)=\big(\frac{\pi}{2},\frac{\pi}{2}\big),\big(\frac{\pi}{2},\frac{\pi}{4}\big),\big(\frac{\pi}{2},\frac{3\pi}{4}\big)$, the Tsirel'son--Landau inequality~\eqref{eq:Landau's inequality for pm1 observables (original)}, the CHSH inequalities~\eqref{eq:CHSH ineq}, and Tsirel'son's inequalities~\eqref{eq:Tsirel'son ineq} are illustrated in Fig.~\ref{fig:cross sections}.
Each region defined by our inequality is contained within the region described by the Tsirel'son--Landau inequality, which, in turn, lies within the region defined by the Tsirel'son inequality. Moreover, within the domain where $|C_{ij}| \leq 1$, the region described by the Tsirel'son--Landau inequality fully encompasses the region defined by the CHSH inequalities.
% Our inequality is represented by the light blue region, the Tsirel'son--Landau inequality by the green region and its interior, and the CHSH inequalities by the region enclosed by red dashed line segments.
The six two-dimensional cross-sections are classified into three categories based on the symmetry under the exchange of $A_1$ and $A_2$, as well as $B_1$ and $B_2$: (i) $C_{11}$--$C_{22}$ and $C_{21}$--$C_{12}$, (ii) $C_{11}$--$C_{21}$ and $C_{12}$--$C_{22}$, (iii) $C_{11}$--$C_{12}$ and $C_{21}$--$C_{22}$. 
Furthermore, if $\alpha=\beta$, categories (ii) and (iii) are equivalent due to the symmetry under the exchange of Alice and Bob.

\subsection{Bell's Original Setting}

Bell's original setting \cite{bell1964einstein} can be interpreted as a special case of the $(2,2,2)$ setup with an additional assumption.
Assume that a hidden variable $\lambda \in \Lambda$ determines the value $A(\bs{a},\lambda) \in \{\pm 1\}$ of Alice's observable and the value $B(\bs{b},\lambda) \in \{\pm 1\}$ of Bob's observable, where $\bs{a},\bs{b}\in \mbb{R}^3$ are vectors that specify the observables.
Letting $w(\lambda)$ be the probability distribution of $\lambda$, the correlation of these observables is given by
\begin{equation}
    C(\bs{a},\bs{b})
    := \int dw(\lambda) A(\bs{a},\lambda) B(\bs{b},\lambda).
\end{equation}
In Bell's original setting \cite{bell1964einstein}, Alice's observable corresponding to a vector $\bs{b}$ and Bob's observable corresponding to the same vector $\bs{b}$ are assumed to be perfectly anti-correlated:
\begin{equation}
    A(\bs{b},\lambda) = - B(\bs{b},\lambda),
    \label{eq:anti-correlated}
\end{equation}
or equivalently
\begin{equation}
    C(\bs{b},\bs{b}) = -1.
    \label{eq:anti-correlated correlation}
\end{equation}
Bell \cite{bell1964einstein} imposed this condition on all vectors $\bs{b}$, but we point out that it is sufficient for deriving Bell's original inequality to assume~\eqref{eq:anti-correlated} for a single specific vector $\bs{b}$.
In quantum mechanics, such a state exists for any given vector $\bs{b}$.
Under the above assumptions, Bell's original inequality takes the following form:
\begin{equation}
    \big| C(\bs{a},\bs{b}) - C(\bs{a},\bs{c})\big| \leq 1 + C(\bs{b},\bs{c}).
    \label{eq:Bell's original inequality}
\end{equation}

To formulate the corresponding quantum problem, we associate the vectors $\bs{a},\bs{b},\bs{b},\bs{c}$ with the observables $A_1,A_2,B_1,B_2$, respectively. 
The assumption~\eqref{eq:anti-correlated correlation} corresponds to the condition $C_{21} = -1$. 
Therefore, given the condition $C(\bs{b},\bs{b})=C_{21} = -1$, there exists a density operator $\rho$ that satisfies $\Tr[\rho A_1 B_1]=C(\bs{a},\bs{b})=C_{11}$, $\Tr[\rho A_1 B_2] = C(\bs{a},\bs{c})=C_{12}$, and $\Tr[\rho A_2 B_2] = C(\bs{b},\bs{c})=\gamma_{22}$ if and only if the following inequality holds:
\begin{gather}
    \sqrt{\tilde{\bs{C}}^\top F_{\alpha,\beta} \tilde{\bs{C}}} 
    +  \sqrt{\tilde{\bs{C}}^\top F_{\alpha,-\beta} \tilde{\bs{C}}} 
    \leq 2,\label{eq:main inequality applied to Bell's setup} \\
    \tilde{\bs{C}} := (C_{11},-1,C_{12},C_{22})^\top.
\end{gather}

For instance, consider the case where $\alpha=\beta=\frac{\pi}{4}$. 
Then, our inequality~\eqref{eq:main inequality applied to Bell's setup} holds with equality when $C_{11}=C_{22}=-\frac{1}{\sqrt{2}}$ and $-1 \leq C_{12} \leq 0$.
However, Bell's original inequality~\eqref{eq:Bell's original inequality} is violated if $C_{11}=C_{22}=-\frac{1}{\sqrt{2}}$ and $-\sqrt{2}+1 < C_{12} \leq 1$.

Note that for an arbitrary set of observables, a quantum state that violates Bell's original inequality~\eqref{eq:Bell's original inequality} does not necessarily exist.
Indeed, when the angles are set to $\alpha = \beta = \frac{\pi}{2}$, all correlations that satisfy our inequality~\eqref{eq:main inequality applied to Bell's setup} also satisfy Bell's original inequality, and hence no violation occurs.

\section{Summary and discussion \label{sec:summary and discussion}}

In this paper, we investigated the problems of determining the set of realizable distributions in quantum mechanics under given conditions.
We introduced two settings: one in which the measurement observables are left unfixed, and another in which they are arbitrarily fixed in advance.
The setting with unfixed observables can be understood as the logical disjunction of all possible fixed observables settings.
% In Section~\ref{sec:preliminaries}, we introduced the necessary mathematical tools for our analysis, including the Fourier transform, marginalization, and quasi-joint probability distributions. The characteristic function, defined as the Fourier transform of a distribution, is particularly useful when marginal distributions are specified.
% Quasi-joint probability distributions allow us to treat the quantum case in close analogy with the local realistic one.
% Furthermore, we presented a symmetry-based technique for reducing the number of degrees of freedom, which enables significant simplification of the problem while preserving its generality.

In Section~\ref{sec:quantum restrictions in the CHSH setting}, we derived a necessary and sufficient condition for the existence of a quantum state that satisfies given correlation constraints for a fixed set of normalized observables in the CHSH setup.
This result has practical significance, as it can be directly applied to real experimental settings where the available observables are limited. 
Note that while we considered the setting in which only the correlation constraints are given, the case where the full marginal distributions are specified remains an open problem.
An important insight for addressing this problem is that the mapping $T_2$ used in the proof of Theorem~\ref{thm:main theorem} reduces six degrees of freedom while preserving all expectation values and correlations.
Moreover, even when only the correlation constraints are specified in the CHSH setup, characterizing the set of quantum correlations for a given set of observables on a Hilbert space of dimension higher than $\mathbb{C}^2 \otimes \mathbb{C}^2$ remains unresolved.

The proof of Theorem~\ref{thm:main theorem} is based on the use of characteristic functions and quasi-joint probability distributions. 
Quasi-joint probability distributions enable a unified treatment of quantum and local realistic scenarios and the use of characteristic function allows us to reformulate the problem in terms of decoupled algebraic constraints.
A symmetry-based reduction technique significantly simplifies the analysis while preserving generality.

As observed in Section~\ref{sec:comparison},
our inequality~\eqref{eq:main inequality (C)} strengthens previously known constraints, such as Tsirel'son's inequalities~\eqref{eq:Tsirel'son ineq} and the Tsirel'son--Landau inequality~\eqref{eq:Landau's inequality for pm1 observables (original)}, by providing an explicit characterization for each given set of binary observables.
The union of the regions of correlations represented by our inequality, taken over all possible sets of observables, must coincide with the region described by the Tsirel'son--Landau inequality when the observables take values in $\{\pm 1\}$.
While allowing a Hilbert space of arbitrarily large dimension and all possible observables ensures that any local realistic distribution can be reproduced in quantum mechanics, our results show that this is not necessarily the case for a fixed set of observables. 
We also observed that our results are applicable to Bell's original setting and that the possibility of violating Bell's original inequality crucially depends on the choice of observables.

There are several directions for future research.
First, deriving explicit inequalities representing the statistical constraints in local realism or quantum mechanics under the general $(n,m,d)$ setup remains an open question.
To this end, combining our method with convex optimization techniques and semi-definite programming could be a promising avenue for further exploration.
As another direction, considering the statistical constraints for a given set of measured observables in local realism could lead to a more comprehensive understanding of the quantum--classical correspondence.
Although there is no canonical way to associate quantum observables with classical observables, finding the statistical constraints for a given set of measured observables in local realism and comparing them with those in quantum mechanics may allow us to systematically identify corresponding observables.

\section*{Acknowledgement}
{R.\,N. would like to thank Shogo Tanimura for his valuable advice and insightful discussions, and Gen Kimura and Yuriko Yamamoto for introducing relevant references and providing helpful comments.
This work is supported by the Make New Standard Program for the Next Generation Researchers (R.\,N.) and JSPS Grant-in-Aid for Scientific Research (KAKENHI), Grant Number JP22K13970 (J.\,L.), which provided financial assitance.}

\appendix

%%%%%%%%%%%%%%%%%%%%%%%%%%%%%%%%%%%%%%%%%%%%%%%%%%%%%%%%%%%%%%%%
%%%%%%%%%%%%%%%%%%%%%%%%%%%%%%%%%%%%%%%%%%%%%%%%%%%%%%%%%%%%%%%%
%%%%%%%%%%%%%%%%%%%%%%%%%%%%%%%%%%%%%%%%%%%%%%%%%%%%%%%%%%%%%%%%

\section{Problems of Determining the Realizable Data \label{sec:realizability problems}}

%%%%%
\subsection{Sets of Probability Distributions}
%%%%%

Let $\vect{\Omega} := \Omega_1 \times \cdots \times \Omega_n$ be the product of outcome spaces corresponding to a tuple
\begin{equation}
\vect{X} := (X_{1},\ldots,X_{n})
\end{equation}
of simultaneously measurable observables.
Let $\mathcal{P}(\vect{\Omega})$ denote the (convex) set of all probability distributions defined on $\vect{\Omega}$.

Define tuples
\begin{equation}
\vect{X}_{\bullet} := (\vect{X}_{1},\ldots,\vect{X}_{N}),\quad \vect{\Omega}_{\bullet} := (\vect{\Omega}_{1} , \ldots , \vect{\Omega}_{N}),
\end{equation}
where each $\vect{X}_i$ is a tuple of simultaneously measurable observables taking values in $\vect{\Omega}_i$.
Let $\mathcal{P}(\vect{\Omega}_{\bullet}):=\bigoplus_{i =1}^N \mathcal{P}(\vect{\Omega}_i)$ denote the (convex) set of all the tuples
\begin{equation}
(P_{\vect{X}_1}, \ldots, P_{\vect{X}_N}),
\end{equation}
where $P_{\vect{X}_i} \in \mathcal{P}(\vect{\Omega}_i)$ for each $i=1,\ldots,N$. 
Such tuples naturally arise from repeated measurements performed under different settings $\vect{X}_i$, each yielding a joint probability distribution over $\vect{\Omega}_i$.

%Note that
%\begin{equation}
%\mathcal{P}(\vect{X}_{\bullet}) \subset \bigoplus_{\vect{X} \in \vect{X}_{\bullet}} \mathcal{P}(\vect{X}).
%\end{equation}

%%%%%
\subsection{Experimental Data}
%%%%%
% \begin{itemize}
%\item The set $\vect{X}_{0}$ of all the observables concerned, and the set of (compatible) combinations of the observables for which one collects experimental data:
%\begin{equation}
%\vect{X}_{\bullet} = \{\vect{X}_{1}, \ldots, \vect{X}_{k}\} \subset \mathfrak{P}(\vect{X}_{0}).
%\end{equation}
%Note that $\vect{X}_{n} \subset \vect{X}_{0}$ with this specific notation (maximum).  Meanwhile, it could be that $\vect{X}_{0} \notin \vect{X}_{\bullet}$.  In general, $\vect{X}_{\bullet}$ may not be a lattice.
% \item 
To specify which aspects of the measured probability distributions are retained as experimental data, we introduce a recording map 
% Recording map (data map):
\begin{equation}
R : \mathcal{P}(\vect{\Omega}_{\bullet}) \to \Gamma,
\end{equation}
where $\Gamma$ is a target space representing the data recorded in an experiment. 

If $R$ is \textit{faithful}, i.e., injective (including the case where $R$ is the identity map), then all information in the joint probability distributions is preserved.  
Otherwise, the recorded data $\gamma \in \Gamma$ represents only partial information extracted from the full distributions.

% \end{itemize}

%%%%%
\subsection{Underlying Statistics}
%%%%%

\subsubsection{Hidden Variable Theory}

In the hidden variable framework, let $\mcal{P}(\Lambda)$ denote the convex set of all probability distributions defined on the space of hidden variables $\Lambda$.
For a tuple of random variables $\vect{X}(\lambda) = (X_{1}(\lambda), \ldots, X_{n}(\lambda))$, we define the associated mapping
\begin{align}
    \Phi_{\vect{X}} :
        {}&\mathcal{P}(\Lambda) \to \mathcal{P}(\vect{\Omega}), \notag \\
        {}&w(\lambda) \mapsto P_{\vect{X}}(\vect{x}) \defeq \int_{\vect{X}^{-1}(\vect{x})} \,dw(\lambda),
\end{align}
which induces a joint probability distribution of the observables $X_1,\ldots,X_n$.

We also consider a tuple of such observable-tuples $\vect{X}_{\bullet}(\lambda) = (\vect{X}_{1}(\lambda), \ldots, \vect{X}_{N}(\lambda))$, and define the corresponding mapping
\begin{align}
    \Phi_{\vect{X}_{\bullet}} :
        {}&\mathcal{P}(\Lambda) \to \mathcal{P}(\vect{\Omega}_{\bullet}), \notag\\
        {}&w(\lambda) \mapsto (P_{\vect{X}_{1}}(\vect{x}_{1}), \ldots, P_{\vect{X}_{N}}(\vect{x}_{N})).
\end{align}

\subsubsection{Quantum Theory}

In the quantum case, let $\mcal{D}(\cal{H})$ denote the convex set of all density operators on a Hilbert space $\mcal{H}$. 
For a tuple of compatible quantum observables $\vect{X} = (X_{1}, \ldots, X_{n})$, represented by commutative self-adjoint operators, the induced mapping is defined as
\begin{align}
    \Phi_{\vect{X}} :
        {}&\mcal{D}(\mcal{H}) \to \mathcal{P}(\vect{\Omega}), \notag \\
        {}&\rho \mapsto P_{\vect{X}}(\vect{x}) \defeq \Tr[X_{1}({x}_1) \cdots X_{n}({x}_n)\rho],
\end{align}
where $E_{X_i}$ denotes the spectral measure of the operator $X_i$.

For a tuple of such tuples $\vect{X}_{\bullet} = (\vect{X}_{1}, \ldots, \vect{X}_{N})$, we define the corresponding mapping
\begin{align}
    \Phi_{\vect{X}_{\bullet}} :
        {}&\mcal{D}(\mcal{H}) \to \mathcal{P}(\vect{\Omega}_{\bullet}), \notag\\
        {}&\rho \mapsto (P_{\vect{X}_{1}}(\vect{x}_{1}), \ldots, P_{\vect{X}_{N}}(\vect{x}_{N})),
\end{align}
which collects the joint distributions corresponding to each measurement context in $\vect{X}_{\bullet}$.

%%%%%
\subsection{Realizability Problems}
%%%%%

Consider a tuple
$\vect{\Omega}_{\bullet}$
of tuples of outcome spaces and a recording map $R : \mathcal{P}(\vect{\Omega}_{\bullet}) \to \Gamma$.  

\begin{problem}{R}[Realism (Hidden Variable Theory)]
Find a necessary and sufficient condition on $\gamma \in \Gamma$ for the existence of a hidden variable model $(\Lambda, \vect{X}_{\bullet}(\lambda))$ with the sample spaces $\bs{\Omega}_{\bullet}$ such that
\begin{equation}
\ran{\left( \Phi_{\vect{X}_{\bullet}} \right)} \cap R^{-1}(\gamma) \neq \emptyset \iff \gamma \in \ran{(R \circ \Phi_{\vect{X}_{\bullet}})}
\end{equation}
holds.  
\end{problem}
This problem can equivalently be formulated in terms of the existence of a joint probability distribution over all the observables involved.
Let 
\begin{equation}
    \bs{X}_0(\lambda) := \bigcup_{i=1}^{N} \bs{X}_i(\lambda)
\end{equation}
denote the set of all the observables under consideration, and let $\bs{\Omega}_{0}$ be the sample space of $\bs{X}_0$.
Define the marginalization map
\begin{equation}
    \mathsf{Marg} : \mathcal{P}(\bs{\Omega}_0) \to \mathcal{P}(\vect{\Omega}_{\bullet}),
\end{equation}
which sends each joint distribution on $\bs{\Omega}_0$ to its family of marginals on each measurement context $\bs{X}_i$.

Then, Problem R is equivalent to the following:
Find a necessary and sufficient condition on $\gamma \in \Gamma$ under which there exists a joint probability distribution $P \in \mathcal{P}(\bs{\Omega}_0)$ such that the corresponding marginals satisfy the data constraint, i.e.,
\begin{equation}
    \ran(\mathsf{Marg}) \cap R^{-1}(\gamma) \neq 0 \iff \gamma \in \ran(R \circ \mathsf{Marg}).
\end{equation}

\begin{problem}{Q}[Quantum Theory]
Find a necessary and sufficient condition on $\gamma \in \Gamma$ for the existence of a quantum model $(\mathcal{H}, \vect{X}_{\bullet})$ with the sample spaces $\bs{\Omega}_{\bullet}$ such that
\begin{equation}
\ran{\left( \Phi_{\vect{X}_{\bullet}} \right)} \cap R^{-1}(\gamma) \neq \emptyset \iff \gamma \in \ran{(R \circ \Phi_{\vect{X}_{\bullet}})}
\end{equation}
holds.
\end{problem}

%%%%%
\subsection{Locality}
%%%%%
{
Locality is an assumption shared by both local realism and quantum mechanics.  
It plays a crucial role in determining which observables are jointly measurable in composite systems involving spatially separated parties.  
For simplicity, we illustrate this in the $(2, m, d)$ setup; the generalization to the $(n, m, d)$ case is straightforward.

Suppose that Alice can measure one of the observables $A_1,\ldots,A_m$, each $A_i$ taking values in a $d$-element set $V_{i}$, and that Bob can measure one of $m$ observables $B_1,\ldots,B_m$, where each $B_j$ takes values in a $d$-element set $W_{j}$.
By the assumption of locality, each pair $\vect{X}_{ij} := (A_i, B_j)$ forms a set of jointly measurable observables for all $i,j \in \{1,\ldots,m\}$.  
In each measurement context, a joint probability distribution $P_{A_i B_j}$ on the space $\vect{\Omega}_{ij}:=V_{i} \times W_{j}$ is obtained.

Collecting the data from all measurement contexts yields the full set of experimentally accessible information, represented by the tuple $(P_{A_i B_j})_{i,j=1}^{m} \in \mcal{P}(\vect{\Omega}_{\bullet})$, where  
\begin{equation}
    \vect{X}_{\bullet} := (\vect{X}_{ij})_{i,j=1}^{m},\quad 
    \vect{\Omega}_{\bullet} := (\vect{\Omega}_{ij})_{i,j=1}^{m}.
    \label{eq:tuple of jointly measurable observables}
\end{equation}
Given a recording map $R:\mcal{P}(\vect{\Omega}_{\bullet}) \to \Gamma$, the recorded data is denoted by  
\begin{equation}
    \gamma := R\left((P_{A_i B_j})_{i,j=1}^{m}\right).
\end{equation}

Based on this structure, we now introduce realizability problems under the locality assumption.
The problems under locality, denoted as Problems {LR} (Local Realism) and {LQ} (Local Quantum Theory), are defined in the same way as Problems {R} and {Q}, respectively, but with the tuple of jointly measurable observables $\vect{X}_{\bullet}$ in the form of \eqref{eq:tuple of jointly measurable observables}.
}

\subsection{Fixed Observables Settings}
{
In the previous subsections, we formulated the problems assuming that the set of observables is not fixed—that is, the problems were defined over all possible choices of observables.  
However, in many situations of practical or theoretical interest, one may wish to analyze the statistical constraints associated with a specific choice of observables.
In this case, the model itself is also fixed, and only the state (probability distribution or density operator) is left to vary.  

In general, the statistical constraints in such fixed observables settings are stronger than those obtained by allowing observables to vary freely.

\begin{problem}{M-FO}[{M} = {R}, {Q}, {LR}, or {LQ} (Fixed Observables)]
    Given data $\gamma \in \Gamma$ and a fixed model $(\Lambda,\vect{X}_{\bullet}(\lambda))$ or $(\mcal{H},\vect{X}_{\bullet})$,  
    find a necessary and sufficient condition such that
    \begin{equation}
        \ran\left( \Phi_{\vect{X}_{\bullet}} \right) \cap R^{-1}(\gamma) \neq \emptyset  \Longleftrightarrow \gamma \in \ran(R \circ \Phi_{\vect{X}_{\bullet}})
    \end{equation}
    holds.
\end{problem}

The fixed-observable problem may be regarded as a refinement of the unfixed-observable problem, since the realizability condition in the latter can be obtained as the disjunction (logical OR) of the realizability conditions for all possible choices of observables.

We illustrate the structure of the realizability problems in Fig. \ref{fig:statistical_constraint}, Fig. \ref{fig:data-region-union}, and Fig. \ref{fig:local-realistic-constraint}.

\begin{figure}
    \centering
    \includegraphics[width=85mm]{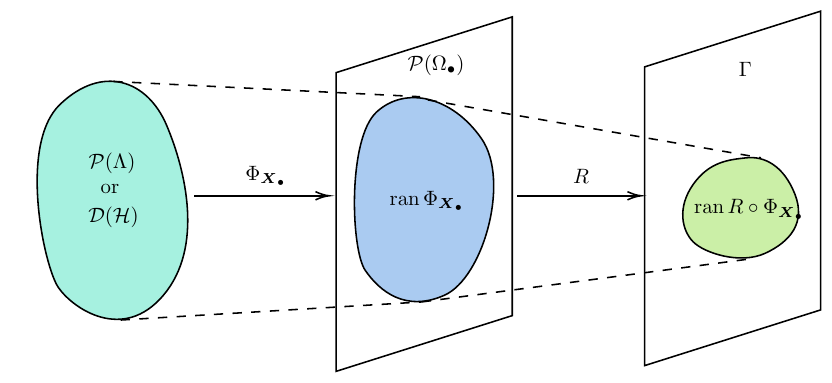}
    \caption{Flow of data and mappings in the realizability problems. A (hidden variable or quantum) state is mapped by $\Phi_{\vect{X}_{\bullet}}$ to a tuple of probability distributions in $\mathcal{P}(\vect{\Omega}_{\bullet})$, which is then transformed by the recording map $R$ to the recorded data in $\Gamma$.}
    \label{fig:statistical_constraint}
\end{figure}
\begin{figure}
    \centering
    \includegraphics[width=80mm]{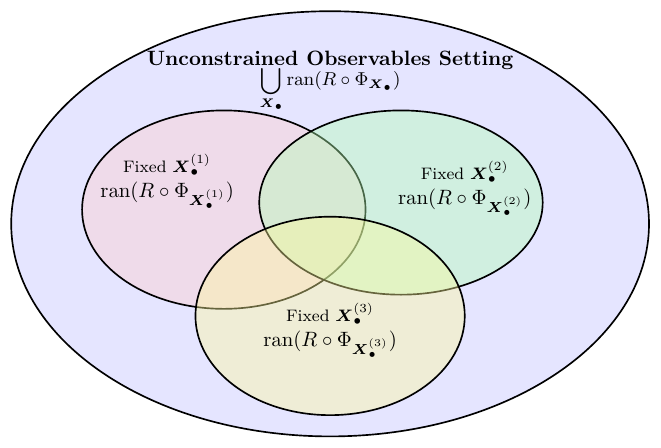}
    \caption{Relationship between the data regions in the fixed and unfixed observable settings. The data region for the unfixed setting is the union of the data regions for all possible fixed observables settings.}
    \label{fig:data-region-union}
\end{figure}

\begin{figure}
    \centering
    \includegraphics[width=85mm]{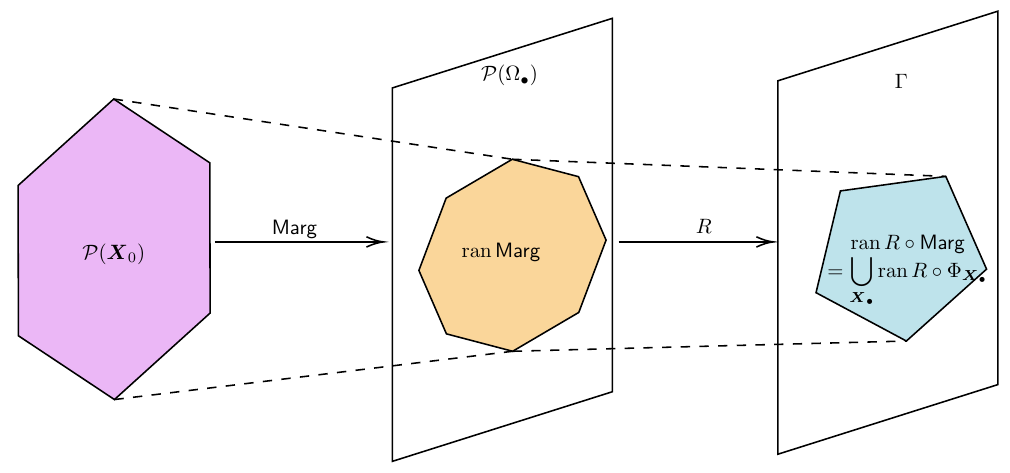}
    \caption{Flow of data and mappings in the formulation based on joint probability distributions.
A joint distribution over all observables in $\bs{X}_0$ is mapped by the marginalization map $\mathsf{Marg}$ to a tuple of marginal distributions in $\mathcal{P}(\vect{\Omega}_{\bullet})$, which is then transformed by the recording map $R$ into the recorded data in $\Gamma$.
The realizable data region $\operatorname{ran}(R \circ \mathsf{Marg})$ coincides with the union of the realizable regions $\operatorname{ran}(R \circ \Phi_{\vect{X}_{\bullet}})$ taken over all possible sets of observables.}
    \label{fig:local-realistic-constraint}
\end{figure}
}

\subsection{Examples of Recording Maps}

\begin{example}{\label{ex:identity constraint}}
	The identity mapping $R:(P_{A_i B_j})_{i,j=1}^m \mapsto (P_{A_i B_j})_{i,j=1}^m$ offers a trivial example of a faithful recording map.
\end{example}
In the $(2,2,2)$ setup, Fine's theorem \cite{fine1982joint,fine1982hidden} provides the solution to the problem under the above constraint.
\begin{theorem}[Fine \cite{fine1982joint,fine1982hidden}]{\label{thm:Fine}}
    Assume the $(2,2,2)$ setup. Then, given joint probability distributions $(P_{A_i B_j})_{i,j=1}^2$, a global joint probability distribution $P$ for all four observables $A_1,A_2,B_1,B_2$ that has $(P_{A_i B_j})_{i,j=1}^2$ as the marginal distributions exists if and only if the following inequality holds for all $i \neq i' \in \{1,2\}, j\neq j' \in \{1,2\}$, and $a_i,a_{i'},b_j,b_{j'} \in \{\pm 1\}$:
    \begin{multline}
        -1 \leq P_{A_i B_j}(a_i, b_j) + P_{A_i B_{j'}}(a_i,b_{j'}) + P_{A_{i'}B_{j'}}(a_{i'},b_{j'}) \\
        - P_{A_{i'}B_{j}}(a_{i'},b_j) - P_{A_i}(a_i) - P_{B_{j'}}(b_{j'}) \leq 0. 
        \label{eq:CHSH ineq in terms of probability}
    \end{multline}
    The set of inequalities~\eqref{eq:CHSH ineq in terms of probability} are also referred to as the CHSH inequalities. 
\end{theorem}

\begin{example}{\label{ex:expectation and correlation constraint}}
	If $d=2$, \ie, $A_i$ and $B_j$ are at most binary observables, the mapping that maps a set of joint distributions to the expectation values and the correlations of the observables concerned $R : (P_{A_i B_j})_{i,j=1}^m \mapsto (\bra A_i \ket, \bra B_j \ket, \bra A_i B_j\ket)_{i,j=1}^m$  offers another example of a faithful recording map.
\end{example}
%If $d=2$, \ie, $A_i$ and $B_j$ are binary observables, the above example represents a faithful constraint.
For binary observables, the recording maps in Example~\ref{ex:identity constraint} and Example~\ref{ex:expectation and correlation constraint} are equivalent. Thus, we can rewrite Fine's theorem as follows:
\begin{theorem}
    Assume the $(2,2,2)$ setup, and let the observables $A_1,A_2,B_1,B_2$ take values in $\{\pm 1\}$. Then, given real numbers $C_{i0},C_{0j},C_{ij} \in \mbb{R}$ satisfying $|C_{i0}|\leq 1$, $|C_{0j}| \leq 1$, and $|C_{i0}+C_{0j}|-1 \leq C_{ij} \leq -|C_{i0}-C_{0j}|+1$ for all $i,j \in \{1,2\}$, a global joint probability distribution $P$ for all four observables $A_1,A_2,B_1,B_2$ that satisfies 
    \begin{subequations}
        \begin{align}
            \bra A_i \ket_P &:= \sum_{a_1,a_2,b_1,b_2} a_i P(a_1,a_2,b_1,b_2) = C_{i0}, \label{eq:expectation constraints A}\\
            \bra B_j \ket_P &:= \sum_{a_1,a_2,b_1,b_2} b_j P(a_1,a_2,b_1,b_2) = C_{0j}, \label{eq:expectation constraints B}\\
            \bra A_i B_j \ket_P  &:= \sum_{a_1,a_2,b_1,b_2} a_i b_j P(a_1,a_2,b_1,b_2)= C_{ij} \label{eq:correlation constraints}
        \end{align}
        \label{eq:expectation and correlation}
    \end{subequations}
    exists if and only if the CHSH inequalities 
    \begin{equation}
        \left|\sum_{i,j \in \{1,2\}} C_{ij} - 2C_{kl}\right| \leq 2 , \quad \forall k,l \in \{1,2\}
        \label{eq:CHSH ineq C}
    \end{equation}
    hold. 
\end{theorem}
{Note that, while~\eqref{eq:CHSH ineq in terms of probability} contains the probabilities $P_{A_i}$ and $P_{B_j}$ for single observables,~\eqref{eq:CHSH ineq C} does not contain the corresponding expectation values $\bra A_i \ket$ and $\bra B_j \ket$.}

\begin{example}\label{ex:correlation}
    The map that records the correlations of the corresponding observables $R: (P_{A_i B_j})_{i,j=1}^m \mapsto (\bra A_i B_j \ket)_{i,j=1}^{m}$ fails to be faithful for $d \geq 2$.
\end{example}
In the $(2,2,2)$ setup, the above recording map is not faithful as it discards the information of the expectation values $\bra A_i \ket$ and $\bra B_j\ket$. {Still,}  a theorem similar to Fine's theorem holds:
\begin{theorem}[\cite{halliwell2014two}]
    Assume the $(2,2,2)$ setup. Then, given real numbers $C_{ij} \in [-1,1]$ for all $i,j \in \{1,2\}$, a global joint probability distribution $P$ for all four observables $A_1,A_2,B_1,B_2$ that satisfies Eq.~\eqref{eq:correlation constraints} exists if and only if the CHSH inequalities~\eqref{eq:CHSH ineq C} hold.
\end{theorem}

Theorem~\ref{thm:QI in CHSH setup} provides a necessary and sufficient condition for the existence of a quantum state that satisfies given correlation constraints in the CHSH setup.

\begin{example}[Moment Recording Map]
    For $d^{\prime} \in \mathbb{N}$, the map $R : (P_{A_i B_j})_{i,j=1}^m \mapsto ((\langle A_i^s B_j^t \rangle)_{s,t =0} ^{d^{\prime}})_{i,j=1}^m$ that records the $(s,t)$-th moments $\langle A_i^s B_j^t \rangle$ of the joint distribution $P_{A_i B_j}$ up to order $d^{\prime}-1$ in both variables
%    This map sends each joint probability distribution to its collection of moments up to order $d'-1$ in both variables. 
is faithful if $d^{\prime} \geq d$. In such cases, the sequences of moments of the observables under consideration uniquely determine the corresponding joint probability distributions.
\end{example}

\section{Preliminaries for the Proof of Theorem~\ref{thm:main theorem} \label{sec:preliminaries}}
\subsection{Fourier Transform}

The Fourier transform of a complex Borel measure $\mu$ on $\mathbb{R}^{N}$ is given by
\begin{equation}
\hat{\mu}(\vect{s}) \defeq \int_{\mbb{R}^N} \epower{- \im \vect{s}\cdot\vect{x}}\, d\mu(\vect{x}),
\end{equation}
which in this paper we refer to as its \textit{characteristic function}.

For a positive integer $d \in \mathbb{N}$, let $[d]:=\{0,\ldots,d-1\}$ denote the set of integers from $0$ to $d-1$.
For a complex measure $m$ defined on the $N$-fold product of a finite sample space with $d$ elements,  where $m(\bs{k})$ denotes the probability of the outcome labeled by $\bs{k} \in [d]^N$, we refer to its discrete Fourier transform
\begin{equation}
    \hat{m}(\bs{s}) := \sum_{\bs{k} \in [d]^N} \omega_{d}^{- \bs{s}\cdot \bs{k}} m(\bs{k}), \quad \forall \bs{s} \in [d]^N,
\end{equation}
as its characteristic function, where $\omega_{d} := \epower{2\pi \im / d}$ denotes a $d$-th root of unity.

For further details on the Fourier transform, see Appendices \ref{sec:FT on LCA groups} and \ref{sec:CFT and DFT}.

\subsection{Marginalization}
Let $\vect{\Omega} := \Omega_1 \times \cdots \times \Omega_n$ be a tuple of outcome spaces, and let $\tilde{\vect{\Omega}} := \Omega_{k_1} \times \cdots \times \Omega_{k_m}$ be its sub-tuple, $1 \leq k_1 < \cdots < k_m \leq n$.  The marginalization map $\tilde{M} : \mathcal{P}(\vect{\Omega}) \to \mathcal{P}(\tilde{\vect{\Omega}})$ is defined by
\begin{equation}
    \tilde{M} : P \mapsto \tilde{P}(\tilde{\vect{x}}) \defeq \int_{\tilde{\pi}^{-1}(\tilde{\vect{x}})} \,d P(\vect{x}),
    \label{def:marginalization}
\end{equation}
where $\tilde{\pi} : \vect{\Omega} \to \tilde{\vect{\Omega}}$ is the natural projection that maps $\vect{x} \in \vect{\Omega}$ to $\tilde{\vect{x}} := (x_{k_1}, \ldots, x_{k_m}) \in \tilde{\vect{\Omega}}$.

If each $\Omega_k$ is $\mathbb{R}$ (or more generally, a locally compact abelian group), the characteristic function $\hat{P}$ of a probability measure $P$ is defined as
\begin{equation}
    \hat{P}(\vect{s}) \coloneqq \int_{\mathbb{R}^n} \epower{-\im \bs{s} \cdot \bs{x}} \, d P(\vect{x}).
\end{equation}
By simple computation, the characteristic function of the marginal distribution is found to satisfy
\begin{equation}
    \widehat{(\tilde{M}P)}(\tilde{\vect{s}}) = \hat{P}(\sigma_{*}^{-1}(\tilde{\bs{s}},\bs{0}))
\end{equation}
for a permutation $\sigma$ of $(1, \ldots, n)$ satisfying $\sigma(i) = k_{i}$, $i = 1, \ldots, m$, where $\sigma_{*}: \mathbb{R}^{n} \to \mathbb{R}^{n},\ (s_{1}, \ldots, s_{n}) \mapsto (s_{\sigma(1)}, \ldots, s_{\sigma(n)})$ denotes the bijection induced by it.  For the case of finite sample spaces, see Appendix \ref{sec: marginalization in finite spaces}.

\subsection{Quasi-Joint Probability Distributions}
Unlike in local realism, in quantum mechanics, a joint probability distribution for non-commuting observables does not exists in general.
However, it is possible to assign even non-commuting observables \textit{quasi-joint probability (QJP) distributions} \cite{lee2017quasi,lee2018general}. 
Quasi-joint probability distributions share the same properties as standard joint probability distributions, except that they can take negative or complex values. 
In the following, we describe the framework of QJP distributions \cite{lee2017quasi,lee2018general} as a preliminary step toward addressing quantum problems.

Let $X$ be a self-adjoint operator on a Hilbert space $\mcal{H}$ and let $E_X$ denote the spectral measure of $X$. The Fourier transform of $E_{X}$ is given by
\begin{equation}
    (\mscr{F}_{\mbb{R}} E_{X})(s) := \int_{\mbb{R}} \epower{- \im s x} dE_X(x) = \epower{- \im s X}, \quad s\in \mbb{R}.
\end{equation}
Self-adjoint operators $X$ and $Y$ are said to \textit{strongly commute} if the spectral measures of $X$ and $Y$ commute:
\begin{equation}
    E_X(\Delta) E_Y(\Delta') = E_Y(\Delta') E_X(\Delta), \quad \forall \Delta,\Delta' \in \mfrak{B}(\mbb{R}),
\end{equation}
where $\mfrak{B}(S)$ denotes the Borel algebra of a topological space $S$.
Two self-adjoint operators $X$ and $Y$ strongly commute if and only if their Fourier transforms commute: 
\begin{equation}
    \epower{- \im s X} \epower{- \im t Y} = \epower{- \im t Y} \epower{- \im s X}, \quad \forall s,t \in \mbb{R}.
\end{equation}
For a tuple $\bs{X} = (X_1,\ldots,X_N)$ of pairwise strongly commuting self-adjoint operators, the \textit{joint spectral measure} $E_{\bs{X}}$ is uniquely defined so that the following equation is satisfied:
\begin{equation}
    E_{\bs{X}}\bigg(\prod_{i=1}^N \Delta_i \bigg)
    = \prod_{i=1}^{N} E_{X_{i}}(\Delta_i) ,\quad \forall \Delta_1,\ldots,\Delta_N \in \mfrak{B}(\mbb{R}).
\end{equation}
The Fourier transform of $E_{\bs{X}}$ is given by
\begin{align*}
    (\mscr{F}_{\mbb{R}^N} E_{\bs{X}})(\bs{s}) 
    &:= \int_{\mbb{R}^N} \epower{- \im \bs{s} \cdot \bs{x}} d E_{\bs{X}}(\bs{x})\\
    &~= \epower{- \im \bs{s} \cdot \bs{X}} \\
    &~= \prod_{i=1}^N \epower{- \im s_i X_i}, \quad \bs{s} \in \mbb{R}^N,
    \numberthis
    \label{eq:Fourier transform of joint spectral measure}
\end{align*}
where the last equality holds since the components of $\bs{X}$ pairwise strongly commute.

For a tuple $\bs{X}$ of self-adjoint operators that are not necessarily commuting, we define a \textit{hashed operator} \cite{lee2017quasi,lee2018general} as a generalization of the Fourier transform of a joint spectral measure as follows:
\begin{equation}
    \hat{\hash}_{\bs{X}}(\bs{s}) :=
    \left[
        \begin{gathered}
            \text{a suitable `mixture' of the `disentangled' } \\
            \text{components of the unitary groups }\\
             \epower{-\im s_1 X_1}, \ldots, \epower{-\im s_N X_N}
        \end{gathered}
    \right].
\end{equation}
\textit{Quasi-joint spectral (QJS) distributions} \cite{lee2017quasi,lee2018general} are given by the inverse Fourier transforms of hashed operators:
\begin{equation}
    \begin{split}
        \hash_{\bs{X}}(\bs{x}) &:= (\mscr{F}_{\mbb{R}^N}^{-1}\hat{\hash})(\bs{x})\\
        &~= \frac{1}{(2\pi)^N} \int_{\mbb{R}^N}\epower{+ \im \bs{s}\cdot \bs{x}} \hat{\hash}(\bs{s})d\bs{s}, 
        \quad \bs{x} \in \mbb{R}^N.
    \end{split}
\end{equation}
Note that this is a formal expression, as a QJS distribution does not necessarily have a density function; in fact, it is generally an operator valued distribution rather than even an operator valued measure.
Hashed operators and thus QJS distributions are generally not unique, as the components $ \epower{-\im s_1 X_1}, \ldots, \epower{-\im s_N X_N}$ do not commute:
\begin{example}
    Let $\bs{X}=(A,B)$. Examples of hashed operators are given by
    \begin{equation}
        \hat{\hash}_{\bs{X}}(s,t)
        = \begin{cases}
            \epower{-\im s A} \epower{-\im t B} , \\
            \epower{-\im t B} \epower{-\im s A}, \\
            \frac{1+\alpha}{2}\cdot \epower{-\im t B} \epower{-\im s A} + \frac{1-\alpha}{2} \epower{-\im s A} \epower{-\im t B}, \quad \alpha \in \mbb{C}, \\
            \big(\epower{-\im s A /N} \epower{-\im t B/N}\big)^{N}, \\
            \epower{- \im \overline{(s A + t B)}} = \lim_{N\to \infty}\big(\epower{-\im s A /N} \epower{-\im t B/N}\big)^{N},
        \end{cases}
    \end{equation}
    where $\overline{X}$ denotes the self-adjoint extension of an operator $X$.
\end{example}
The hashed operator $\hat{\hash}_{\bs{X}}$ and QJS distribution $\hash_{\bs{X}}$ are unique if and only if $X_1,\ldots,X_N$ are pairwise strongly commuting. In such cases, the QJS distribution $\hash_{\bs{X}}$ coincides with the joint spectral measure $E_{\bs{X}}$.

Given a QJS distribution $\hash_{\bs{X}}$, the \textit{quasi-joint probability (QJP) distribution} $Q_{\bs{X}}$ \cite{lee2017quasi,lee2018general} on a density operator $\rho$ is given by
\begin{equation}
    Q_{\bs{X}}(\bs{x}) := \Tr[\rho \hash_{\bs{X}}(\bs{x})],\quad \bs{x} \in \mbb{R}^N.
\end{equation}
A QJP distribution is generally a tempered distribution.
The characteristic function of $Q_{{\bs{X}}}$ is expressed as
\begin{equation}
    \hat{Q}_{{\bs{X}}}(\bs{s}) := (\mscr{F}_{\mbb{R}^N} Q_{{\bs{X}}})(\bs{s}) = \Tr[\rho \hat{\hash}_{\bs{X}}(\bs{s})], \quad \bs{s} \in \mbb{R}^N.
\end{equation}

For the case of finite sample spaces, see Appendix \ref{sec:QJP with finite supports}.

\subsection{Reducing the Degrees of Freedom Using Symmetry \label{subsec:reducing the degrees of freedom}}

In many cases, problems can be simplified by reducing the number of variables. 
Motivated by the observation in \cite{halliwell2014two}, we introduce a symmetry that preserves the relevant constraints.

%Recall that a map $S : X \to Y$ is said to be invariant under a map $T : X \to X$ if it satisfies $S \circ T = S$.  
In what follows, we seek for a map $T$ between the space of trace-class operators on $\mathcal{H}$ under which both the composite map $S := R \circ \Phi_{\vect{X}_{\bullet}}$ and the set $D(\mathcal{H})$ of density operators are invariant, namely,
\begin{equation}
S \circ T = S, \quad T(D(\mathcal{H})) \subset D(\mathcal{H}).
\end{equation}
In such a case, we have the equivalence of the following two conditions:
    \begin{enumerate}
        \item There exists a density operator $\rho \in \mcal{D}(\mcal{H})$ such that $S(\rho) = \gamma$.
        \item There exists a density operator $\rho \in T(\mcal{D}(\mcal{H})) \subset \mcal{D}(\mcal{H})$ such that $S(\rho) = \gamma$.
    \end{enumerate}
This implies that if $T(\mcal{D}(\mcal{H}))$ can be parameterized by a lower-dimensional subset of $\mcal{D}(\mcal{H})$, then the original realizability problem can be simplified accordingly, as one may restrict the search for a solution $\rho$ satisfying $S(\rho) = \gamma$ on the reduced space.

%%%%%%%%%%%%%%%%%%%%%%%%%%%%%%%%%%%%%%%%%%%%%%%%%%%%%%%%%%%%%%%%%%%%%%

\section{Proof of Theorem~\ref{thm:main theorem}  \label{sec:proof of main thm}}
We prove Theorem~\ref{thm:main theorem} by considering the cases (i), (ii), and (iii) separately.
We omit the proof of the case (iii), as it can be proved in a similar manner to the case (ii).

For a Hilbert space $\mcal{H}$, $\mcal{B}(\mcal{H})$ represents the Banach space of all the bounded operators on $\mcal{H}$\delete{, and $\mcal{D}(\mcal{H})$ denotes the set of density operators on $\mcal{H}$}.

\subsection{When $\bs{a}_1 \nparallel \bs{a}_2$ and $\bs{b}_1 \nparallel \bs{b}_2$}
First, we prove the case (i).
We assume that $A_i$ and $B_j$ take values from $\{\pm 1\}$, \ie, $A_i^2=B_j^2=I_2$ for all $i,j \in \{1,2\}$. In the general case, replace $A_i$ with $\tilde{A}_i$ and $B_j$ with $\tilde{B}_j$.

% We derive the expression for $\hatmcal{Q}^{\bs{X}}_{2,2,2}$, \ie, 
We express the condition for an operator $\rho$ to be a density operator in terms of the characteristic function $\bs{\varphi}=(\varphi_{\bs{s}})_{\bs{s}\in[2]^4}$.
The operators $A_i$ and $B_j$ are expressed as 
\begin{subequations}
    \begin{align}
        A_i &= \bs{a}_i \cdot \bs{\sigma}, \quad |\bs{a}_i|=1, \quad i \in \{1,2\}, \\
        B_j &= \bs{b}_j \cdot \bs{\sigma}, \quad |\bs{b}_j|=1, \quad j \in \{1,2\}.
    \end{align}
\end{subequations}
Thus, the following equations hold:
\begin{subequations}
    \begin{gather}
        A_1 A_2 = \bs{a}_1 \cdot \bs{a}_2 I_2 + \im (\bs{a}_1 \times \bs{a}_2) \cdot \bs{\sigma} , \\
        B_1 B_2 = \bs{b}_1 \cdot \bs{b}_2 I_2 + \im (\bs{b}_1 \times \bs{b}_2) \cdot \bs{\sigma}.
    \end{gather}
\end{subequations}
Define 
\begin{equation}
    A_3 := \frac{\bs{a}_1 \times \bs{a}_2}{|\bs{a}_1 \times \bs{a}_2|}\cdot\bs{\sigma
    },\quad 
    B_3 := \frac{\bs{b}_1 \times \bs{b}_2}{|\bs{b}_1 \times \bs{b}_2|}\cdot \bs{\sigma}.
\end{equation}
Then, we have
\begin{subequations}
    
    \begin{gather}
        A_i A_3 = - A_3 A_i, ~ B_j B_3 = -B_3 B_j,\quad i,j \in \{1,2\}, \label{eq:A_i and A_3 anti-commute} \\
        A_3^\dagger = A_3 ,\quad B_3^\dagger = B_3 , \label{eq:A_3,B_3 are Hermitian}\\
        A_3^2 = B_3^2 = I_2. \label{eq:A_3^2=B_3^2=I}
    \end{gather}
\end{subequations}
In addition, we define
\begin{subequations}
    \begin{gather}
        E_0 := I_2, \, E_1:=A_1, \, E_3:=A_3, \, E_2 := -\im E_3 E_1, \\
        F_0 := I_2, \, F_1 := B_1, \, F_3 := B_3, \, F_2:= -\im F_3 F_1,
    \end{gather}
\end{subequations}
so that $\{E_i\}_{i=1,2,3}$ and $\{F_j\}_{j=1,2,3}$ satisfy the algebra of the Pauli matrices $\bs{\sigma}$, respectively. We have
\begin{subequations}
    \begin{gather}
        E_{2} = \frac{A_2 - A_1 \cos \alpha}{\sin \alpha}, \quad 
        E_{3} = \frac{A_1 A_2 - I_2 \cos \alpha}{\im \sin \alpha} , \\
        F_{2} = \frac{B_2 - B_1 \cos \beta}{\sin \beta}, \quad F_{3} = \frac{B_1 B_2 -I_2 \cos \beta}{\im \sin \beta}.
    \end{gather}
\end{subequations}

The Fourier transform of {the discrete QJS density} that gives the Kirkwood-Dirac distribution for $\bs{X}=(X_1,X_2,X_3,X_4) :=(A_1 \otimes I_2, A_2 \otimes I_2, I_2 \otimes B_1, I_2 \otimes B_2)$ is expressed as
\begin{equation}
    (\hat{\hash}_{\bs{X}}^{\text{KD}})^{[2]} (\bs{s})
    := \prod_{i=1}^{4} \omega_2^{- s_i \theta^{-1}(X_i) } 
    = X_1^{s_1} X_2^{s_2} X_{3}^{s_3} X_{4}^{s_4},
\end{equation}
where the function $\theta:[2] \to \{\pm 1\}$ is defined by $k \mapsto 1-2k$. 
The corresponding characteristic function for a density operator $\rho$ is given by
\begin{equation}
    \varphi_{\bs{s}} := \Tr[\rho( \hat{\hash}_{\bs{X}}^{\text{KD}})^{[d]}(\bs{s})] = \bra X_1^{s_1} X_2^{s_2} X_{3}^{s_3} X_{4}^{s_4} \ket_{\rho}, \quad \bs{s} \in [2]^4.
\end{equation}

As the set $\{E_{\mu} \otimes F_{\nu}\}_{\mu,\nu=0}^{3}$ forms a basis of $\mcal{B}(\mbb{C}^2\otimes \mbb{C}^2)$, the operator $\rho$ can be expanded as
\begin{equation}
    \rho = \frac{1}{4} \sum_{\mu,\nu=0}^{3} w_{\mu \nu} E_\mu \otimes F_\nu,
    \label{eq:rho extended by E_mu F_nu}
\end{equation}
where
\begin{equation}
    w_{\mu \nu} := \Tr[\rho E_{\mu} \otimes F_\nu], \quad  \forall \mu,\nu \in \{0,1,2,3\}.
\end{equation}
It follows from a straightforward calculation that
\begin{subequations}
    \begin{align}
        w_{00} &  =\varphi_{0000}, \label{eq:r00}\\
        w_{10} &  =  \varphi_{1000}  ,\\
        w_{20} &  =  \frac{\varphi_{0100} - \varphi_{1000}\cos\alpha }{\sin \alpha}  ,\\
        w_{30} &  =   \frac{\varphi_{1100} - \varphi_{0000}\cos \alpha}{\im \sin \alpha} ,\label{eq:r30}\\
        w_{01} &  =   \varphi_{0010} ,\\
        w_{11} &  = \varphi_{1010}   ,\\
        w_{21} &  = \frac{\varphi_{0110} - \varphi_{1010}\cos\alpha}{\sin \alpha}   ,\\
        w_{31} &  =  \frac{\varphi_{1110} - \varphi_{0100} \cos \alpha}{\im \sin \alpha}  ,\\
        w_{02} &  =   \frac{\varphi_{0001} - \varphi_{0010} \cos \beta}{\sin \beta} ,\\
        w_{12} &  =  \frac{\varphi_{1001}-\varphi_{1010}\cos \beta}{\sin \beta}  ,\\
        w_{22} &  = \frac{1}{\sin \alpha \sin \beta} \big(
                \varphi_{0101} - \varphi_{1001}\cos\alpha \notag \\  
                &\qquad\qquad-\varphi_{0110}\cos\beta + \varphi_{1010} \cos \alpha \cos \beta
            \big)  ,\\
        w_{32} &  =  \frac{1}{\im \sin \alpha \sin \beta}  \big(\varphi_{1101} - \varphi_{0001} \cos \alpha \notag \\ 
        &\qquad\qquad  - \varphi_{1110} \cos \beta + \varphi_{0010}\cos\alpha\cos\beta
            \big),\\
        w_{03} &  =  \frac{\varphi_{0011} - \varphi_{0000}\cos \beta}{\im \sin \beta}  ,\label{eq:r03}\\
        w_{13} &  =  \frac{\varphi_{1011} - \varphi_{1000}\cos \beta}{\im \sin \beta}  ,\\
        w_{23} &  =  \frac{1}{\im \sin \alpha \sin \beta}  \big(\varphi_{0111} - \varphi_{1011} \cos\alpha \notag \\ 
        &\qquad \qquad - \varphi_{0100} \cos\beta +\varphi_{1000} \cos \alpha\cos \beta
            \big),\\
        w_{33} & =   \frac{1}{\sin\alpha \sin \beta} \big(-\varphi_{1111} + \varphi_{0011}\cos\alpha \notag \\ 
        &\qquad \qquad + \varphi_{1100} \cos\beta - \varphi_{0000}\cos\alpha \cos\beta 
        \big).  \label{eq:r33}
    \end{align}
    \label{eq:w_ij represented by varphi}
\end{subequations}
By substituting~\eqref{eq:w_ij represented by varphi} into~\eqref{eq:rho extended by E_mu F_nu}, $\rho$ can be computed from the characteristic function $\bs{\varphi}=(\varphi_{\bs{s}})_{\bs{s}\in[2]^4}$. 
Thus, we can express the condition for $\rho$ to satisfy $\Tr[\rho]=1$ and $\rho \geq 0$ in terms of the characteristic function $\bs{\varphi}$. The condition $\Tr[\rho]=1$ is equivalent to $\varphi_{0000}=1$.
By assumption, $\gamma_{11} = \varphi_{1010}$, $\gamma_{21} = \varphi_{0110}$, $\gamma_{12} = \varphi_{1001}$, and $\gamma_{22}=\varphi_{0101}$ are known. 
We derive a necessary and sufficient condition for the existence of the other components $\varphi_{\bs{s}}$ such that $\rho$ is a density operator.

Now, we reduce the degrees of freedom using the technique explained in Subsection~\ref{subsec:reducing the degrees of freedom} to simplify the problem. 
We introduce new notation.
Let $x_i := \bra X_i\ket_\rho$, $x_{ij} := \bra X_i X_j \ket_{\rho}$, $x_{ijk}:=\bra X_i X_j X_k\ket_{\rho}$, and $x_{1234} := \bra X_1 X_2 X_3 X_4 \ket_\rho$. For example, $x_1 = \varphi_{1000}$, $x_{23}=\varphi_{0110}$, $x_{124} = \varphi_{1101}$, and $x_{1234}=\varphi_{1111}$. 
The known variables are $x_{i(j+2)} = \gamma_{ij}$ for $i,j \in \{1,2\}$. 
The problem under consideration is to find the necessary and sufficient condition on $x_{i(j+2)} \, (i,j \in \{1,2\})$ that ensures the existence of the other components $x_i$, $x_{12}$, $x_{34}$, $x_{ijk}$, and $x_{1234}$ such that $\rho$ is a density operator.
Since the existence of $x_{12} (=\varphi_{1100})$, $x_{34}(=\varphi_{0011})$, and $x_{1234} (=\varphi_{1234})$ is equivalent to the existence of $w_{30}$, $w_{03}$, and $w_{33}$, which can be assumed to be real numbers due to the self-adjoint property of $\rho$, we consider the latter.
We can find a transformation $T$ that preserves the constraints $x_{i(j+2)}\,(i,j\in\{1,2\})$.
\begin{lemma}\label{lem:transformation}
    A linear mapping $T$ on $\mcal{B}(\mbb{C}^2 \otimes \mbb{C}^2)$ that satisfies the following conditions exists:
    \begin{itemize}
        \item If $\rho$ is a density operator, then $T(\rho)$ is also a density operator. 
        \item $T$ preserves $x_{i(j+2)}$ for $i,j \in \{1,2\}$ and $w_{33}$ while transforming $x_i$, $x_{ijk}$, $w_{03}$, and $w_{30}$ to zero for all $i,j,k$:
        \begin{subequations}
            \begin{align}
                \tilde{x}_{i(j+2)} &:= \bra X_i X_{j+2} \ket_{T(\rho)} \notag \\ 
                &~= \bra X_i X_{j+2} \ket_\rho = x_{i(j+2)}, \quad \forall i,j \in \{1,2\}, \\
                \tilde{w}_{33} &:= \bra E_{3} \otimes F_{3} \ket_{T(\rho)} = \bra E_3 \otimes F_3 \ket_{\rho} = w_{33}, \\
                \tilde{x}_{i} &:= \bra X_i \ket_{T(\rho)} = 0, \quad \forall i,\\
                \tilde{x}_{ijk} &:= \bra X_i X_j X_k \ket_{T(\rho)} = 0 , \quad \forall i,j,k, \\
                \tilde{w}_{03} &:= \bra I_2 \otimes F_3 \ket_{T(\rho)} = 0, \\
                \tilde{w}_{30} &:= \bra E_3 \otimes I_2 \ket_{T(\rho)} = 0.
            \end{align}
        \end{subequations}
    \end{itemize}
\end{lemma}
The mapping $S:\mcal{D}(\mbb{C}^2\otimes \mbb{C}^2)\to \mbb{C}^4, \rho \mapsto (x_{i(j+2)})_{i,j}$ is invariant under the transformation $T$, \ie, $S\circ T = S$.
Therefore, a density operator satisfying $x_{i(j+2)} = \gamma_{ij}$ for $i,j \in \{1,2\}$ exists if and only if a density operator satisfying the same conditions while additionally imposing $x_i=x_{ijk}=w_{03}=w_{30}=0$ for all $i,j,k$ also exists. 
If we assume that $x_i=x_{ijk}=w_{03}=w_{30}=0$ for all $i,j,k$, the eigenvalues of $\rho$ are given by
\begin{subequations}
    \begin{align}
        &\frac{1}{4} \left(1+w_{33} \pm \sqrt{(w_{11}-w_{22})^2  +(w_{12}+w_{21})^2} \right) \notag\\
        &\quad = \frac{1}{4} \left(1 + w_{33} \pm \sqrt{\bs{\gamma}^\top F_{\alpha,\beta} \bs{\gamma}}\right), \\
        &\frac{1}{4} \left(1-w_{33} \pm \sqrt{(w_{11}+w_{22})^2 +(w_{12}-w_{21})^2}\right) \notag\\ 
        &\quad = \frac{1}{4} \left(1 - w_{33} \pm \sqrt{\bs{\gamma}^\top F_{\alpha,-\beta}\bs{\gamma}} \right).
    \end{align}
\end{subequations}
The condition that all these four eigenvalues are non-negative can be expressed as
\begin{equation}
    \left\{
    \begin{aligned}
        &1 + w_{33} \pm \sqrt{\bs{\gamma}^\top F_{\alpha,\beta} \bs{\gamma}} \geq 0, \\
        &1 - w_{33} \pm \sqrt{\bs{\gamma}^\top F_{\alpha,-\beta}\bs{\gamma}}  \geq 0,
    \end{aligned}\right.
\end{equation}
or equivalently
\begin{equation}
    -1 \pm \sqrt{\bs{\gamma}^\top F_{\alpha,\beta} \bs{\gamma}} \le w_{33} \le 1 \pm \sqrt{\bs{\gamma}^\top F_{\alpha,-\beta}\bs{\gamma}}.\label{eq:eigenvalues are non-negative}
\end{equation}
There exists a real number $w_{33} \in \mbb{R}$ satisfying~\eqref{eq:eigenvalues are non-negative} if and only if the following inequality holds:
\begin{equation}
    -1 + \sqrt{\bs{\gamma}^\top F_{\alpha,\beta} \bs{\gamma}} \le 1 - \sqrt{\bs{\gamma}^\top F_{\alpha,-\beta} \bs{\gamma}},
\end{equation}
which is equivalent to~\eqref{eq:main inequality (i)}.
Note that $x_i=0$ for all $i \in \{1,2,3,4\}$ implies $\bra A_i \ket_\rho=\bra B_j \ket_\rho=0$ for all $i, j\in \{1,2\}$.

Now, we complete the proof by establishing Lemma~\ref{lem:transformation}.

First, we define a linear mapping $T_1$ on $\mcal{B}(\mbb{C}^2 \otimes \mbb{C}^{2})$ as
\begin{equation}
    T_1 (\rho) := \frac{1}{2}\rho + \frac{1}{2}\Ad{E_3 \otimes F_3}(\rho), \quad \rho \in \mcal{B}(\mbb{C}^2 \otimes \mbb{C}^2),
\end{equation}
where, for an invertible operator $U \in \mcal{B}(\mbb{C}^2 \otimes \mbb{C}^2)$, a similarity transformation $\Ad{U}$ is defined as
\begin{equation}
    \Ad{U} (\, \cdot \,) := U (\, \cdot \,)U^{-1}.
\end{equation}
Since a similarity transformation leaves eigenvalues invariant, if $\rho$ is a density operator, then $\tilde{\rho}:= \Ad{E_3 \otimes F_3}(\rho)$ is also a density operator. 
Therefore, if $\rho$ is a density operator, then $T_1(\rho)$ is also a density operator due to the convexity of the set $\mcal{D}(\mbb{C}^2 \otimes \mbb{C}^2)$ of density operators.
It follows from~\eqref{eq:A_i and A_3 anti-commute} that
\begin{subequations}
    \begin{align}
        \bra X_i \ket_{\tilde{\rho}} &= - \bra X_i \ket_{\rho}, \quad \forall i, \\
        \bra X_i X_j \ket_{\tilde{\rho}} &=  \bra X_i X_j \ket_{\rho}, \quad \forall i,j,\\
        \bra X_i X_j X_k \ket_{\tilde{\rho}} &= - \bra X_i X_i X_k \ket_{\rho},\quad \forall i,j,k,\\
        \bra X_1 X_2 X_3 X_4 \ket_{\tilde{\rho}} &= \bra X_1 X_2 X_3 X_4 \ket_{\rho}.
    \end{align}
\end{subequations}
Thus, we obtain
\begin{subequations}
    \begin{align}
        x_{i}' &:= \bra X_i \ket_{T_1(\rho)} = 0 , \quad \forall i,\\
        x_{ij}' &:= \bra X_i X_j \ket_{T_1(\rho)} = \bra X_i X_j \ket_{\rho} = x_{ij}, \quad \forall i,j,\\
        x_{ijk}' &:= \bra X_i X_j X_k \ket_{T_1(\rho)} = 0 , \quad \forall i,j,k,\\
        x_{1234}' &:= \bra X_1 X_2 X_3 X_4 \ket_{T_1(\rho)} = \bra X_1 X_2 X_3 X_4 \ket_{\rho} = x_{1234},
    \end{align}
\end{subequations}
which implies that $T_1$ preserves $x_{ij}$ and $x_{1234}$ (and thus $w_{03}$, $w_{30}$, and $w_{33}$) invariant while transforming $x_{i}$ and $x_{ijk}$ to $0$ for all $i,j,k$.

Further, we define a linear mapping $(\,\cdot\,)^{\top_{\mrm{A}}}$ on $\mcal{B}(\mbb{C}^2)$ as 
\begin{multline}
    (\,\cdot \,)^{\top_{\mrm{A}}} :  x_0 I_2 + x_1 E_1 + x_2 E_2 + x_3 E_3 \\
    \mapsto x_0 I_2 + x_1 E_1 + x_2 E_2 - x_3 E_3.
\end{multline}
The linear mapping  $(\,\cdot\,)^{\top_{\mrm{A}}}$ can be expressed as
\begin{equation}
    (\,\cdot\,)^{\top_{\mrm{A}}}
    : \sum_{ij} M_{ij} |i\ket \bra j| \mapsto  \sum_{ij} M_{ji}|i\ket \bra j|,
\end{equation}
where $|i\ket$ denotes an eigenvector of $E_1$ belonging to the eigenvalue $i \in \{\pm 1\}$, with an arbitrary but fixed choice of phase. 
In other words, $(\,\cdot\,)^{\top_{\mrm{A}}}$ is the transpose mapping in the eigenbasis $\{|i\ket\}$. 
Similarly, we can define a linear mapping $(\, \cdot \,)^{\top_{\mrm{B}}}$ as 
\begin{multline}
    (\,\cdot\,)^{\top_{\mrm{B}}} :  x_0 I_2 + x_1 F_1 + x_2 F_2 + x_3 F_3 \\
    \mapsto x_0 I_2 + x_1 F_1 + x_2 F_2 - x_3 F_3,
\end{multline}
which is the transpose mapping in an eigenbasis of $F_1$.
Therefore, the tensor product $(\, \cdot \,)^\top := (\, \cdot \,)^{\top_{\mrm{A}}} \otimes (\, \cdot \,)^{\top_{\mrm{B}}}$ is the transpose mapping on $\mcal{B}(\mbb{C}^2 \otimes \mbb{C}^2)$ in an eigenbasis of $E_{1} \otimes F_{1}$.
Then, we define a linear mapping $T_2$ on $\mcal{B}(\mbb{C}^2 \otimes \mbb{C}^2)$ as
\begin{equation}
    T_2(\rho) := \frac{1}{2}\rho + \frac{1}{2}\rho^\top, \quad \forall \rho \in \mcal{B}(\mbb{C}^2 \otimes \mbb{C}^2).
\end{equation}
Since a transpose mapping leaves eigenvalues invariant, if $\rho$ is a density operator, then $\rho^{\top}$ is also a density operator. 
Therefore, if $\rho$ is a density operator, then $T_2(\rho)$ is also a density operator due to the convexity of the set $\mcal{D}(\mbb{C}^2 \otimes C^2)$ of density operators.
Noting that 
\begin{equation}
    \bra E_{\mu} \otimes F_\nu \ket_{\rho^\top} = \begin{cases}
        - \bra E_\mu \otimes F_\nu \ket_{\rho} & (\mu = 3) \veebar ( \nu = 3)  ,\\
        \bra E_\mu \otimes F_\nu \ket_{\rho} & (\text{otherwise}),
    \end{cases}
\end{equation}
we obtain 
\begin{equation}
    \begin{split}
        {w}''_{\mu \nu} &:= \bra E_\mu \otimes F_\nu \ket_{T_2(\rho)} \\
        &~= \begin{cases}
            0 &  (\mu = 3) \veebar (\nu = 3), \\
            \bra E_\mu \otimes F_\nu \ket_{\rho} = w_{\mu\nu} & (\text{otherwise}),
        \end{cases}
    \end{split}
\end{equation}
where $\veebar$ denotes the exclusive OR operation.
This implies that $T_2$ transforms $w_{\mu \nu}$ to $0$ for $(\mu,\nu) = (0,3), (1,3), (2,3), (3,0), (3,1), (3,2)$ and preserves $w_{\mu \nu}$ for all other values of $(\mu,\nu)$.

Finally, we define a mapping $T$ as $T := T_2 \circ T_1$. 
The mapping $T$ possesses the desired properties.

\subsection{When $\bs{a}_1 \nparallel \bs{a}_2$ and $\bs{b}_1 \parallel \bs{b}_2$}
When $\bs{b}_1 \parallel \bs{b}_2$, the angle $\beta$ equals $0$ or $\pi$, and therefore~\eqref{eq:w_ij represented by varphi} does not hold.  
Hence, the case (ii) must be addressed differently from the case (i). 
Again, we assume that $A_i$ and $B_j$ take values from $\{\pm 1\}$ for $i,j \in \{1,2\}$. In the general case, replace $A_i$ with $\tilde{A}_i$ and $B_j$ with $\tilde{B}_j$.
If $\beta=0$, then $B_1 = B_2$, and if $\beta=\pi$, then $B_1 = -B_2$, leading to $\gamma_{i1} = s_b \gamma_{i2}$ for $i \in \{1,2\}$, where $s_b := \mrm{sgn}(\bs{b}_1 \cdot \bs{b}_2) = \mrm{sgn}(\cos \beta)$.
Thus, it is sufficient to consider the condition on $\gamma_{11}$ and $\gamma_{21}$.
We find a necessary and sufficient condition for a density operator $\rho$ satisfying $\gamma_{i1} = \Tr[\rho A_i \otimes B_1]$ for $i \in \{1,2\}$ to exist.

We choose a self-adjoint operator $B_{\star} \, (\neq \pm B_1)$ on $\mcal{B}(\mbb{C}^2)$ such that $B_{\star}^2=I_2$, and $B_{\star}$ anti-commutes with $B_1$. 
It follows from the case (i) that if $\gamma_{i\star} := \bra A_i  \otimes B_{\star}\ket_{\rho}$ is given for $i\in \{1,2\}$, a density operator $\rho$ satisfying $\gamma_{i1} = \bra A_i \otimes B_1 \ket_{\rho}$ for $i \in \{1,2\}$ exists if and only if the following inequality holds:
\begin{equation}
    \sqrt{\bs{\gamma}_{\star}^\top F_{\alpha,\beta} \bs{\gamma}_{\star}} + \sqrt{\bs{\gamma}_{\star}^\top F_{\alpha,-\beta} \bs{\gamma}_{\star}} \leq 2,
    \label{eq:inequality for new B2}
\end{equation}
where $\bs{\gamma}_{\star} := (\gamma_{11}, \gamma_{21}, \gamma_{1\star}, \gamma_{2\star})^\top$.

Moreover, we define a linear mapping $T$ on $\mcal{B}(\mbb{C}^2 \otimes \mbb{C}^2)$ as
\begin{equation}
    T(\rho) := \frac{1}{2}\rho + \frac{1}{2} \Ad{I_2 \otimes B_1}(\rho), \quad \rho \in \mcal{B}(\mbb{C}^2 \otimes \mbb{C}^2).
\end{equation}
Noting that 
\begin{subequations}
    \begin{align}
        \Tr[\Ad{I_2 \otimes B_1}(\rho) A_i \otimes B_1] &= \Tr[\rho A_i \otimes B_1^3] \notag\\
        & = \Tr[\rho A_i \otimes B_1], \\
        \Tr[\Ad{I_2 \otimes B_1}(\rho) A_i \otimes B_{\star}] &= \Tr[\rho A_i \otimes B_1 B_{\star} B_1] \notag \\
        &= - \Tr[\rho A_i \otimes B_{\star}],
    \end{align}
\end{subequations}
for $i \in \{1,2\}$, we obtain
\begin{subequations}
    \begin{align}
        \gamma_{i1}' &:= \bra A_i \otimes B_1 \ket_{T(\rho)} = \bra A_i \otimes B_1 \ket_{\rho} = \gamma_{i1}, \\
        \gamma_{i\star}' &:= \bra A_i \otimes B_{\star} \ket_{T(\rho)} = 0,
    \end{align}
\end{subequations}
for $i \in \{1,2\}$.
In other words, the mapping $T$ preserves $\gamma_{11}$ and $\gamma_{21}$ while transforming $\gamma_{1\star}$ and $\gamma_{2\star}$ to zero. 
Therefore, a density operator $\rho$ satisfying $\gamma_{i1} = \bra A_i \otimes B_1 \ket_{\rho}$ for $i \in \{1,2\}$ exists if and only if a density operator $\rho$ satisfying $\gamma_{i1} = \bra A_i \otimes B_1 \ket_{\rho}$ and $\gamma_{i\star} = \bra A_i \otimes B_{\star} \ket_{\rho} = 0$ for $i \in \{1,2\}$ also exists.
We can obtain the necessary and sufficient condition for the latter by substituting $\gamma_{1\star}=\gamma_{2\star}=0$ into~\eqref{eq:inequality for new B2}, which leads to 
\begin{equation}
    \sqrt{\gamma_{11}^2 + \gamma_{21}^2 - 2 \cos(\alpha) \gamma_{11} \gamma_{21}} \leq \sin (\alpha).
\end{equation}

%%%%%%%%%%%%%%%%%%
\section{Fourier Transform on Locally Compact Abelian Groups \label{sec:FT on LCA groups}}
In this appendix, we provide a brief introduction to Fourier analysis on Locally Compact Abelian (LCA) groups. 
\subsection{Definition and Examples}

\begin{definition}
    A \textit{Locally Compact Abelian (LCA) group} is a topological group that is both locally compact and abelian. 
    A topological group $G$ is said to be \textit{locally compact} if it is Hausdorff, and every point $x \in G$ has a compact neighborhood.
    A group is called \textit{abelian} if its group operation is commutative.
\end{definition}

\begin{example}{\hfill}
    \begin{itemize}
        \item $\mbb{R}^N$ for a positive integer $N \in \mbb{N}$.
        \item The finite cyclic group $\mbb{Z}_d := \mbb{Z}/d\mbb{Z}$ for a positive integer $d \in \mbb{N}$.
        \item The circle group $\mbb{T} := \{z \in \mbb{C} \mid |z| = 1\} \cong \mbb{R}/\mbb{Z}$. 
    \end{itemize}
\end{example}

\subsection{The Dual Group and Pontryagin Duality}
\begin{definition}
    For an LCA group $G$, the \textit{dual group} $\widehat{G}$ is the group of continuous group homomorphisms (characters) $\chi: G \to \mbb{T}$, where the group operation is given by pointwise multiplication of functions. 
\end{definition}
Equiped with the compact-open topology, $\widehat{G}$ also forms an LCA group. 
\begin{example}{\hfill}
    \begin{itemize}
        \item $\widehat{\mbb{R}^N} \cong \mbb{R}^N$, with characters given by $\chi_{\bs{y}} (\bs{x}) = \epower{\im \bs{x}\cdot\bs{y}}$.
        \item $\widehat{\mbb{Z}_d} \cong \mbb{Z}_d$, with characters given by $\chi_{k}(m)=\epower{  \frac{2\pi\im}{d} mk}$.
        \item $\widehat{\mbb{T}} \cong \mbb{Z}$, with characters given by $\chi_n (\epower{2\pi \im \theta}) = \epower{2\pi \im n \theta}$.
    \end{itemize}
\end{example}
\begin{theorem}[Pontryagin Duality]
    The canonical evaluation map
    \begin{equation}
        G \to \widehat{\widehat{G}}, \quad x \mapsto (\chi \mapsto \chi(x))
    \end{equation}
    is a topological isomorphism.
\end{theorem}

\subsection{Fourier Transform on LCA Groups}
A measure on a locally compact topological group that is invariant under both left and right translations is called the \textit{Haar measure}. 
Every LCA group $G$ admits a Haar measure, denoted by $\mu_{G}$, which is unique up to scaling. 
\begin{example}{\hfill}
    \begin{itemize}
        \item The Haar measure of $\mbb{R}^N$ is the Lebesgue measure. 
        \item The Haar measure of a discrete group is the counting measure.
    \end{itemize}
\end{example}
\begin{definition}
    For an integrable function $f \in L^1(G, \mu_{G})$, the \textit{Fourier transform} of $f$ is the function $\hat{f}:\widehat{G} \to \mbb{C}$ defined by
    \begin{equation}
        \hat{f}(\chi) := \int_{G} f(x) \overline{\chi(x)} \,d \mu_{G}(x), \quad \chi \in \widehat{G}.
    \end{equation}
\end{definition}
Given a Harr measure $\mu_{\widehat{G}}$ on $\widehat{G}$, there exists a normalization constant $C$ such that the following Fourier inversion formula holds:
\begin{equation}
    f(x) = C \int_{\widehat{G}} \hat{f}(\chi) \chi(x) \, d\mu_{\widehat{G}}(\chi), \quad x \in G.
\end{equation}

\begin{example}{\hfill}
    \begin{itemize}
        \item The classical Fourier transform on $\mbb{R}^N$: 
        \begin{align}
            \hat{f}(\bs{y}) &= \int_{\mbb{R}^N} f(\bs{x}) \epower{- \im \bs{x} \cdot \bs{y}} \,d\bs{x},  \quad \forall \bs{y} \in \mbb{R}^N \\
            f(\bs{x}) &= \frac{1}{(2\pi)^N} \int_{\mbb{R}^N} \hat{f}(\bs{y}) \epower{\im \bs{x} \cdot \bs{y}} \, d\bs{y}, \quad \forall \bs{x} \in \mbb{R}^N.
        \end{align}
        \item The discrete Fourier transform on $\mbb{Z}_d$:
        \begin{align}
            \hat{f}(k) &= \sum_{m \in \mbb{Z}_d} f(m) \epower{-\frac{2\pi \im}{d} mk} , \quad  \forall k \in \mbb{Z}_d ,\\
            f(m) &= \frac{1}{d} \sum_{k \in \mbb{Z}_d} \hat{f}(k) \epower{\frac{2\pi \im}{d} mk}, \quad \forall m \in \mbb{Z}_d.
        \end{align}
    \end{itemize}
\end{example}

The Fourier transform on $\mbb{R}^N$ extends naturally to the space $\mcal{S}'(\mbb{R}^N)$ of tempered distributions. For a tempered distribution $T \in \mcal{S}'(\mbb{R}^N)$, its Fourier transform $\hat{T}$ is defined by
\begin{equation}
    \bra \hat{T},\phi \ket = \bra T, \hat{\phi} \ket, \quad \forall \phi \in \mcal{S}(\mbb{R}^N),
\end{equation}
where $\mcal{S}(\mbb{R}^N)$ denotes the set of Schwartz functions on $\mbb{R}^N$.

\section{Relations Between the Continuous Fourier Transform and Discrete Fourier Transform \label{sec:CFT and DFT}}

Consider a function $f$ on a finite set $\prod_{i=1}^N V_i \subset \mbb{R}^N$, where $V_i = \{v_{i,k}\}_{k\in[d]}$ for all $i \in \{1,\ldots,N\}$. 
We denote the pullback $\bs{v}^{*}f(\bs{k}) := f(\bs{v}(\bs{k}))$ by $\bs{v}(\bs{k}) := (v_{1,k_1},\ldots,v_{N,k_N})$, simply as $f$.
The function $f$ is identified with a distribution supported on $\prod_{i=1}^N V_i$, defined as 
\begin{equation}
        {F}(\bs{x}) = \sum_{\bs{k} \in [d]^N} f(\bs{k}) \delta_{\bs{v}(\bs{k})}(\bs{x}), \quad \forall \bs{x} \in \mbb{R}^N.
        \label{eq:a function F supported on a finite set}
\end{equation}
We investigate the relationship between the continuous Fourier transform of ${F}$ and the discrete Fourier transform of $f$.

The set 
\begin{equation}
    \mcal{D}\left(\prod_{i=1}^N V_i\right) 
    := \left\{\sum_{\bs{k} \in [d]^N} f(\bs{k}) \delta_{\bs{v}(\bs{k})}(\bs{x}) \,\middle|\, f : [d]^N \to \mbb{C} \right\}
\end{equation}
of complex distributions supported on $\prod_{i=1}^N V_i$ forms a finite-dimensional linear space.
The Fourier transform of $F$ is given by
\begin{align}
    \hat{F}(\bs{s}) &= \int_{\mbb{R}^N} \sum_{\bs{k} \in [d]^N} f(\bs{k}) \delta_{\bs{v}(\bs{k})}(\bs{x}) \epower{- \im \bs{s} \cdot \bs{x}} \, d\bs{x}\notag \\
    &= \sum_{\bs{k} \in [d]^N} f(\bs{k}) \epower{- \im \bs{s}\cdot \bs{v}(\bs{k})}, \quad \bs{s} \in [d]^N.
    \label{eq:FT of a function supported on a finite set}
\end{align}
The set 
\begin{equation}
\hat{\mcal{D}}\left(\prod_{i=1}^N V_i\right)
:= \left\{\sum_{\bs{k} \in [d]^N} f(\bs{k}) \epower{- \im \bs{s}\cdot \bs{v}(\bs{k})}  \,\middle| \, f:[d]^N \to \mbb{C}\right\}
\end{equation}
of the Fourier transforms of complex distributions supported on $\prod_{i=1}^N V_i$ also forms a finite-dimensional linear space.
The Fourier transform $\mscr{F}_{\mbb{R}^N}$ (restricted to this domain) is a linear mapping from $\mcal{D}\big(\prod_{i=1}^N V_i\big)$ to $\hat{\mcal{D}}\big(\prod_{i=1}^N V_i\big)$.

If we take $\{\delta_{\bs{v}(\bs{k})}\}_{\bs{k}\in[d]^N}$ as a basis of $\mcal{D}\big(\prod_{i=1}^N V_i\big)$, and $\{\epower{- \im \bs{s}\cdot \bs{v}(\bs{k})}\}_{\bs{k} \in [d]^N}$ as a basis of $\hat{\mcal{D}}\big(\prod_{i=1}^N V_i\big)$, then both $F$, given by~\eqref{eq:a function F supported on a finite set}, and $\hat{F}$, given by~\eqref{eq:FT of a function supported on a finite set}, are represented by the vector $\bs{f}:= (f(\bs{k}))_{\bs{k}\in[d]^N}$, with multi-dimensional indices, in these bases. 
This implies that the matrix representation of the Fourier transform in these bases is the identity matrix.

Now, consider transforming the basis $\{\epower{- \im \bs{s}\cdot \bs{v}(\bs{k})}\}_{\bs{k} \in [d]^N}$ with a regular matrix 
\begin{equation}
    G_{N,d}:=\Big(\frac{1}{d}\omega_d^{\bs{s}\cdot \bs{k}}\Big)_{\bs{k},\bs{s}}, 
\end{equation}
where $\omega_d$ is a primitive $d$-th root of unity:
\begin{equation}
    \omega_d := \exp \left(\frac{2\pi \im}{d}\right).
\end{equation}
Then, the matrix representation of the Fourier transform $\mscr{F}_{\mbb{R}^N}$ takes the form
\begin{equation}
    F_{N,d} := G_{N,d}^{-1} 
    = (\omega_d^{-\bs{s}\cdot \bs{k}})_{\bs{s},\bs{k}}.
\end{equation}
For $N=1$, $F_{1,d}$ is explicitly expressed as
\begin{equation}
    F_{1,d} = \begin{pmatrix}
        \omega_{d}^{-0 \cdot 0} & \omega_d^{-0 \cdot 1} & \cdots & \omega_{d}^{-0\cdot (d-1)} \\
        \omega_{d}^{-1 \cdot 0} & \omega_{d}^{-1 \cdot 1} & \cdots & \omega_d^{-1 \cdot (d-1)} \\
        \vdots & \vdots & \ddots & \vdots \\
        \omega_d^{-(d-1) \cdot 0} & \omega_{d}^{-(d-1)\cdot 1} & \cdots & \omega_d^{-(d-1)\cdot (d-1)}
        \end{pmatrix}.
\end{equation}

The vector that represents $\hat{F}$ in the transformed basis is given by $\bs{\phi} = (\phi(\bs{s}))_{\bs{s} \in [d]^N} := F_{N,d}\bs{f}$, i.e., 
\begin{equation}
    \phi(\bs{s}) = \sum_{\bs{k} \in [d]^N} \omega_d^{- \bs{s}\cdot \bs{k}} f(\bs{k}),
\end{equation}
which coincides with the discrete Fourier transform of $f$ as a function on $\mbb{Z}_d^N \cong [d]^N$.
In other words, the discrete Fourier transform appears as the matrix representation of the continuous Fourier transform with appropriate bases.
% A clearer explanation of these bases is provided in Appendix~\ref{sec: marginalization in finite spaces}.

\section{Marginalization as Projection}

Let $\vect{X} := (X_1,\ldots,X_n)$ be a tuple of observables taking values in $\vect{\Omega} := \Omega_1 \times \cdots \times \Omega_n$, and let $\tilde{\vect{X}} := (X_{k_1},\ldots,X_{k_m})$ be a sub-tuple taking values in $\tilde{\vect{\Omega}} := \Omega_{k_1} \times \cdots \times \Omega_{k_m}$, where $1 \leq k_1 < \cdots < k_m \leq n$.

Consider a map $\pi : \vect{x} \mapsto (\tilde{\vect{x}},\vect{0})$, where $\tilde{\vect{x}} := (x_{k_1}, \ldots, x_{k_m}) \in \tilde{\vect{\Omega}}$ and $\vect{0} \defeq (0,\ldots,0)$ denotes the zero tuple of cardinality $n-m$. We define a projection on $D(\vect{\Omega})$ by
\begin{equation}\label{def:marginalization_projection}
M : p_{\vect{X}} \mapsto p_{\tilde{\vect{X}}}(\vect{x}) = \int_{\pi^{-1}(\vect{x})}\, dp_{\vect{X}}(\vect{x}').
\end{equation}
Since both the map $\tilde{M}$ in \eqref{def:marginalization} and the map $M$ in \eqref{def:marginalization_projection} are constant on each other's fibers, the surjectivity of $\tilde{M}$ implies the existence of a unique linear injection $T : D(\tilde{\vect{\Omega}}) \to D(\vect{\Omega})$ such that $M = T \circ \tilde{M}$.
This establishes the identification $D(\tilde{\vect{\Omega}}) = \ran \tilde{M} \simeq \ran M \subset D(\vect{\Omega})$, allowing us to view the distribution space over the subset $\tilde{\vect{\Omega}}$ as a subspace of that over the full set $\vect{\Omega}$.

By construction, $M$ is idempotent:
\begin{equation}
M^2 = M,
\end{equation}
and hence a projection operator.  
This is equivalent to the statement that $T$ is a (linear) section of $\tilde{M}$:
\begin{equation}
\tilde{M} \circ T = \operatorname{id}.
\end{equation}

It follows that the space $D(\vect{\Omega})$ decomposes as the algebraic direct sum:
\begin{equation}
D(\vect{\Omega}) \simeq D(\tilde{\vect{\Omega}}) \oplus \ker M,
\end{equation}
which provides a geometric interpretation of the marginalization process.

\section{Marginalization in Finite Spaces \label{sec: marginalization in finite spaces}}
We derive marginalization on finite sample spaces as a restriction of marginalization on an encompassing continuous space.

Let us consider a random variable $\bs{Z} := (Z_1, \ldots, Z_N)$ taking values in $\mbb{R}^N$, and a probability distribution $P$ over $\bs{Z}$.
Define $X_k := Z_{i_k}$ for $k \in \{1,\ldots,K\}$ and $Y_l := Z_{j_l}$ for $l \in \{1,\ldots,L\}$, where $K+L=N$ and $\{i_1,\ldots,i_K\} \cup \{j_1,\ldots,j_L\} = \{1,\ldots,N\}$. 
The marginal probability distribution $P_{\bs{X}}$ for $\bs{X}:=(X_1,\ldots,X_K)$ obtained by integrating $P$ over $\bs{Y}:=(Y_1,\ldots,Y_L)$ is given by
\begin{equation}
    P_{\bs{X}}(\bs{x}) = \int_{\mbb{R}^L} P([\bs{x},\bs{y}])\, d\bs{y}, \quad \forall \bs{x} \in \mbb{R}^K,
    \label{eq:marginalization (continuous)}
\end{equation}
where the sorting map $[\,\cdot\,,\,\cdot\,]: \mbb{R}^K \times \mbb{R}^L \to \mbb{R}^N$ is defined as follows:
\begin{subequations}
    \begin{align}
        [\bs{x},\bs{y}] &:= \bs{z}, \\
         z_{i_k} &:= x_k, \quad \forall k\in\{1,\ldots,K\}, \\
          z_{j_l} &:= y_l, \quad \forall l \in \{1,\ldots,L\}.
    \end{align}
\end{subequations}

The Fourier transforms of $P$ and $P_{\bs{X}}$ are given by
\begin{gather}
    \hat{P}(\bs{u}) = (\mscr{F}_{\mbb{R}^N}P)(\bs{u}) := \int_{\mbb{R}^N} \epower{-\im \bs{u}\cdot \bs{z}} P(\bs{z})\, d\bs{z}, \quad \forall \bs{u}\in \mbb{R}^N, \label{eq:FT of P (continuous)} \\
    \hat{P}_{\bs{X}} = (\mscr{F}_{\mbb{R}^K}P_{\bs{X}})(\bs{s}) := \int_{\mbb{R}^K} \epower{- \im \bs{s}\cdot \bs{x}} P_{\bs{X}}(\bs{x}) \, d\bs{x}, \quad \forall \bs{s} \in \mbb{R}^K,  \label{eq:FT of P_X (continuous)}
\end{gather}
respectively. 
It follows from~\eqref{eq:marginalization (continuous)},~\eqref{eq:FT of P (continuous)}, and~\eqref{eq:FT of P_X (continuous)} that 
\begin{equation}
    \hat{P}_{\bs{X}}(\bs{s}) = \hat{P}([\bs{s},\bs{0}]), \quad \forall \bs{s} \in \mbb{R}^K.
\end{equation}

If the distribution $P$ is supported on a finite set $\prod_{i=1}^N V_i$, where $V_i := \{v_{i,k}\}_{k\in[d]}$ for $i \in \{1,\ldots,N\}$, $P$ can be
expressed in the form
\begin{equation}
    P(\bs{z}) = \sum_{\bs{n} \in [d]^N} p(\bs{n}) \delta_{\bs{Z}(\bs{n})}(\bs{z}), \quad \forall \bs{z} \in \mbb{R}^N, 
\end{equation}
where $\bs{Z}(\bs{k}):=(v_{1,k_1}, \ldots, v_{N,k_N})$. Then, the marginal distribution $P_{\bs{X}}$ is given by
\begin{equation}
    P_{\bs{X}}(\bs{x}) = \sum_{\bs{k}\in[d]^L} p_{\bs{X}}(\bs{k}) \delta_{\bs{X}(\bs{k})} (\bs{x}),\quad \forall \bs{x} \in \mbb{R}^K,
\end{equation}
where 
\begin{align}
    p_{\bs{X}}(\bs{k}) &:= \sum_{\bs{l}\in [d]^L} p([\bs{k},\bs{l}]),\quad \forall \bs{k} \in [d]^K, \label{eq:marginal distribution vector}\\
    \bs{X}(\bs{k}) &:= (v_{i_1, k_{1}}, \ldots, v_{i_K, k_K}), \quad \forall \bs{k}\in[d]^K.
\end{align}

If we chose transformed bases for the spaces $\hatmcal{D}\big(\prod_{i=1}^N V_{i}\big)$ and $\hatmcal{D}\big(\prod_{k=1}^K V_{i_k}\big)$ as in Appendix~\ref{sec:CFT and DFT}, the vectors that represent $\hat{P}$ and $\hat{P}_{\bs{X}}$ are given by
\begin{align}
    \bs{\phi} &= (\phi(\bs{u}))_{\bs{u} \in [d]^N}, \quad \phi(\bs{u}) := \sum_{\bs{n}\in [d]^N} \omega_d^{-\bs{u}\cdot \bs{n}} p(\bs{n}),\\
    \bs{\phi}_{\bs{X}} &= (\phi_{\bs{X}}(\bs{s}))_{\bs{s}\in[d]^K}, \quad
    \phi_{\bs{X}}(\bs{s}) := \sum_{\bs{k}\in [d]^K}\omega_d^{-\bs{s}\cdot \bs{k}} p_{\bs{X}}(\bs{k}), 
\end{align}
respectively. It follows from~\eqref{eq:marginal distribution vector} that 
\begin{equation}
    \phi_{\bs{X}}(\bs{s}) = \phi([\bs{s},\bs{0}]), \quad \forall \bs{s} \in [d]^K.
    \label{eq:marginal characteristic function vector}
\end{equation}
% Equations~\eqref{eq:marginal distribution vector} and~\eqref{eq:marginal characteristic function vector} correspond to~\eqref{eq:marginal distribution (discrete)} and~\eqref{eq:marginal characteristic function (discrete)}, respectively.

% Therefore, the marginalization of $P$ corresponds to a projection $\msf{Proj}_{\bs{X}}:\mbb{C}^{d^N} \to \mbb{C}^{d^N}$ defined by
% \begin{equation}
%     \bs{\psi} = \begin{pmatrix}
%         (\psi([\bs{s},\bs{0}])_{\bs{s}\in[d]^K}) \\
%         (\psi([\bs{s},\bs{t}]))_{\bs{s}\in [d]^K, \bs{t} \neq \bs{0}}
%     \end{pmatrix}
%     \mapsto \begin{pmatrix}
%         (\psi([\bs{s},\bs{0}])_{\bs{s}\in[d]^K}) \\
%         \bs{0}
%     \end{pmatrix}.
% \end{equation}
% This shows that the matrix representation of the projection associated with the marginalization of $P$ is diagonal in the bases introduced in Appendix~\ref{sec:CFT and DFT}.

\section{Quasi-Joint Probability Distributions with Finite Supports \label{sec:QJP with finite supports}}
\subsection{Discrete QJP and QJS Densities}
If a quasi-joint probability or spectral distributions is supported on a finite set, it is sufficient to know a finite number of components to fully specify the distribution. We demonstrate it below.

Consider a QJS distribution $\hash_{\bs{X}}$ supported on a finite set $\prod_{i=1}^N V_i \subset \mbb{R}^N$, where $V_i := \{v_{i,k}\}_{k \in [d]}$. Then, $\hash_{\bs{X}}$ can be expressed as 
\begin{equation}
    \hash_{\bs{X}}(\bs{x}) = \sum_{\bs{k}\in[d]^N} \hash_{\bs{X}}(\bs{v}({\bs{k}})) \delta_{\bs{v}(\bs{k})}(\bs{x}),
    \quad \bs{x}\in \mbb{R}^N,
    \label{eq:discrete QJS density}
\end{equation}
where $\bs{v}(\bs{k}):=(v_{1,k_1},\ldots,v_{N,k_N})$ is the index function, and, for $\bs{a} \in \mbb{R}^N$, $\delta_{\bs{a}}$ is defined as 
\begin{equation}
    \delta_{\bs{a}}(\bs{x}) := \delta (\bs{x} - \bs{a}).
\end{equation}
% We call $\hash_{\bs{X}}^{[d]}$ the {\textit{discrete quasi-joint spectral (QJS) density}}. 
Using the pullback of the QJS distribution, defined as $\hash_{\bs{X}}^{[d]}(\bs{k}) := \hash_{\bs{X}}(\bs{v}(\bs{k}))$, we can express~\eqref{eq:discrete QJS density} as
\begin{equation}
    \hash_{\bs{X}}(\bs{x}) = \sum_{\bs{k}\in[d]^N} \hash_{\bs{X}}^{[d]}({\bs{k}}) \delta_{\bs{v}(\bs{k})}(\bs{x}),
    \quad \bs{x}\in \mbb{R}^N.
\end{equation}
% For simplicity, we write $\hash_{\bs{X}}^{[d]}(\bs{k})$ instead of $\bs{v}^{*}\hash_{\bs{X}}^{[d]}(\bs{k})$. 
We call $\hash_{\bs{X}}^{[d]}$ the \textit{discrete quasi-joint spectral (QJS) density}.

The {Fourier transform of the discrete QJS density $\hash_{\bs{X}}^{[d]}$} is defined as:
\begin{equation}
    \hat{\hash}_{\bs{X}}^{[d]}(\bs{s}) := (\mscr{F}_{[d]^N}\hash_{\bs{X}}^{[d]})(\bs{s})
    = \sum_{\bs{k} \in [d]^N} \omega_d^{- \bs{s} \cdot \bs{k}} \hash_{\bs{X}}^{[d]}({\bs{k}}).
\end{equation}
{Here, the hat symbol $\hat{\,}$} denotes the discrete Fourier transform of $\hash_{\bs{X}}^{[d]}$, not the hashed operator in the continuous case.
The QJP distribution corresponding to $\hash_{\bs{X}}$ is given by
\begin{multline}
    Q_{{\bs{X}}}(\bs{x}) = \Tr[\rho \hash_{\bs{X}}(\bs{x})] \\
    = \sum_{\bs{k} \in [d]^N} Q_{{\bs{X}}}(\bs{v}({\bs{k}})) \delta_{\bs{v}(\bs{k})}(\bs{x}), \quad 
    \bs{x} \in \mbb{R}^N.
\end{multline}
% where the {\textit{discrete quasi-joint probability density}} is defined as
% \begin{equation}
%     \rho_{\hash_{\bs{X}}}^{[d]}(\bs{v}({\bs{k}})) := \Tr[\rho \hash_{\bs{X}}^{[d]}(\bs{v}({\bs{k}}))], \quad \bs{k} \in [d]^N.
% \end{equation}
We call the pullback of the QJP distribution $Q_{{\bs{X}}}^{[d]}({\bs{k}}) := Q_{{\bs{X}}}(\bs{v}({\bs{k}}))$ the \textit{discrete quasi-joint probability (QJP) density}.
The characteristic function of the discrete QJP density $Q_{{\bs{X}}}^{[d]}$ is given by
\begin{equation}
    \hat{Q}_{{\bs{X}}}^{[d]}(\bs{s}) := (\mscr{F}_{[d]^N}Q_{{\bs{X}}}^{[d]})(\bs{s})
    = \Tr[\rho \hat{\hash}_{\bs{X}}^{[d]}(\bs{s})], \quad \bs{s}\in[d]^N.
\end{equation}
Therefore, to describe a QJP distribution on $[d]^N$, it is sufficient to know $\hash_{\bs{X}}^{[d]}$ and $\hat{\hash}_{\bs{X}}^{[d]}$.

The marginalization of {the discrete QJS densities} and their Fourier transforms proceeds in a similar manner to the continuous case. 
In the same setting as the previous subsubsection, the following relations hold:
\begin{align}
    \hash_{\bs{X}}^{[d]}({\bs{k}}) &= \sum_{\bs{l} \in [d]^L}\hash_{\bs{Z}}^{[d]}([{\bs{k}},{\bs{l}}]), \quad \bs{k}\in[d]^K\\
    \hat{\hash}_{\bs{X}}^{[d]}(\bs{s}) &= \hat{\hash}_{\bs{Z}}^{[d]}([\bs{s},\bs{0}]), \quad \bs{s}\in[d]^K.
\end{align}

\subsection{The Kirkwood-Dirac Distribution}
In the following, we consider specifically the Kirkwood-Dirac distribution $Q_{{\bs{X}}}^{\text{KD}}$:
\begin{equation}
    \hat{\hash}_{\bs{X}}^{\text{KD}}(\bs{s}) := \prod_{k=1}^N \epower{- \im s_k X_k}.
\end{equation}
Let $\sigma(X)$ denote the spectrum of an operator $X$. Since the Kirkwood-Dirac distribution is supported on $\prod_{i=1}^N \sigma(X_i)$ \cite{umekawa2024advantages}, if the distribution $Q_{{\bs{X}}}^{\text{KD}}$ is supported on $\prod_{i=1}^N V_i$, we can assume that $\sigma(X_i) = V_i$ for all $i \in \{1,\ldots,N\}$. Then, in the sense of distributions, the spectral measure of $X_i$ can be expressed as 
\begin{equation}
    E_{X_i} (x) = \sum_{k \in [d]} E_{i,k} \delta_{v_{i,k}}(x), \quad x\in\mbb{R},
\end{equation}
The spectral decomposition of $X_i$ is given by
\begin{equation}
    X_i = \sum_{k \in [d]} v_{i,k} E_{i,k}.
\end{equation}
The QJS distribution is expressed as
\begin{multline}
    \hash_{\bs{X}}^{\text{KD}}(\bs{x}) = \prod_{i=1}^{N}E_{X_i}(x_i)
    = \prod_{i=1}^{N} \sum_{k \in [d]} E_{i,k} \delta_{v_{i,k}}(x_i) \\
    = \sum_{\bs{k} \in [d]^N} \left(\prod_{i=1}^{N} E_{i,k_i}\right)\delta_{\bs{v}(\bs{k})}(\bs{x}), 
    \quad \bs{x} \in \mbb{R}^N.
    \label{eq:KD QJSD supported on V^N}
\end{multline}
% Thus, we obtain
% \begin{equation}
%     \rho_{\hash_{\bs{X}}^{\text{KD}}}(\bs{x}) = \sum_{\bs{k}\in[d]^N} \Tr\left[\rho \prod_{i=1}^{N}E_{i,k_i} \right] \delta(\bs{x}-\bs{k}).
%     \label{eq:KD distribution on [d]^N}
% \end{equation}
Comparing~\eqref{eq:KD QJSD supported on V^N} with~\eqref{eq:discrete QJS density} yields
\begin{equation}
    (\hash_{\bs{X}}^{\text{KD}})^{[d]}({\bs{k}}) = \prod_{i= 1}^{N} E_{i,k_i}, \quad \bs{k} \in [d]^N.
\end{equation}
The Fourier transform of the discrete QJS density $(\hash_{\bs{X}}^{\text{KD}})^{[d]}$ is given by
\begin{align*}
    (\hat{\hash}_{\bs{X}}^{\text{KD}})^{[d]}(\bs{s})
    &= \sum_{\bs{k}\in[d]^N}\omega_d^{- \bs{k} \cdot \bs{s}}  \prod_{i= 1}^{N} E_{i,k_i} \\
    &= \prod_{i=1}^N \left(\sum_{k_i \in [d]} \omega_d^{- k_i s_i} E_{i,k_i}\right) \\
    &= \prod_{i=1}^{N} \omega_d^{- s_i \theta_i^{-1} (X_i)},\quad \bs{s} \in [d]^N, \numberthis
\end{align*}
where the function $\theta_i:[d] \to V_i$ is defined by $\theta_i(k) := v_{i,k}$ and thus 
\begin{equation}
    \theta_i^{-1}(X_i) = \sum_{k \in [d]} k E_{i,k}.
\end{equation}

\section{Geometric Interpretation of the Problems}

\subsection{Geometric Interpretation of Problem LR}
We present a geometric interpretation of Problem LR.
For simplicity, we explain in the $(2,m,d)$ setup. 
The generalization to the $(n,m,d)$ setup is straightforward. 
% Suppose that Alice's observables $A_1,\ldots, A_m$ and Bob's observables $B_1,\ldots,B_m$ take values from $[d]$.
{By using the pullbacks of joint probability distributions, we can treat the observables $A_1,\ldots,A_m,B_1,\ldots,B_m$ as if each of them took values in $[d]$.}

If there exists a joint probability distribution $P$ for all $2m$ observables that reproduces the marginal distributions $(P_{A_i B_j})_{i,j}$, it must satisfy the following equations: 
\begin{multline}
    P_{A_i B_j} (a_i,b_j) = \sum_{\bs{a}_i^\times , \bs{b}_j^\times \in [d]^{m-1}} P(a_1,\ldots,a_m,b_1,\ldots,b_m),\\
     \forall a_i,b_j \in [d], \quad \forall i,j \in \{1,\ldots,m\},
    \label{eq:marginal distribution P_AiBj}
\end{multline}
where
\begin{subequations}
    \begin{align}
        \bs{a}_i^\times &:= (a_1,\ldots,a_{i-1},a_{i+1},\ldots,a_m), \\
        \quad \bs{b}_j^\times &:= (b_1,\ldots,b_{j-1},b_{j+1},\ldots,b_m).
    \end{align}
\end{subequations}
Defining $[\,\cdot\,,\,\cdot\,]_i:[d] \times [d]^{m-1} \to [d]^m, (a_i,\bs{a}_i^\times) \mapsto (a_1,\ldots,a_m)$, we can rewrite~\eqref{eq:marginal distribution P_AiBj} as
\begin{multline}
    P_{A_i B_j}(a,b) = \sum_{\bs{a}^\times, \bs{b}^\times \in [d]^{m-1}} P([a,\bs{a}^\times]_i, [b, \bs{b}^\times]_j),\\
    \quad \forall a,b \in [d], \forall i,j \in \{1,\ldots,m\}.
    \label{eq:marginal distribution P_AiBj 2}
\end{multline}

The characteristic functions of $P$ and $P_{A_i B_j}$ are defined as follows:
\begin{align}
    \hat{P}(\bs{s},\bs{t}) &:= \sum_{\bs{a},\bs{b} \in [d]^{m}} \omega_d^{-\bs{s}\cdot \bs{a}} \omega_{d}^{-\bs{t}\cdot \bs{b}} P({\bs{a}},{\bs{b}}), \\
    \hat{P}_{A_i B_j}(s, t) &:= \sum_{a,b \in [d]} \omega_d^{- sa} \omega_d^{-tb} P_{A_iB_j}(a,b).
\end{align}
Using~\eqref{eq:marginal distribution P_AiBj 2}, we obtain
\begin{align*}
    \hat{P}_{A_i B_j} (s,t)&= \hat{P}\big(\underbrace{0,\ldots,\overset{i \text{-th}}{s},\ldots,0}_{m},\underbrace{0,\ldots,\overset{j \text{-th}}{t},\ldots,0}_{m}\big) \\
    &= \hat{P}([s,\bs{0}]_i, [t,\bs{0}]_j) , \\
    & \qquad \forall s,t\in[d], \forall i,j \in \{1,\ldots,m\}. \numberthis
\end{align*}
Equivalently, 
\begin{equation}
    (\phi_{A_i B_j})_{s,t} = \phi_{[s,\bs{0}]_i, [t,\bs{0}]_j} , \quad \forall s,t \in [d],\forall i,j \in \{1,\ldots,m\},
    \label{eq:marginal characteristic function vector phi_AiBj}
\end{equation}
where $(\phi_{A_i B_j})_{s,t}:=\hat{P}_{A_i B_j}(s,t)$ and $\phi_{\bs{s},\bs{t}} := \hat{P}(\bs{s},\bs{t})$. 

Let $\msf{Proj}$ denote a projection in the space of characteristic function vectors defined as
\begin{multline}
    \msf{Proj}:\mbb{C}^{d^{2m}} \to \mbb{C}^{d^{2m}}, \\
    \bs{\psi}=\begin{pmatrix}
        (\psi_{\bs{s},\bs{t}})_{(\bs{s},\bs{t})\in I \times I} \\ (\psi_{\bs{s,\bs{t}}})_{(\bs{s},\bs{t}) \notin I\times I}
    \end{pmatrix} \mapsto \begin{pmatrix}
        (\psi_{\bs{s},\bs{t}})_{(\bs{s},\bs{t})\in I \times I} \\ \bs{0}
    \end{pmatrix},\label{eq:Proj}
\end{multline}
where
\begin{equation}
  I := \{[s,\bs{0}]_i \in [d]^{m} \mid  i \in \{1,\ldots,m\}, s\in [d]\}.
\end{equation}

As the map $\ran(\msf{Proj}) \to \mbb{C}^{m^2d^2}, \left(\begin{smallmatrix}
    (\psi_{\bs{s},\bs{t}})_{(\bs{s},\bs{t})\in I\times I} \\ \bs{0}
\end{smallmatrix}\right) \mapsto (\psi_{\bs{s},\bs{t}})_{(\bs{s},\bs{t})\in I\times I}$ is bijective, from~\eqref{eq:marginal characteristic function vector phi_AiBj}, we have
\begin{equation}
    \bs{\phi}_{\bs{AB}}
 = (\phi_{\bs{s},\bs{t}})_{(\bs{s},\bs{t})\in I\times I}
    \cong \msf{Proj} (\bs{\phi}),
    \label{eq:marginal characteristic function vector phi_AiBj 2}
\end{equation}
where
\begin{align}
    \bs{\phi}_{\bs{AB}} &:= (\bs{\phi}_{A_i B_j})_{i,j=1}^{m}, \\
    \bs{\phi}_{A_i B_j} &:= ((\phi_{A_i B_j})_{s,t})_{s,t \in [d]}.
\end{align}

Note that the projection in \eqref{eq:Proj} is not equivalent to the collection of projections defined by \eqref{def:marginalization_projection} for each pair $(A_i, B_j)$.
This discrepancy arises from the difference in how each marginal distribution is embedded into the original space prior to projection.

\begin{figure}[htbp]
    \centering
    \includegraphics[width=75mm]{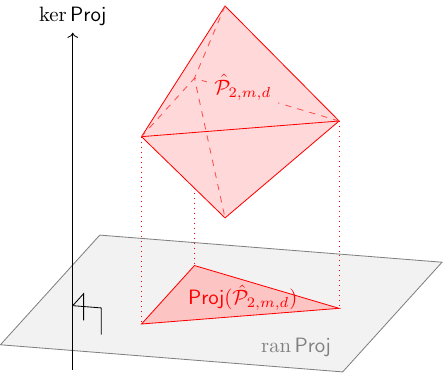}
    \caption{A characteristic function vector $\bs{\phi}$ that is compatible with $\bs{\phi}_{\bs{AB}}$ exists if and only if $\bs{\phi}_{\bs{AB}}$ belongs to the image $\msf{Proj}(\hatmcal{P}_{2,m,d})$ of the set $\hatmcal{P}_{2,m,d}$ of characteristic function vectors under the projection $\msf{Proj}$. The inequalities characterizing the `shade' $\msf{Proj}(\hatmcal{P}_{2,m,d})$ correspond to `all the Bell inequalities'.}
    \label{fig:proj_LR}
\end{figure}

First, consider the case where the constraint $R$ is faithful, \ie, all information about joint the probability distributions $(P_{A_i B_j})_{i,j=1}^{m}$ is available.
Under this condition, the components $({\phi}_{A_i B_j})_{s,t}$ are determined for all $s,t\in[d]$ and $i,j\in\{1,\ldots,m\}$.  
Therefore, the problem reduces to finding a necessary and sufficient condition for a vector $\bs{\phi}$ satisfying~\eqref{eq:marginal characteristic function vector phi_AiBj 2} to exist. 
The necessary and sufficient condition is that the vector $\bs{\gamma} := \bs{\phi}_{\bs{AB}}$ belongs to the image $\msf{Proj}(\hatmcal{P}_{2,m,d})$ of the projection $\msf{Proj}$, applied to the set $\hatmcal{P}_{2,m,d}$, which consists of characteristic function vectors of probability distributions on $[d]^{2m}$. 
In other words, the inequalities characterizing the `shade' $\msf{Proj}(\hatmcal{P}_{2,m,d})$ are the statistical constraints we want to find, which correspond to `all the Bell inequalities' (Fig.~\ref{fig:proj_LR}).

The set $\hatmcal{P}_{2,m,d}$ of characteristic function vectors is obtained as follows.
% A vector $\bs{p} = (p_{\bs{a},\bs{b}})_{\bs{a},\bs{b}}$ represents a probability distribution if and only if the following conditions are satisfied:
The set $\mcal{P}_{2,m,d}$ of probability distribution vectors on $[d]^{2m}$ is given by
\begin{multline}
    \mcal{P}_{2,m,d} 
    = \Bigg\{ \bs{p}= (P({\bs{a},\bs{b}}))_{\bs{a},\bs{b}\in [d]^m} \in \mbb{C}^{d^{2m}} 
    \,\Bigg|\, \\ \sum_{\bs{a},\bs{b} \in [d]^m} P({\bs{a},\bs{b}}) = 1,  
    \quad \forall \bs{a},\bs{b} \in [d]^m,
    P({\bs{a},\bs{b}}) \ge 0 \Bigg\}.
\end{multline}
Obviously, $\mcal{P}_{2,m,d}$ is a convex set.
Since the inverse Fourier transform is given by
\begin{equation}
    P({\bs{a}},{\bs{b}}) = \frac{1}{d^{2m}} \sum_{\bs{s},\bs{t} \in [d]^m} \omega_d^{\bs{s}\cdot\bs{a}} \omega_{d}^{\bs{t}\cdot\bs{b}} \phi_{\bs{s},\bs{t}},\quad \forall \bs{a},\bs{b} \in[d]^m,
\end{equation}
the set $\hatmcal{P}_{2,m,d}$ is expressed as follows:
\begin{multline}
    \hatmcal{P}_{2,m,d}
    = \Bigg\{ \bs{\phi}=(\phi_{\bs{s},\bs{t}})_{\bs{s},\bs{t}\in[d]^m} \in \mbb{C}^{d^{2m}} 
    \,\Bigg|\,\\ \phi_{\bs{0},\bs{0}}=1,\quad
    \forall \bs{k},\bs{l} \in [d]^m, \sum_{\bs{s},\bs{t} \in [d]^m} \omega_d^{\bs{s}\cdot\bs{k}} \omega_{d}^{\bs{t}\cdot\bs{l}} \phi_{\bs{s},\bs{t}} \ge 0 \Bigg\}.
\end{multline}
The set $\hatmcal{P}_{2,m,d}$ is also convex.

Next, consider a general constraint $R$. Suppose that partial information $\bs{\gamma} = R(\bs{\phi}_{\bs{AB}})$ about the joint probability distributions $(P_{A_i B_j})_{i,j=1}^{m}$ is given. Then, the problem we adress is whether a vector $\bs{\phi}$ satisfying the following equation exists:
\begin{equation}
    \bs{\gamma} = R(\msf{Proj}(\bs{\phi})).
\end{equation}
Thus, the set of possible values of $\bs{\gamma}$ corresponds to the image $R(\msf{Proj}(\hatmcal{P}_{2,m,d}))$ of the `shade' $\msf{Proj}(\hatmcal{P}_{2,m,d})$ under $R$. 
In particular, when $R$ is a map that returns some components $(\phi_{A_i B_j})_{s,t}$, \ie, $R$ is expressed as 
\begin{equation}
    R : ((\phi_{A_i B_j})_{s,t})_{s,t;i,j} \mapsto ((\phi_{A_i B_j})_{s,t})_{(s,t) \in J_{ij};i,j}
    \label{eq:constraint that gives some components}
\end{equation}
for some sets $J_{ij} \, (i,j \in [d])$, $R$ can also be interpreted as a projection. 
Consequently, the composition
\begin{multline}
    R \circ \msf{Proj} :\mbb{C}^{d^{2m}} \to \mbb{C}^{d^{2m}}, \\
     \bs{\psi} = \begin{pmatrix}
        (\psi_{\bs{s},\bs{t}})_{(\bs{s},\bs{t})\in J} \\ (\psi_{\bs{s},\bs{t}})_{(\bs{s},\bs{t})\notin J}
    \end{pmatrix}\mapsto \begin{pmatrix}
        (\psi_{\bs{s},\bs{t}})_{(\bs{s},\bs{t})\in J} \\ \bs{0}
    \end{pmatrix},
    \label{eq:composition of R and Proj}
\end{multline}
where
\begin{multline}
    J := \Big\{ ([s,\bs{0}]_i,[t,\bs{0}]_j) \in [d]^{m}\times [d]^{m} \, \Big| \, \\
    i,j\in \{1,\ldots,m\}, (s,t) \in J_{ij} \Big\},
    \label{eq:definition of the set J}
\end{multline}
is also a projection.
Hence, $R(\msf{Proj}(\hatmcal{P}_{2,m,d}))$ can also be regarded as the `shade' of $\hatmcal{P}_{2,m,d}$ under the projection $R \circ \msf{Proj}$.

\subsection{Geometric Interpretation of Problem LQ}
We consider Problem LQ in the $(2,m,d)$ setup. The generalization to the $(n,m,d)$ setup is straightforward. 

We first consider the fixed observables setting.
Let $A_1,\ldots,A_m,B_1,\ldots, B_m$ be $d$-valued self-adjoint operators without degeneracy. 
% We can assume each has the spectrum $[d]$ without loss of generality.
The quantum state of the composite system is specified by a density operator $\rho$ on the Hilbert space $\mcal{H}:=\mbb{C}^d \otimes \mbb{C}^d$. 
We simply write $A_i$ and $B_j$ for the operators $A_i \otimes I_d$ and $I_d \otimes B_j$, respectively, on $\mcal{H}$, where $I_d$ denotes the identity operator on $\mbb{C}^d$. 
As $A_i \otimes I_d$ and $I_d \otimes B_j$ strongly commute, the joint spectral measure $E_{(A_i,B_j)}$ exists. The joint probability distributions for $A_i$ and $B_j$ are given by
\begin{equation}
    P_{A_i B_j}(a,b) = \Tr[\rho E_{(A_i,B_j)}(a,b)] = \Tr[\rho E_{A_i}(a) \otimes E_{B_j}(b)].
\end{equation}
 {By using the pullbacks of (quasi-)probability distributions and (quasi-)joint spectral distributions, we can treat the observables $A_1,\ldots,A_m,B_1,\ldots,B_m$ as if each of them took values in $[d]$.}

\begin{figure}[htbp]
    \centering
    \includegraphics[width=75mm]{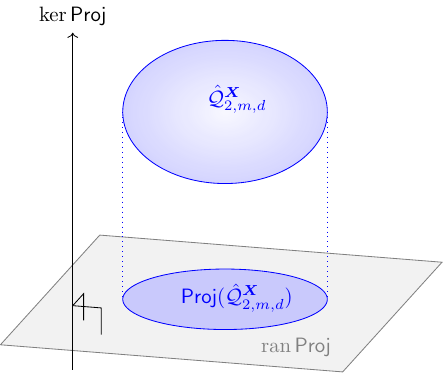}
    \caption{A density operator $\rho$ that is compatible with $\bs{\phi}_{\bs{AB}}$ and $(A_i,B_j)_{i,j=1}^{m}$ exists if and only if $\bs{\phi}_{\bs{AB}}$ belongs to the image $\msf{Proj}(\hatmcal{Q}^{\bs{X}}_{2,m,d})$ of the set $\hatmcal{Q}^{\bs{X}}_{2,m,d}$ under the projection $\msf{Proj}$.}
    \label{fig:proj_Q}
\end{figure}

Given the joint probability distributions $(P_{A_i B_j})_{i,j}$, there exists a global QJP distribution $Q^{[d]}_{{\bs{X}}}$ on $[d]^{2m}$ that satisfies the marginal property
% \begin{equation}
%     P_{A_i B_j}(a_i,b_j) = \sum_{\bs{a}_i^{\times},\bs{b}_j^{\times} \in [d]^{m-1}} Q_{\hash_{\bs{X}}}([a_i,\bs{a}_i^{\times}]_i, [b_j,\bs{b}_j^{\times}]_j),\quad 
%     \forall a_i,b_j \in [d],\forall i,j \in \{1,\ldots,m\},
% \end{equation}

\begin{multline}
    P_{A_i B_j}(a,b) = \sum_{\bs{a}^{\times},\bs{b}^{\times} \in [d]^{m-1}} Q^{[d]}_{{\bs{X}}}([a,\bs{a}^{\times}]_i, [b,\bs{b}^{\times}]_j),\\
    \forall a,b \in [d],\forall i,j \in \{1,\ldots,m\},
\end{multline}

where $\bs{X}:=(A_1,\ldots,A_m,B_1,\ldots,B_m)$.
Then, similar reasoning to that in the local realistic case is applicable.
The only difference from the local realistic case is that $Q^{[d]}_{{\bs{X}}}$ is a QJP distribution corresponding to a quantum state rather than a joint probability distribution.
Let $\bs{\phi}_{A_i B_j} = ((\phi_{A_i B_j})_{s,t})_{s,t \in[d]}:=(\hat{P}_{A_i B_j}(s,t))_{s,t \in[d]}$ denote the characteristic function vector of $P_{A_i B_j}$, and let $\bs{\varphi} = (\varphi_{\bs{s},\bs{t}})_{\bs{s},\bs{t} \in[d]} := (\hat{Q}^{[d]}_{{\bs{X}}}(\bs{s},\bs{t}))_{\bs{s},\bs{t}\in [d]}$ denote the characteristic function vector of $Q_{{\bs{X}}}$. 
Similarly to Eq.~\eqref{eq:marginal characteristic function vector phi_AiBj 2}, the following equation holds:
\begin{equation}
    \bs{\phi}_{\bs{AB}} := 
    (\bs{\phi}_{A_i B_j})_{i,j=1}^{m} = (\varphi_{\bs{s},\bs{t}})_{(\bs{s},\bs{t})\in I\times I} \cong \msf{Proj}(\bs{\varphi}).
\end{equation}
The necessary and sufficient condition for a vector $\bs{\varphi}$ corresponding to a quantum state to exist is that $(\bs{\phi}_{\bs{AB}})$ belongs to the image $\msf{Proj}(\hatmcal{Q}^{\bs{X}}_{2,m,d})$, where $\hatmcal{Q}^{\bs{X}}_{2,m,d}$ denotes the set of the characteristic function vectors of QJP distributions on $[d]^{2m}$ corresponding to quantum states (Fig.~\ref{fig:proj_Q}).
If we consider a constraint $R$ and a set of data $\bs{\gamma}$ is given, the following equation holds:
\begin{equation}
    \bs{\gamma} = R (\msf{Proj}(\bs{\varphi})).
\end{equation}
Thus, the set of possible values of $\bs{\gamma}$ corresponds to the image $R(\msf{Proj}(\hatmcal{Q}^{\bs{X}}_{2,m,d}))$ of the `shade' $\msf{Proj}(\hatmcal{Q}^{\bs{X}}_{2,m,d})$ under $R$.

The set $\hatmcal{Q}^{\bs{X}}_{2,m,d}$ is derived as follows. Let $\Phi_{\hash_{\bs{X}}}$ denote the map $\rho \mapsto \hat{Q}^{[d]}_{{\bs{X}}}$. The set $\hatmcal{Q}^{\bs{X}}_{2,m,d}$ can be expressed as 
\begin{equation}
    \hatmcal{Q}^{\bs{X}}_{2,m,d} = \{ \Phi_{\hash_{\bs{X}}}(\rho) \in \mbb{C}^{d^{2m}} \mid \Tr[\rho]=1,\quad  \rho \geq 0\}.
\end{equation}
Since the condition $\Tr[\rho]=1$ is equivalent to $\varphi_{\bs{0},\bs{0}}=1$, it follows that $\hatmcal{Q}^{\bs{X}}_{2,m,d}$ can be rewritten as
\begin{equation}
    \hatmcal{Q}^{\bs{X}}_{2,m,d} = \{\bs{\varphi} \in \mbb{C}^{d^{2m}} \mid \varphi_{\bs{0},\bs{0}}=1, \quad \exists \rho \geq 0 , \Phi_{\hash_{\bs{X}}}(\rho)=\bs{\varphi}\}.
\end{equation}
If the QJS distribution $\hash_{\bs{X}}$ is faithful, \ie, an injective map, the inverse mapping $\Phi_{\hash_{\bs{X}}}^{-1}$ of $\Phi_{\hash_{\bs{X}}}$ exists, and then $\hatmcal{Q}^{\bs{X}}_{2,m,d}$ is given by
\begin{equation}
    \hatmcal{Q}^{\bs{X}}_{2,m,d} = \{\bs{\varphi} \in \mbb{C}^{d^{2m}}  \mid  \varphi_{\bs{0},\bs{0}}=1, \quad \Phi_{\hash_{\bs{X}}}^{-1}(\bs{\varphi}) \geq 0\}.
\end{equation}
The condition $\Phi_{\hash_{\bs{X}}}^{-1}(\bs{\varphi}) \geq 0$ is equivalent to the non-negativity of all the eigenvalues of $\Phi_{\hash_{\bs{X}}}^{-1}(\bs{\varphi})$, or alternatively other criteria \cite{kimura2003bloch,dangniam2015quantum} can be used.

Note that although the set $\hatmcal{Q}^{\bs{X}}_{2,m,d}$ depends on the selection of the QJS distribution $\hash_{\bs{X}}$, the projected image $\msf{Proj}(\hatmcal{Q}^{\bs{X}}_{2,m,d})$ does not depend on the selection of $\hash_{\bs{X}}$. Let us prove this.  
Consider two QJS distributions $\hash_{\bs{X}}$ and $\hash_{\bs{X}}'$. Given a density operator $\rho$, both the images of $\bs{\varphi}:=\Phi_{\hash_{\bs{X}}}(\rho)$ and $\bs{\varphi}' := \Phi_{\hash_{\bs{X}}}'(\rho)$ under the projection $\msf{Proj}$ coincide with $(\bs{\phi}_{A_i B_j})_{i,j}$. Therefore, the sets $\msf{Proj}(\hatmcal{Q}^{\bs{X}}_{2,m,d})$  and $\msf{Proj}({\hatmcal{Q}^{\bs{X}\prime}_{2,m,d}})$, representing the possible values of $\msf{Proj}(\Phi_{\hash_{\bs{X}}}(\rho))$ and $\msf{Proj}(\Phi'_{\hash_{\bs{X}}}(\rho))$, respectively, are identical.

The set of possible values of $\bs{\gamma}$ in the unfixed observable setting is given by the union of the images of the `shades' $\msf{Proj}(\hatmcal{Q}^{\bs{X}}_{2,m,d})$ taken over all possible observables $\bs{X}$.

\subsection{Replacement of Measured Observables}
% \memo{generalize $\bs{Y}$}

Further, we discuss a geometric interpretation of measurements of different sets of observables. 
We consider two sets of observables to be measured. 
In one case, the observables measured by Alice and Bob are $A_1,\ldots,A_m$ and $B_1,\ldots,B_{m}$, respectively. 
In the other case, the observables measured by Alice and Bob are $C_1,\ldots,C_{m'}$ and $D_1,\ldots,D_{m'}$, respectively. 
In both cases, Alice and Bob conduct measurements on a quantum state $\rho$.
Let $\bs{X}:=(A_1,\ldots,A_m,B_1,\ldots,B_m)$ and $\bs{Y}:=(C_1,\ldots,C_{m'},D_1,\ldots,D_{m'})$.

\begin{figure*}[htbp]
    \centering
    \includegraphics[width=130mm]{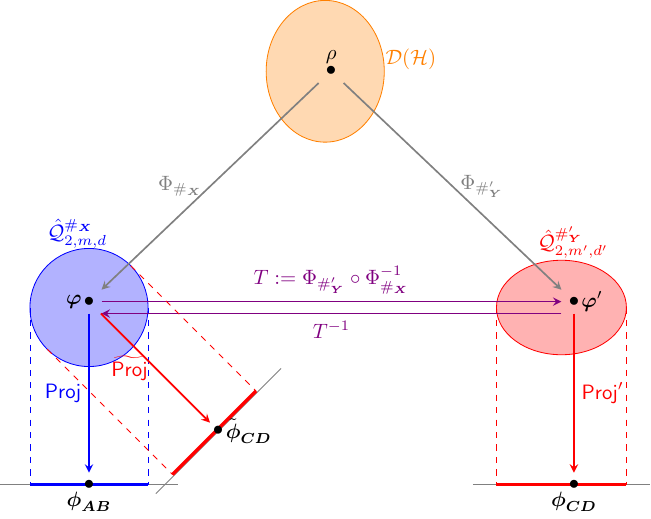}
    \caption{{The quasi-joint spectral distributions for different sets of observables and the corresponding marginalized probability distributions. The projection associated with the measurement of $\bs{Y}$ is interpreted as a non-orthogonal projection $\widetilde{\msf{Proj}}$ in the space of the characteristic functions of QJP distributions for $\bs{X}$}.}
    \label{fig:replacing observables}
\end{figure*}

Choosing two QJS distributions $\hash_{\bs{X}}$ and $\hash'_{\bs{Y}}$, we define $\bs{\varphi} := \Phi_{\hash_{\bs{X}}} (\rho)$ and $\bs{\varphi}' := \Phi_{\hash'_{\bs{Y}}} (\rho)$. 
Then, the characteristic functions $\bs{\phi}_{\bs{AB}}:=(\bs{\phi}_{A_i B_j})_{i,j=1}^{m}$ and $\bs{\phi}_{\bs{CD}}:=(\bs{\phi}_{C_i D_j})_{i,j=1}^{m'}$ are given by
\begin{align}
    \bs{\phi}_{\bs{AB}} &= \msf{Proj}(\bs{\varphi}) ,\\
    \bs{\phi}_{\bs{CD}} &= \msf{Proj}'(\bs{\varphi}'),
\end{align}
where $\msf{Proj}$ and $\msf{Proj}'$ are the projections corresponding to the marginalization in the $(2,m,d)$ setup and $(2,m',d')$ setup, respectively.
If $\hash_{\bs{X}}$ and $\hash'_{\bs{Y}}$ are faithful, the transformation $T$ from the space of characteristic function vectors for $\bs{X}$ to that for $\bs{Y}$ exists:
\begin{equation}
    T := \Phi_{\hash'_{\bs{Y}}} \circ \Phi_{\hash_{\bs{X}}}^{-1}.
\end{equation}
Then, the relation $\bs{\varphi}' = T(\bs{\varphi})$ holds, and the vector $\bs{\phi}_{\bs{CD}}$ corresponds to the vector $\tilde{\bs{\phi}}_{\bs{CD}} := T^{-1}(\bs{\phi}_{\bs{CD}})$ in the space of characteristic function vectors for $\bs{X}$. 
Therefore, it follows that
\begin{equation}
    \tilde{\bs{\phi}}_{\bs{CD}} = T^{-1} \circ \msf{Proj}' \circ T (\bs{\varphi})
    = \widetilde{\msf{Proj}}(\bs{\varphi}).
\end{equation}
Here, $\widetilde{\msf{Proj}}$ is the projection defined as $\widetilde{\msf{Proj}} := T^{-1} \circ \msf{Proj}' \circ T$, which, in general, is not orthogonal. These relations are illustrated in Fig.~\ref{fig:replacing observables}.

This result can be interpreted as follows. When we fix a faithful QJS distribution $\hash_{\bs{X}}$, the vector $\bs{\varphi}$ represents the quantum state $\rho$ uniquely, and the projected vector $\bs{\phi}_{\bs{AB}}=\msf{Proj}(\bs{\varphi})$ represents the joint distributions for $A_i$ and $B_j$. 
The vector $\tilde{\bs{\phi}}_{\bs{CD}} = \widetilde{\msf{Proj}}(\bs{\varphi})$, which is obtained by applying another projection to $\bs{\varphi}$, represents the joint distributions for $C_i$ and $D_j$. 
In other words, replacing the observables to be measured corresponds to changing the subspace on which the vector $\bs{\varphi}$ is projected. 
Only partial information about the quantum state can be obtained from the `shade' $\msf{Proj}(\bs{\varphi})$, which is obtained through the measurements of one set of observables. 
However, additional information about the quantum state can be obtained from another `shade', which is obtained through the measurements of another set of observables.

\section{More on the Unfixed Observables Quantum Problem}\label{sec:more on type I}

\begin{theorem}[{\cite[Theorem 2.1]{tsirelson1987quantum}}]\label{thm:QI in (2,m,2) setup}
    Given $C_{ij} \in \mbb{R}$ for $i \in \{1,\ldots,m\}$ and $j \in \{1,\ldots,m'\}$, the following five statements are equivalent:
    \begin{enumerate}[label={(\roman{*})}]
        \item There exist a $\mrm{C}^{*}$-algebra $\mscr{A}$ equipped with an identity element $I$, Hermitian elements $A_1,\ldots,A_m,B_1,\ldots,B_{m'}$, and a state $f$ on $\mscr{A}$ that satisfy the following conditions for all $i \in \{1,\ldots,m\}$ and $j \in \{1,\ldots,m'\}$: 
        \begin{gather}
            A_i B_j = B_j A_i ,\\
            -I \leq A_i \leq I, \quad -I \leq B_j \leq I, \\
            f(A_i B_j) = C_{ij}.
        \end{gather}
        \item There exist self-adjoint operators $A_1,\ldots,A_m, B_1,\ldots,B_n$, and a density operator $\rho$ on a finite-dimensional or separable Hilbert space $\mcal{H}$ that satisfy the following conditions for all $i \in \{1,\ldots,m\}$ and $j \in \{1,\ldots,m'\}$:
        \begin{gather}
            A_i B_j = B_j A_i , \\
            \sigma(A_i) \subset [-1,1], \quad \sigma(B_j) \subset [-1,1], \\
            \Tr[\rho A_i B_j] = C_{ij},
        \end{gather}
        where $\sigma(X)$ denotes the spectrum of an operator $X$.
        \item There exist self-adjoint operators $A_1,\ldots,A_m, B_1,\ldots,B_n$, and a density operator $\rho$ on a finite-dimensional Hilbert space $\mcal{H}=\mcal{H}_1 \otimes \mcal{H}_2$ that satisfy the same conditions as in (ii) and the following conditions: 
        \begin{itemize}
            \item $A_i^2=B_j^2 = I$ for all $i \in \{1,\ldots,m\}$ and $j \in \{1,\ldots,m'\}$.
            \item $A_i = A_i^{(1)} \otimes I^{(2)}$ for all $i \in \{1,\ldots,m\}$ and $B_j = I^{(1)} \otimes B_j^{(2)}$ for all $j \in \{1,\ldots,m'\}$, where $A_i^{(1)}$ and $B_j^{(2)}$ are respectively operators on $\mcal{H}_1$ and $\mcal{H}_2$, and $I^{(1)}$ and $I^{(2)}$ are respectively the identity operators on $\mcal{H}_1$ and $\mcal{H}_2$.
            \item $A_{i_1}^{(1)} A_{i_2}^{(1)} + A_{i_2}^{(1)} A_{i_1}^{(1)}$ is proportional to the identity operator for all $i_1,i_2 \in \{1,\ldots,m\}$, and $B_{j_1}^{(2)} B_{j_2}^{(2)} + B_{j_{2}}^{(2)} B_{j_1}^{(2)}$ is proportional to the identity operator for all $j_1 , j_2 \in \{1,\ldots,m'\}$.
            \item The dimensions of $\mcal{H}_1$ and $\mcal{H}_2$ are bounded by the following inequalities:
            \begin{gather}
                2 \log_2 \dim \mcal{H}_1 \leq \begin{cases}
                    m & (\text{for even $m$}) \\
                    m+1 & (\text{for odd $m$})
                \end{cases} ,\\
                2 \log_2 \dim \mcal{H}_2 \leq \begin{cases}
                    m' & (\text{for even $m'$}) \\
                    m' + 1 & (\text{for odd $m'$})
                \end{cases}.
            \end{gather}
        \end{itemize}
        \item There exist unit vectors $\bs{x}_1, \ldots, \bs{x}_m, \bs{y}_1, \ldots,\bs{y}_{m'} \in \mbb{R}^{m+m'}$ that satisfy $\bs{x}_i \cdot \bs{y}_j = C_{ij}$ for all $i \in \{1,\ldots,m\}$ and $j \in \{1,\ldots,m'\}$.
        \item There exist vectors $\bs{x}_1, \ldots, \bs{x}_m, \bs{y}_1, \ldots,\bs{y}_{m'} \in \mbb{R}^{\min \{m,m'\}}$ that satisfy $|\bs{x}_i|\leq 1$, $|\bs{y}_j|\leq 1$, and $\bs{x}_i \cdot \bs{y}_j = C_{ij}$ for all $i \in \{1,\ldots,m\}$ and $j \in \{1,\ldots,m'\}$.

    \end{enumerate}
\end{theorem}
While (iii)$\Rightarrow$(ii) is trivial, (ii)$\Rightarrow$(iii) seems non-trivial as (iii) imposes additional conditions. 
The point is that the density operator $\rho$ and self-adjoint operators $A_1,\ldots,A_m,B_1,\ldots,B_{m'}$ in (iii) can be different from those in (ii). 
In (iii), it is not necessarily possible to take $\rho$ and $A_1,\ldots,A_m,B_1,\ldots,B_{m'}$ as a representation of the $\mrm{C}^{*}$-algebra $\mscr{A}$ in (i), whereas this is possible in (ii).

Theorem~\ref{thm:QI in CHSH setup} is obtained by a direct application of Theorem~\ref{thm:QI in (2,m,2) setup} to the CHSH setting \cite{tsirelson1987quantum}.

Any local realistic correlations $\{C_{ij}\}$ can be described by a quantum state and observables given by~\eqref{eq:density operator that reproduces LR} and~\eqref{eq:observables that reproduce LR}, which implies the validity of (ii) in Theorem~\ref{thm:QI in (2,m,2) setup}. Since (ii) is equivalent to (iii), especially in the $(2,2,2)$ setup, the correlations $\{C_{ij}\}$ can also be described by a density operator and self-adjoint operators on a Hilbert space of dimension $\dim \mcal{H}_1 \times \dim \mcal{H}_2 = 2 \times 2 = 4$.

\section{On the Equivalence of Different Representations of the Tsirel'son--Landau Inequality \label{sec:equivalence of Tsirel'son--Landau}}

We explain the equivalence between the conditions stated in Theorem~\ref{thm:QI in CHSH setup} and Landau's inequality~\eqref{eq:Landau's inequality for pm1 observables (original)} in the CHSH setup.

Landau \cite{landau1988empirical} pointed out that, in quantum mechanics, the following matrix must be positive semi-definite:
\begin{equation}
    \begin{pmatrix}
        \bra A_1 A_1 \ket & \bra A_1 A_2 \ket & \bra A_1 B_1 \ket & \bra A_1 B_2 \ket \\
        \bra A_1 A_2 \ket^* & \bra A_2 A_2 \ket & \bra A_2 B_1 \ket & \bra A_2 B_2 \ket \\
        \bra A_1 B_1 \ket^* & \bra A_2 B_1 \ket^* & \bra B_1 B_1 \ket & \bra B_1 B_2 \ket \\
        \bra A_1 B_2 \ket^* & \bra A_2 B_2 \ket^* & \bra B_1 B_2 \ket^* & \bra B_2 B_2 \ket
    \end{pmatrix}.
\end{equation}
Therefore, there exist complex numbers $x,y \in \mbb{C}$ such that the following matrix is positive semi-definite:
\begin{equation}
    {\Gamma}:= \begin{pmatrix}
        \bra A_1 A_1 \ket & x & \bra A_1 B_1 \ket & \bra A_1 B_2 \ket \\
        x^* & \bra A_2 A_2 \ket &  \bra A_2 B_1 \ket & \bra A_2 B_2 \ket \\
        \bra A_1 B_1 \ket^* & \bra A_2 B_1 \ket^* & \bra B_1 B_1 \ket & y \\
        \bra A_1 B_2 \ket^* & \bra A_2 B_2 \ket^* & y^* & \bra B_2 B_2 \ket
    \end{pmatrix},
\end{equation}
which is equivalent to the validity of~\eqref{eq:Landau's inequality (original)} \cite{landau1988empirical}.

On the other hand, the condition $\Gamma \geq 0$ is equivalent to the existence of vectors $\bs{v}_1,\ldots,\bs{v}_4 \in \mbb{R}^4$ such that $\Gamma_{i,j} = \bs{v}_i \cdot \bs{v}_j$. 
Thus, defining $\bs{x}_1 := \bs{v}_1, \bs{x}_2 := \bs{v}_2, \bs{y}_1 := \bs{v}_3, \bs{y}_2 := \bs{v}_4$ establishes the equivalence between the validity of Theorem~\ref{thm:QI in (2,m,2) setup} (iv) and the existence of $x,y\in\mbb{C}$ that make $\Gamma$ positive semi-definite, which is also equivalent to the condition stated in Theorem~\ref{thm:QI in CHSH setup} under the CHSH setup \cite{tsirelson1987quantum}.
Note that $|\bs{v}_i|^2 = \Gamma_{i,i}=1$ holds for $i\in\{1,\ldots,4\}$ when the observables take values in $\{\pm 1\}$.

% \bibliographystyle{unsrt}
% \bibliographystyle{apsrev4-2}

\bibliography{main}

\begin{thebibliography}{60}%
\makeatletter
\providecommand \@ifxundefined [1]{%
 \@ifx{#1\undefined}
}%
\providecommand \@ifnum [1]{%
 \ifnum #1\expandafter \@firstoftwo
 \else \expandafter \@secondoftwo
 \fi
}%
\providecommand \@ifx [1]{%
 \ifx #1\expandafter \@firstoftwo
 \else \expandafter \@secondoftwo
 \fi
}%
\providecommand \natexlab [1]{#1}%
\providecommand \enquote  [1]{``#1''}%
\providecommand \bibnamefont  [1]{#1}%
\providecommand \bibfnamefont [1]{#1}%
\providecommand \citenamefont [1]{#1}%
\providecommand \href@noop [0]{\@secondoftwo}%
\providecommand \href [0]{\begingroup \@sanitize@url \@href}%
\providecommand \@href[1]{\@@startlink{#1}\@@href}%
\providecommand \@@href[1]{\endgroup#1\@@endlink}%
\providecommand \@sanitize@url [0]{\catcode `\\12\catcode `\$12\catcode
  `\&12\catcode `\#12\catcode `\^12\catcode `\_12\catcode `\%12\relax}%
\providecommand \@@startlink[1]{}%
\providecommand \@@endlink[0]{}%
\providecommand \url  [0]{\begingroup\@sanitize@url \@url }%
\providecommand \@url [1]{\endgroup\@href {#1}{\urlprefix }}%
\providecommand \urlprefix  [0]{URL }%
\providecommand \Eprint [0]{\href }%
\providecommand \doibase [0]{https://doi.org/}%
\providecommand \selectlanguage [0]{\@gobble}%
\providecommand \bibinfo  [0]{\@secondoftwo}%
\providecommand \bibfield  [0]{\@secondoftwo}%
\providecommand \translation [1]{[#1]}%
\providecommand \BibitemOpen [0]{}%
\providecommand \bibitemStop [0]{}%
\providecommand \bibitemNoStop [0]{.\EOS\space}%
\providecommand \EOS [0]{\spacefactor3000\relax}%
\providecommand \BibitemShut  [1]{\csname bibitem#1\endcsname}%
\let\auto@bib@innerbib\@empty
%</preamble>
\bibitem [{\citenamefont {Einstein}\ \emph {et~al.}(1935)\citenamefont
  {Einstein}, \citenamefont {Podolsky},\ and\ \citenamefont
  {Rosen}}]{einstein1935can}%
  \BibitemOpen
  \bibfield  {author} {\bibinfo {author} {\bibfnamefont {A.}~\bibnamefont
  {Einstein}}, \bibinfo {author} {\bibfnamefont {B.}~\bibnamefont {Podolsky}},\
  and\ \bibinfo {author} {\bibfnamefont {N.}~\bibnamefont {Rosen}},\ }\bibfield
   {title} {\bibinfo {title} {Can quantum-mechanical description of physical
  reality be considered complete?},\ }\href
  {https://doi.org/10.1103/PhysRev.47.777} {\bibfield  {journal} {\bibinfo
  {journal} {Phys. Rev.}\ }\textbf {\bibinfo {volume} {47}},\ \bibinfo {pages}
  {777} (\bibinfo {year} {1935})}\BibitemShut {NoStop}%
\bibitem [{\citenamefont {Bell}(1964)}]{bell1964einstein}%
  \BibitemOpen
  \bibfield  {author} {\bibinfo {author} {\bibfnamefont {J.~S.}\ \bibnamefont
  {Bell}},\ }\bibfield  {title} {\bibinfo {title} {On the {E}instein {P}odolsky
  {R}osen paradox},\ }\href
  {https://doi.org/10.1103/PhysicsPhysiqueFizika.1.195} {\bibfield  {journal}
  {\bibinfo  {journal} {Physics Physique Fizika}\ }\textbf {\bibinfo {volume}
  {1}},\ \bibinfo {pages} {195} (\bibinfo {year} {1964})}\BibitemShut {NoStop}%
\bibitem [{\citenamefont {Clauser}\ \emph {et~al.}(1969)\citenamefont
  {Clauser}, \citenamefont {Horne}, \citenamefont {Shimony},\ and\
  \citenamefont {Holt}}]{clauser1969proposed}%
  \BibitemOpen
  \bibfield  {author} {\bibinfo {author} {\bibfnamefont {J.~F.}\ \bibnamefont
  {Clauser}}, \bibinfo {author} {\bibfnamefont {M.~A.}\ \bibnamefont {Horne}},
  \bibinfo {author} {\bibfnamefont {A.}~\bibnamefont {Shimony}},\ and\ \bibinfo
  {author} {\bibfnamefont {R.~A.}\ \bibnamefont {Holt}},\ }\bibfield  {title}
  {\bibinfo {title} {Proposed experiment to test local hidden-variable
  theories},\ }\href {https://doi.org/10.1103/PhysRevLett.23.880} {\bibfield
  {journal} {\bibinfo  {journal} {Physical Review Letters}\ }\textbf {\bibinfo
  {volume} {23}},\ \bibinfo {pages} {880} (\bibinfo {year} {1969})}\BibitemShut
  {NoStop}%
\bibitem [{\citenamefont {Clauser}\ and\ \citenamefont
  {Shimony}(1978)}]{clauser1978bell}%
  \BibitemOpen
  \bibfield  {author} {\bibinfo {author} {\bibfnamefont {J.~F.}\ \bibnamefont
  {Clauser}}\ and\ \bibinfo {author} {\bibfnamefont {A.}~\bibnamefont
  {Shimony}},\ }\bibfield  {title} {\bibinfo {title} {Bell's theorem.
  {E}xperimental tests and implications},\ }\href
  {https://doi.org/10.1088/0034-4885/41/12/002} {\bibfield  {journal} {\bibinfo
   {journal} {Reports on Progress in Physics}\ }\textbf {\bibinfo {volume}
  {41}},\ \bibinfo {pages} {1881} (\bibinfo {year} {1978})}\BibitemShut
  {NoStop}%
\bibitem [{\citenamefont {Fine}(1982{\natexlab{a}})}]{fine1982joint}%
  \BibitemOpen
  \bibfield  {author} {\bibinfo {author} {\bibfnamefont {A.}~\bibnamefont
  {Fine}},\ }\bibfield  {title} {\bibinfo {title} {Joint distributions, quantum
  correlations, and commuting observables},\ }\href
  {https://doi.org/10.1063/1.525514} {\bibfield  {journal} {\bibinfo  {journal}
  {Journal of Mathematical Physics}\ }\textbf {\bibinfo {volume} {23}},\
  \bibinfo {pages} {1306} (\bibinfo {year} {1982}{\natexlab{a}})}\BibitemShut
  {NoStop}%
\bibitem [{\citenamefont {Cirel'son}(1980)}]{cirel1980quantum}%
  \BibitemOpen
  \bibfield  {author} {\bibinfo {author} {\bibfnamefont {B.~S.}\ \bibnamefont
  {Cirel'son}},\ }\bibfield  {title} {\bibinfo {title} {Quantum generalizations
  of {B}ell's inequality},\ }\href {https://doi.org/10.1007/BF00417500}
  {\bibfield  {journal} {\bibinfo  {journal} {Letters in Mathematical Physics}\
  }\textbf {\bibinfo {volume} {4}},\ \bibinfo {pages} {93} (\bibinfo {year}
  {1980})}\BibitemShut {NoStop}%
\bibitem [{\citenamefont {Aspect}\ \emph {et~al.}(1981)\citenamefont {Aspect},
  \citenamefont {Grangier},\ and\ \citenamefont
  {Roger}}]{aspect1981experimental}%
  \BibitemOpen
  \bibfield  {author} {\bibinfo {author} {\bibfnamefont {A.}~\bibnamefont
  {Aspect}}, \bibinfo {author} {\bibfnamefont {P.}~\bibnamefont {Grangier}},\
  and\ \bibinfo {author} {\bibfnamefont {G.}~\bibnamefont {Roger}},\ }\bibfield
   {title} {\bibinfo {title} {{Experimental tests of realistic local theories
  via Bell's theorem}},\ }\href {https://doi.org/10.1103/PhysRevLett.47.460}
  {\bibfield  {journal} {\bibinfo  {journal} {Physical Review Letters}\
  }\textbf {\bibinfo {volume} {47}},\ \bibinfo {pages} {460} (\bibinfo {year}
  {1981})}\BibitemShut {NoStop}%
\bibitem [{\citenamefont {Aspect}\ \emph
  {et~al.}(1982{\natexlab{a}})\citenamefont {Aspect}, \citenamefont
  {Grangier},\ and\ \citenamefont {Roger}}]{aspect1982experimentalEPR}%
  \BibitemOpen
  \bibfield  {author} {\bibinfo {author} {\bibfnamefont {A.}~\bibnamefont
  {Aspect}}, \bibinfo {author} {\bibfnamefont {P.}~\bibnamefont {Grangier}},\
  and\ \bibinfo {author} {\bibfnamefont {G.}~\bibnamefont {Roger}},\ }\bibfield
   {title} {\bibinfo {title} {Experimental realization of
  {E}instein-{P}odolsky-{R}osen-{B}ohm {G}edankenexperiment: {A} new violation
  of {B}ell's inequalities},\ }\href
  {https://doi.org/10.1103/PhysRevLett.49.91} {\bibfield  {journal} {\bibinfo
  {journal} {Physical Review Letters}\ }\textbf {\bibinfo {volume} {49}},\
  \bibinfo {pages} {91} (\bibinfo {year} {1982}{\natexlab{a}})}\BibitemShut
  {NoStop}%
\bibitem [{\citenamefont {Aspect}\ \emph
  {et~al.}(1982{\natexlab{b}})\citenamefont {Aspect}, \citenamefont
  {Dalibard},\ and\ \citenamefont {Roger}}]{aspect1982time}%
  \BibitemOpen
  \bibfield  {author} {\bibinfo {author} {\bibfnamefont {A.}~\bibnamefont
  {Aspect}}, \bibinfo {author} {\bibfnamefont {J.}~\bibnamefont {Dalibard}},\
  and\ \bibinfo {author} {\bibfnamefont {G.}~\bibnamefont {Roger}},\ }\bibfield
   {title} {\bibinfo {title} {Experimental test of {B}ell's inequalities using
  time-varying analyzers},\ }\href
  {https://doi.org/10.1103/PhysRevLett.49.1804} {\bibfield  {journal} {\bibinfo
   {journal} {Physical Review Letters}\ }\textbf {\bibinfo {volume} {49}},\
  \bibinfo {pages} {1804} (\bibinfo {year} {1982}{\natexlab{b}})}\BibitemShut
  {NoStop}%
\bibitem [{\citenamefont {Freedman}\ and\ \citenamefont
  {Clauser}(1972)}]{freedman1972experimental}%
  \BibitemOpen
  \bibfield  {author} {\bibinfo {author} {\bibfnamefont {S.~J.}\ \bibnamefont
  {Freedman}}\ and\ \bibinfo {author} {\bibfnamefont {J.~F.}\ \bibnamefont
  {Clauser}},\ }\bibfield  {title} {\bibinfo {title} {Experimental test of
  local hidden-variable theories},\ }\href
  {https://doi.org/10.1103/PhysRevLett.28.938} {\bibfield  {journal} {\bibinfo
  {journal} {Physical Review Letters}\ }\textbf {\bibinfo {volume} {28}},\
  \bibinfo {pages} {938} (\bibinfo {year} {1972})}\BibitemShut {NoStop}%
\bibitem [{\citenamefont {Weihs}\ \emph {et~al.}(1998)\citenamefont {Weihs},
  \citenamefont {Jennewein}, \citenamefont {Simon}, \citenamefont
  {Weinfurter},\ and\ \citenamefont {Zeilinger}}]{weihs1998violation}%
  \BibitemOpen
  \bibfield  {author} {\bibinfo {author} {\bibfnamefont {G.}~\bibnamefont
  {Weihs}}, \bibinfo {author} {\bibfnamefont {T.}~\bibnamefont {Jennewein}},
  \bibinfo {author} {\bibfnamefont {C.}~\bibnamefont {Simon}}, \bibinfo
  {author} {\bibfnamefont {H.}~\bibnamefont {Weinfurter}},\ and\ \bibinfo
  {author} {\bibfnamefont {A.}~\bibnamefont {Zeilinger}},\ }\bibfield  {title}
  {\bibinfo {title} {Violation of {B}ell's inequality under strict {E}instein
  locality conditions},\ }\href {https://doi.org/10.1103/PhysRevLett.81.5039}
  {\bibfield  {journal} {\bibinfo  {journal} {Physical Review Letters}\
  }\textbf {\bibinfo {volume} {81}},\ \bibinfo {pages} {5039} (\bibinfo {year}
  {1998})}\BibitemShut {NoStop}%
\bibitem [{\citenamefont {Fine}(1982{\natexlab{b}})}]{fine1982hidden}%
  \BibitemOpen
  \bibfield  {author} {\bibinfo {author} {\bibfnamefont {A.}~\bibnamefont
  {Fine}},\ }\bibfield  {title} {\bibinfo {title} {Hidden variables, joint
  probability, and the {B}ell inequalities},\ }\href
  {https://link.aps.org/doi/10.1103/PhysRevLett.48.291} {\bibfield  {journal}
  {\bibinfo  {journal} {Physical Review Letters}\ }\textbf {\bibinfo {volume}
  {48}},\ \bibinfo {pages} {291} (\bibinfo {year}
  {1982}{\natexlab{b}})}\BibitemShut {NoStop}%
\bibitem [{\citenamefont {Peres}(1999)}]{peres1999all}%
  \BibitemOpen
  \bibfield  {author} {\bibinfo {author} {\bibfnamefont {A.}~\bibnamefont
  {Peres}},\ }\bibfield  {title} {\bibinfo {title} {All the bell
  inequalities},\ }\href {https://doi.org/10.1023/A:1018816310000} {\bibfield
  {journal} {\bibinfo  {journal} {Foundations of Physics}\ }\textbf {\bibinfo
  {volume} {29}},\ \bibinfo {pages} {589} (\bibinfo {year} {1999})}\BibitemShut
  {NoStop}%
\bibitem [{\citenamefont {Froissart}(1981)}]{froissart1981constructive}%
  \BibitemOpen
  \bibfield  {author} {\bibinfo {author} {\bibfnamefont {M.}~\bibnamefont
  {Froissart}},\ }\bibfield  {title} {\bibinfo {title} {{Constructive
  generalization of Bell's inequalities}},\ }\href
  {https://doi.org/10.1007/BF02903286} {\bibfield  {journal} {\bibinfo
  {journal} {Nuovo Cimento B;(Italy)}\ }\textbf {\bibinfo {volume} {64}}
  (\bibinfo {year} {1981})}\BibitemShut {NoStop}%
\bibitem [{\citenamefont {Garg}\ and\ \citenamefont
  {Mermin}(1984)}]{garg1984farkas}%
  \BibitemOpen
  \bibfield  {author} {\bibinfo {author} {\bibfnamefont {A.}~\bibnamefont
  {Garg}}\ and\ \bibinfo {author} {\bibfnamefont {N.~D.}\ \bibnamefont
  {Mermin}},\ }\bibfield  {title} {\bibinfo {title} {Farkas's lemma and the
  nature of reality: Statistical implications of quantum correlations},\ }\href
  {https://doi.org/10.1007/BF00741645} {\bibfield  {journal} {\bibinfo
  {journal} {Foundations of Physics}\ }\textbf {\bibinfo {volume} {14}},\
  \bibinfo {pages} {1} (\bibinfo {year} {1984})}\BibitemShut {NoStop}%
\bibitem [{\citenamefont {Pitowsky}(1991)}]{pitowsky1991correlation}%
  \BibitemOpen
  \bibfield  {author} {\bibinfo {author} {\bibfnamefont {I.}~\bibnamefont
  {Pitowsky}},\ }\bibfield  {title} {\bibinfo {title} {Correlation polytopes:
  their geometry and complexity},\ }\href {https://doi.org/10.1007/BF01594946}
  {\bibfield  {journal} {\bibinfo  {journal} {Mathematical Programming}\
  }\textbf {\bibinfo {volume} {50}},\ \bibinfo {pages} {395} (\bibinfo {year}
  {1991})}\BibitemShut {NoStop}%
\bibitem [{\citenamefont {Pitowsky}\ and\ \citenamefont
  {Svozil}(2001)}]{pitowsky2001optimal}%
  \BibitemOpen
  \bibfield  {author} {\bibinfo {author} {\bibfnamefont {I.}~\bibnamefont
  {Pitowsky}}\ and\ \bibinfo {author} {\bibfnamefont {K.}~\bibnamefont
  {Svozil}},\ }\bibfield  {title} {\bibinfo {title} {Optimal tests of quantum
  nonlocality},\ }\href {https://doi.org/10.1103/PhysRevA.64.014102} {\bibfield
   {journal} {\bibinfo  {journal} {Physical Review A}\ }\textbf {\bibinfo
  {volume} {64}},\ \bibinfo {pages} {014102} (\bibinfo {year}
  {2001})}\BibitemShut {NoStop}%
\bibitem [{\citenamefont {Kaszlikowski}\ \emph {et~al.}(2002)\citenamefont
  {Kaszlikowski}, \citenamefont {Kwek}, \citenamefont {Chen}, \citenamefont
  {\ifmmode~\dot{Z}\else \.{Z}\fi{}ukowski},\ and\ \citenamefont
  {Oh}}]{Kaszlikowski2002clauser}%
  \BibitemOpen
  \bibfield  {author} {\bibinfo {author} {\bibfnamefont {D.}~\bibnamefont
  {Kaszlikowski}}, \bibinfo {author} {\bibfnamefont {L.~C.}\ \bibnamefont
  {Kwek}}, \bibinfo {author} {\bibfnamefont {J.-L.}\ \bibnamefont {Chen}},
  \bibinfo {author} {\bibfnamefont {M.}~\bibnamefont {\ifmmode~\dot{Z}\else
  \.{Z}\fi{}ukowski}},\ and\ \bibinfo {author} {\bibfnamefont {C.~H.}\
  \bibnamefont {Oh}},\ }\bibfield  {title} {\bibinfo {title} {{Clauser-Horne
  inequality for three-state systems}},\ }\href
  {https://doi.org/10.1103/PhysRevA.65.032118} {\bibfield  {journal} {\bibinfo
  {journal} {Phys. Rev. A}\ }\textbf {\bibinfo {volume} {65}},\ \bibinfo
  {pages} {032118} (\bibinfo {year} {2002})}\BibitemShut {NoStop}%
\bibitem [{\citenamefont {Bacon}\ and\ \citenamefont
  {Toner}(2003)}]{bacon2003bell}%
  \BibitemOpen
  \bibfield  {author} {\bibinfo {author} {\bibfnamefont {D.}~\bibnamefont
  {Bacon}}\ and\ \bibinfo {author} {\bibfnamefont {B.~F.}\ \bibnamefont
  {Toner}},\ }\bibfield  {title} {\bibinfo {title} {Bell inequalities with
  auxiliary communication},\ }\href
  {https://doi.org/10.1103/PhysRevLett.90.157904} {\bibfield  {journal}
  {\bibinfo  {journal} {Physical review letters}\ }\textbf {\bibinfo {volume}
  {90}},\ \bibinfo {pages} {157904} (\bibinfo {year} {2003})}\BibitemShut
  {NoStop}%
\bibitem [{\citenamefont {Collins}\ and\ \citenamefont
  {Gisin}(2004)}]{collins2004relevant}%
  \BibitemOpen
  \bibfield  {author} {\bibinfo {author} {\bibfnamefont {D.}~\bibnamefont
  {Collins}}\ and\ \bibinfo {author} {\bibfnamefont {N.}~\bibnamefont
  {Gisin}},\ }\bibfield  {title} {\bibinfo {title} {{A relevant two qubit Bell
  inequality inequivalent to the CHSH inequality}},\ }\href
  {https://doi.org/10.1088/0305-4470/37/5/021} {\bibfield  {journal} {\bibinfo
  {journal} {Journal of Physics A: Mathematical and General}\ }\textbf
  {\bibinfo {volume} {37}},\ \bibinfo {pages} {1775} (\bibinfo {year}
  {2004})}\BibitemShut {NoStop}%
\bibitem [{\citenamefont {{\'S}liwa}(2003)}]{sliwa2003symmetries}%
  \BibitemOpen
  \bibfield  {author} {\bibinfo {author} {\bibfnamefont {C.}~\bibnamefont
  {{\'S}liwa}},\ }\bibfield  {title} {\bibinfo {title} {{Symmetries of the Bell
  correlation inequalities}},\ }\href
  {https://doi.org/10.1016/S0375-9601(03)01115-0} {\bibfield  {journal}
  {\bibinfo  {journal} {Physics Letters A}\ }\textbf {\bibinfo {volume}
  {317}},\ \bibinfo {pages} {165} (\bibinfo {year} {2003})}\BibitemShut
  {NoStop}%
\bibitem [{\citenamefont {Ito}\ \emph {et~al.}(2006)\citenamefont {Ito},
  \citenamefont {Imai},\ and\ \citenamefont {Avis}}]{ito2006bell}%
  \BibitemOpen
  \bibfield  {author} {\bibinfo {author} {\bibfnamefont {T.}~\bibnamefont
  {Ito}}, \bibinfo {author} {\bibfnamefont {H.}~\bibnamefont {Imai}},\ and\
  \bibinfo {author} {\bibfnamefont {D.}~\bibnamefont {Avis}},\ }\bibfield
  {title} {\bibinfo {title} {{Bell inequalities stronger than the
  Clauser-Horne-Shimony-Holt inequality for three-level isotropic states}},\
  }\href {https://doi.org/10.1103/PhysRevA.73.042109} {\bibfield  {journal}
  {\bibinfo  {journal} {Physical Review A—Atomic, Molecular, and Optical
  Physics}\ }\textbf {\bibinfo {volume} {73}},\ \bibinfo {pages} {042109}
  (\bibinfo {year} {2006})}\BibitemShut {NoStop}%
\bibitem [{\citenamefont {Wie{\'s}niak}\ \emph {et~al.}(2007)\citenamefont
  {Wie{\'s}niak}, \citenamefont {Badzi{\k{a}}g},\ and\ \citenamefont
  {{\.Z}ukowski}}]{wiesniak2007explicit}%
  \BibitemOpen
  \bibfield  {author} {\bibinfo {author} {\bibfnamefont {M.}~\bibnamefont
  {Wie{\'s}niak}}, \bibinfo {author} {\bibfnamefont {P.}~\bibnamefont
  {Badzi{\k{a}}g}},\ and\ \bibinfo {author} {\bibfnamefont {M.}~\bibnamefont
  {{\.Z}ukowski}},\ }\bibfield  {title} {\bibinfo {title} {{Explicit form of
  correlation-function three-setting tight Bell inequalities for three
  qubits}},\ }\href {https://doi.org/10.1103/PhysRevA.76.012110} {\bibfield
  {journal} {\bibinfo  {journal} {Physical Review A—Atomic, Molecular, and
  Optical Physics}\ }\textbf {\bibinfo {volume} {76}},\ \bibinfo {pages}
  {012110} (\bibinfo {year} {2007})}\BibitemShut {NoStop}%
\bibitem [{\citenamefont {Brunner}\ and\ \citenamefont
  {Gisin}(2008)}]{brunner2008partial}%
  \BibitemOpen
  \bibfield  {author} {\bibinfo {author} {\bibfnamefont {N.}~\bibnamefont
  {Brunner}}\ and\ \bibinfo {author} {\bibfnamefont {N.}~\bibnamefont
  {Gisin}},\ }\bibfield  {title} {\bibinfo {title} {{Partial list of bipartite
  Bell inequalities with four binary settings}},\ }\href
  {https://doi.org/10.1016/j.physleta.2008.01.052} {\bibfield  {journal}
  {\bibinfo  {journal} {Physics Letters A}\ }\textbf {\bibinfo {volume}
  {372}},\ \bibinfo {pages} {3162} (\bibinfo {year} {2008})}\BibitemShut
  {NoStop}%
\bibitem [{\citenamefont {P{\'a}l}\ and\ \citenamefont
  {V{\'e}rtesi}(2009)}]{pal2009quantum}%
  \BibitemOpen
  \bibfield  {author} {\bibinfo {author} {\bibfnamefont {K.~F.}\ \bibnamefont
  {P{\'a}l}}\ and\ \bibinfo {author} {\bibfnamefont {T.}~\bibnamefont
  {V{\'e}rtesi}},\ }\bibfield  {title} {\bibinfo {title} {{Quantum bounds on
  Bell inequalities}},\ }\href {https://doi.org/10.1103/PhysRevA.79.022120}
  {\bibfield  {journal} {\bibinfo  {journal} {Physical Review A—Atomic,
  Molecular, and Optical Physics}\ }\textbf {\bibinfo {volume} {79}},\ \bibinfo
  {pages} {022120} (\bibinfo {year} {2009})}\BibitemShut {NoStop}%
\bibitem [{\citenamefont {Werner}\ and\ \citenamefont
  {Wolf}(2001)}]{werner2001all}%
  \BibitemOpen
  \bibfield  {author} {\bibinfo {author} {\bibfnamefont {R.~F.}\ \bibnamefont
  {Werner}}\ and\ \bibinfo {author} {\bibfnamefont {M.~M.}\ \bibnamefont
  {Wolf}},\ }\bibfield  {title} {\bibinfo {title} {{All-multipartite
  Bell-correlation inequalities for two dichotomic observables per site}},\
  }\href {https://doi.org/10.1103/PhysRevA.64.032112} {\bibfield  {journal}
  {\bibinfo  {journal} {Physical Review A}\ }\textbf {\bibinfo {volume} {64}},\
  \bibinfo {pages} {032112} (\bibinfo {year} {2001})}\BibitemShut {NoStop}%
\bibitem [{\citenamefont {{\.Z}ukowski}\ and\ \citenamefont
  {Brukner}(2002)}]{zukowski2002bell}%
  \BibitemOpen
  \bibfield  {author} {\bibinfo {author} {\bibfnamefont {M.}~\bibnamefont
  {{\.Z}ukowski}}\ and\ \bibinfo {author} {\bibfnamefont
  {{\v{C}}.}~\bibnamefont {Brukner}},\ }\bibfield  {title} {\bibinfo {title}
  {Bell's theorem for general $n$-qubit states},\ }\href
  {https://doi.org/10.1103/PhysRevLett.88.210401} {\bibfield  {journal}
  {\bibinfo  {journal} {Physical review letters}\ }\textbf {\bibinfo {volume}
  {88}},\ \bibinfo {pages} {210401} (\bibinfo {year} {2002})}\BibitemShut
  {NoStop}%
\bibitem [{\citenamefont {Paterek}\ \emph {et~al.}(2006)\citenamefont
  {Paterek}, \citenamefont {Laskowski},\ and\ \citenamefont
  {\.{Z}ukowski}}]{paterek2006series}%
  \BibitemOpen
  \bibfield  {author} {\bibinfo {author} {\bibfnamefont {T.}~\bibnamefont
  {Paterek}}, \bibinfo {author} {\bibfnamefont {W.}~\bibnamefont {Laskowski}},\
  and\ \bibinfo {author} {\bibfnamefont {M.}~\bibnamefont {\.{Z}ukowski}},\
  }\bibfield  {title} {\bibinfo {title} {{ON SERIES OF MULTIQUBIT BELL'S
  INEQUALITIES}},\ }\href {https://doi.org/10.1142/S0217732306019414}
  {\bibfield  {journal} {\bibinfo  {journal} {Modern Physics Letters A}\
  }\textbf {\bibinfo {volume} {21}},\ \bibinfo {pages} {111} (\bibinfo {year}
  {2006})},\ \Eprint
  {https://arxiv.org/abs/https://doi.org/10.1142/S0217732306019414}
  {https://doi.org/10.1142/S0217732306019414} \BibitemShut {NoStop}%
\bibitem [{\citenamefont {Zukowski}(2006)}]{zukowski2006tight}%
  \BibitemOpen
  \bibfield  {author} {\bibinfo {author} {\bibfnamefont {M.}~\bibnamefont
  {Zukowski}},\ }\bibfield  {title} {\bibinfo {title} {{On tight multiparty
  Bell inequalities for many settings}},\ }\href
  {https://doi.org/10.1007/s11128-006-0020-7} {\bibfield  {journal} {\bibinfo
  {journal} {Quantum Information Processing}\ }\textbf {\bibinfo {volume}
  {5}},\ \bibinfo {pages} {287} (\bibinfo {year} {2006})}\BibitemShut {NoStop}%
\bibitem [{\citenamefont {Wu}\ \emph {et~al.}(2008)\citenamefont {Wu},
  \citenamefont {Badziag}, \citenamefont {Wie{\'s}niak},\ and\ \citenamefont
  {{\.Z}ukowski}}]{wu2008extending}%
  \BibitemOpen
  \bibfield  {author} {\bibinfo {author} {\bibfnamefont {Y.-C.}\ \bibnamefont
  {Wu}}, \bibinfo {author} {\bibfnamefont {P.}~\bibnamefont {Badziag}},
  \bibinfo {author} {\bibfnamefont {M.}~\bibnamefont {Wie{\'s}niak}},\ and\
  \bibinfo {author} {\bibfnamefont {M.}~\bibnamefont {{\.Z}ukowski}},\
  }\bibfield  {title} {\bibinfo {title} {{Extending Bell inequalities to more
  parties}},\ }\href {https://doi.org/10.1103/PhysRevA.77.032105} {\bibfield
  {journal} {\bibinfo  {journal} {Physical Review A—Atomic, Molecular, and
  Optical Physics}\ }\textbf {\bibinfo {volume} {77}},\ \bibinfo {pages}
  {032105} (\bibinfo {year} {2008})}\BibitemShut {NoStop}%
\bibitem [{\citenamefont {Collins}\ \emph {et~al.}(2002)\citenamefont
  {Collins}, \citenamefont {Gisin}, \citenamefont {Linden}, \citenamefont
  {Massar},\ and\ \citenamefont {Popescu}}]{collins2002bell}%
  \BibitemOpen
  \bibfield  {author} {\bibinfo {author} {\bibfnamefont {D.}~\bibnamefont
  {Collins}}, \bibinfo {author} {\bibfnamefont {N.}~\bibnamefont {Gisin}},
  \bibinfo {author} {\bibfnamefont {N.}~\bibnamefont {Linden}}, \bibinfo
  {author} {\bibfnamefont {S.}~\bibnamefont {Massar}},\ and\ \bibinfo {author}
  {\bibfnamefont {S.}~\bibnamefont {Popescu}},\ }\bibfield  {title} {\bibinfo
  {title} {{Bell inequalities for arbitrarily high-dimensional systems}},\
  }\href {https://doi.org/10.1103/PhysRevLett.88.040404} {\bibfield  {journal}
  {\bibinfo  {journal} {Physical review letters}\ }\textbf {\bibinfo {volume}
  {88}},\ \bibinfo {pages} {040404} (\bibinfo {year} {2002})}\BibitemShut
  {NoStop}%
\bibitem [{\citenamefont {Massar}\ \emph {et~al.}(2002)\citenamefont {Massar},
  \citenamefont {Pironio}, \citenamefont {Roland},\ and\ \citenamefont
  {Gisin}}]{massar2002resistant}%
  \BibitemOpen
  \bibfield  {author} {\bibinfo {author} {\bibfnamefont {S.}~\bibnamefont
  {Massar}}, \bibinfo {author} {\bibfnamefont {S.}~\bibnamefont {Pironio}},
  \bibinfo {author} {\bibfnamefont {J.}~\bibnamefont {Roland}},\ and\ \bibinfo
  {author} {\bibfnamefont {B.}~\bibnamefont {Gisin}},\ }\bibfield  {title}
  {\bibinfo {title} {Bell inequalities resistant to detector inefficiency},\
  }\href {https://doi.org/10.1103/PhysRevA.66.052112} {\bibfield  {journal}
  {\bibinfo  {journal} {Phys. Rev. A}\ }\textbf {\bibinfo {volume} {66}},\
  \bibinfo {pages} {052112} (\bibinfo {year} {2002})}\BibitemShut {NoStop}%
\bibitem [{\citenamefont {Nagata}\ \emph {et~al.}(2006)\citenamefont {Nagata},
  \citenamefont {Laskowski},\ and\ \citenamefont {Paterek}}]{nagata2006bell}%
  \BibitemOpen
  \bibfield  {author} {\bibinfo {author} {\bibfnamefont {K.}~\bibnamefont
  {Nagata}}, \bibinfo {author} {\bibfnamefont {W.}~\bibnamefont {Laskowski}},\
  and\ \bibinfo {author} {\bibfnamefont {T.}~\bibnamefont {Paterek}},\
  }\bibfield  {title} {\bibinfo {title} {Bell inequality with an arbitrary
  number of settings and its applications},\ }\href
  {https://doi.org/10.1103/PhysRevA.74.062109} {\bibfield  {journal} {\bibinfo
  {journal} {Physical Review A—Atomic, Molecular, and Optical Physics}\
  }\textbf {\bibinfo {volume} {74}},\ \bibinfo {pages} {062109} (\bibinfo
  {year} {2006})}\BibitemShut {NoStop}%
\bibitem [{\citenamefont {Masanes}(2003)}]{masanes2003tight}%
  \BibitemOpen
  \bibfield  {author} {\bibinfo {author} {\bibfnamefont {L.}~\bibnamefont
  {Masanes}},\ }\bibfield  {title} {\bibinfo {title} {{Tight Bell inequality
  for $d$-outcome measurements correlations}},\ }\href@noop {} {\bibfield
  {journal} {\bibinfo  {journal} {Quantum Info. Comput.}\ }\textbf {\bibinfo
  {volume} {3}},\ \bibinfo {pages} {345} (\bibinfo {year} {2003})}\BibitemShut
  {NoStop}%
\bibitem [{\citenamefont {Ji}\ \emph {et~al.}(2008)\citenamefont {Ji},
  \citenamefont {Lee}, \citenamefont {Lim}, \citenamefont {Nagata},\ and\
  \citenamefont {Lee}}]{ji2008multisetting}%
  \BibitemOpen
  \bibfield  {author} {\bibinfo {author} {\bibfnamefont {S.-W.}\ \bibnamefont
  {Ji}}, \bibinfo {author} {\bibfnamefont {J.}~\bibnamefont {Lee}}, \bibinfo
  {author} {\bibfnamefont {J.}~\bibnamefont {Lim}}, \bibinfo {author}
  {\bibfnamefont {K.}~\bibnamefont {Nagata}},\ and\ \bibinfo {author}
  {\bibfnamefont {H.-W.}\ \bibnamefont {Lee}},\ }\bibfield  {title} {\bibinfo
  {title} {{Multisetting Bell inequality for qudits}},\ }\href
  {https://doi.org/10.1103/PhysRevA.78.052103} {\bibfield  {journal} {\bibinfo
  {journal} {Physical Review A—Atomic, Molecular, and Optical Physics}\
  }\textbf {\bibinfo {volume} {78}},\ \bibinfo {pages} {052103} (\bibinfo
  {year} {2008})}\BibitemShut {NoStop}%
\bibitem [{\citenamefont {Liang}\ \emph {et~al.}(2009)\citenamefont {Liang},
  \citenamefont {Lim},\ and\ \citenamefont {Deng}}]{liang2009reexamination}%
  \BibitemOpen
  \bibfield  {author} {\bibinfo {author} {\bibfnamefont {Y.-C.}\ \bibnamefont
  {Liang}}, \bibinfo {author} {\bibfnamefont {C.-W.}\ \bibnamefont {Lim}},\
  and\ \bibinfo {author} {\bibfnamefont {D.-L.}\ \bibnamefont {Deng}},\
  }\bibfield  {title} {\bibinfo {title} {{Reexamination of a multisetting Bell
  inequality for qudits}},\ }\href {https://doi.org/10.1103/PhysRevA.80.052116}
  {\bibfield  {journal} {\bibinfo  {journal} {Physical Review A—Atomic,
  Molecular, and Optical Physics}\ }\textbf {\bibinfo {volume} {80}},\ \bibinfo
  {pages} {052116} (\bibinfo {year} {2009})}\BibitemShut {NoStop}%
\bibitem [{\citenamefont {{IQOQI Institute for Quantum Optics and Quantum
  Information Vienna}}()}]{open_quantum_problems}%
  \BibitemOpen
  \bibfield  {author} {\bibinfo {author} {\bibnamefont {{IQOQI Institute for
  Quantum Optics and Quantum Information Vienna}}},\ }\href
  {https://oqp.iqoqi.univie.ac.at/open-quantum-problems} {\bibinfo {title}
  {Open quantum problems}},\ \bibinfo {note} {accessed: 2024-12-25}\BibitemShut
  {NoStop}%
\bibitem [{\citenamefont {Pitowsky}(1989)}]{pitowsky1989quantum}%
  \BibitemOpen
  \bibfield  {author} {\bibinfo {author} {\bibfnamefont {I.}~\bibnamefont
  {Pitowsky}},\ }\bibfield  {title} {\bibinfo {title} {Quantum logic},\ }\href
  {https://doi.org/10.1007/BFb0021186} {\bibfield  {journal} {\bibinfo
  {journal} {Quantum Probability—Quantum Logic}\ ,\ \bibinfo {pages} {100}}
  (\bibinfo {year} {1989})}\BibitemShut {NoStop}%
\bibitem [{\citenamefont {D{\"u}r}(2001)}]{dur2001multipartite}%
  \BibitemOpen
  \bibfield  {author} {\bibinfo {author} {\bibfnamefont {W.}~\bibnamefont
  {D{\"u}r}},\ }\bibfield  {title} {\bibinfo {title} {{Multipartite bound
  entangled states that violate Bell's inequality}},\ }\href
  {https://doi.org/10.1103/PhysRevLett.87.230402} {\bibfield  {journal}
  {\bibinfo  {journal} {Physical Review Letters}\ }\textbf {\bibinfo {volume}
  {87}},\ \bibinfo {pages} {230402} (\bibinfo {year} {2001})}\BibitemShut
  {NoStop}%
\bibitem [{\citenamefont {Ac{\'\i}n}(2001)}]{acin2001distillability}%
  \BibitemOpen
  \bibfield  {author} {\bibinfo {author} {\bibfnamefont {A.}~\bibnamefont
  {Ac{\'\i}n}},\ }\bibfield  {title} {\bibinfo {title} {{Distillability, Bell
  inequalities, and multiparticle bound entanglement}},\ }\href
  {https://doi.org/10.1103/PhysRevLett.88.027901} {\bibfield  {journal}
  {\bibinfo  {journal} {Physical review letters}\ }\textbf {\bibinfo {volume}
  {88}},\ \bibinfo {pages} {027901} (\bibinfo {year} {2001})}\BibitemShut
  {NoStop}%
\bibitem [{\citenamefont {Ac{\'\i}n}\ \emph
  {et~al.}(2002{\natexlab{a}})\citenamefont {Ac{\'\i}n}, \citenamefont
  {Scarani},\ and\ \citenamefont {Wolf}}]{acin2002bell}%
  \BibitemOpen
  \bibfield  {author} {\bibinfo {author} {\bibfnamefont {A.}~\bibnamefont
  {Ac{\'\i}n}}, \bibinfo {author} {\bibfnamefont {V.}~\bibnamefont {Scarani}},\
  and\ \bibinfo {author} {\bibfnamefont {M.~M.}\ \bibnamefont {Wolf}},\
  }\bibfield  {title} {\bibinfo {title} {{Bell's inequalities and
  distillability in $N$-quantum-bit systems}},\ }\href
  {https://doi.org/10.1103/PhysRevA.66.042323} {\bibfield  {journal} {\bibinfo
  {journal} {Physical Review A}\ }\textbf {\bibinfo {volume} {66}},\ \bibinfo
  {pages} {042323} (\bibinfo {year} {2002}{\natexlab{a}})}\BibitemShut
  {NoStop}%
\bibitem [{\citenamefont {Ac{\'\i}n}\ \emph
  {et~al.}(2002{\natexlab{b}})\citenamefont {Ac{\'\i}n}, \citenamefont
  {Scarani},\ and\ \citenamefont {Wolf}}]{acin2002violation}%
  \BibitemOpen
  \bibfield  {author} {\bibinfo {author} {\bibfnamefont {A.}~\bibnamefont
  {Ac{\'\i}n}}, \bibinfo {author} {\bibfnamefont {V.}~\bibnamefont {Scarani}},\
  and\ \bibinfo {author} {\bibfnamefont {M.~M.}\ \bibnamefont {Wolf}},\
  }\bibfield  {title} {\bibinfo {title} {{Violation of Bell's inequalities and
  distillability for $N$ qubits}},\ }\href
  {https://doi.org/10.1088/0305-4470/36/2/101} {\bibfield  {journal} {\bibinfo
  {journal} {Journal of Physics A: Mathematical and General}\ }\textbf
  {\bibinfo {volume} {36}},\ \bibinfo {pages} {L21} (\bibinfo {year}
  {2002}{\natexlab{b}})}\BibitemShut {NoStop}%
\bibitem [{\citenamefont {Wehner}(2006)}]{wehner2006tsirelson}%
  \BibitemOpen
  \bibfield  {author} {\bibinfo {author} {\bibfnamefont {S.}~\bibnamefont
  {Wehner}},\ }\bibfield  {title} {\bibinfo {title} {{Tsirelson bounds for
  generalized Clauser-Horne-Shimony-Holt inequalities}},\ }\href
  {https://doi.org/10.1103/PhysRevA.73.022110} {\bibfield  {journal} {\bibinfo
  {journal} {Physical Review A—Atomic, Molecular, and Optical Physics}\
  }\textbf {\bibinfo {volume} {73}},\ \bibinfo {pages} {022110} (\bibinfo
  {year} {2006})}\BibitemShut {NoStop}%
\bibitem [{\citenamefont {Zohren}\ and\ \citenamefont
  {Gill}(2008)}]{zohren2008maximal}%
  \BibitemOpen
  \bibfield  {author} {\bibinfo {author} {\bibfnamefont {S.}~\bibnamefont
  {Zohren}}\ and\ \bibinfo {author} {\bibfnamefont {R.~D.}\ \bibnamefont
  {Gill}},\ }\bibfield  {title} {\bibinfo {title} {{Maximal Violation of the
  Collins-Gisin-Linden-Massar-Popescu Inequality for Infinite Dimensional
  States}},\ }\href {https://doi.org/10.1103/PhysRevLett.100.120406} {\bibfield
   {journal} {\bibinfo  {journal} {Phys. Rev. Lett.}\ }\textbf {\bibinfo
  {volume} {100}},\ \bibinfo {pages} {120406} (\bibinfo {year}
  {2008})}\BibitemShut {NoStop}%
\bibitem [{\citenamefont {Doherty}\ \emph {et~al.}(2008)\citenamefont
  {Doherty}, \citenamefont {Liang}, \citenamefont {Toner},\ and\ \citenamefont
  {Wehner}}]{doherty2008quantum}%
  \BibitemOpen
  \bibfield  {author} {\bibinfo {author} {\bibfnamefont {A.~C.}\ \bibnamefont
  {Doherty}}, \bibinfo {author} {\bibfnamefont {Y.-C.}\ \bibnamefont {Liang}},
  \bibinfo {author} {\bibfnamefont {B.}~\bibnamefont {Toner}},\ and\ \bibinfo
  {author} {\bibfnamefont {S.}~\bibnamefont {Wehner}},\ }\bibfield  {title}
  {\bibinfo {title} {The quantum moment problem and bounds on entangled
  multi-prover games},\ }in\ \href {https://doi.org/10.1109/CCC.2008.26} {\emph
  {\bibinfo {booktitle} {2008 23rd Annual IEEE Conference on Computational
  Complexity}}}\ (\bibinfo {organization} {IEEE},\ \bibinfo {year} {2008})\
  pp.\ \bibinfo {pages} {199--210}\BibitemShut {NoStop}%
\bibitem [{\citenamefont {Li}(2020)}]{li2020exact}%
  \BibitemOpen
  \bibfield  {author} {\bibinfo {author} {\bibfnamefont {B.}~\bibnamefont
  {Li}},\ }\bibfield  {title} {\bibinfo {title} {{Exact values of quantum
  violations in low-dimensional Bell correlation inequalities}},\ }\href
  {https://doi.org/10.1016/j.laa.2020.05.040} {\bibfield  {journal} {\bibinfo
  {journal} {Linear Algebra and its Applications}\ }\textbf {\bibinfo {volume}
  {603}},\ \bibinfo {pages} {289} (\bibinfo {year} {2020})}\BibitemShut
  {NoStop}%
\bibitem [{\citenamefont {Tsirel'son}(1987)}]{tsirelson1987quantum}%
  \BibitemOpen
  \bibfield  {author} {\bibinfo {author} {\bibfnamefont {B.~S.}\ \bibnamefont
  {Tsirel'son}},\ }\bibfield  {title} {\bibinfo {title} {Quantum analogues of
  the {B}ell inequalities. {T}he case of two spatially separated domains},\
  }\href {https://doi.org/10.1007/BF01663472} {\bibfield  {journal} {\bibinfo
  {journal} {Journal of Soviet mathematics}\ }\textbf {\bibinfo {volume}
  {36}},\ \bibinfo {pages} {557} (\bibinfo {year} {1987})},\ \bibinfo {note}
  {(Translated from a source in Russian of 1985)}\BibitemShut {NoStop}%
\bibitem [{\citenamefont {Landau}(1988)}]{landau1988empirical}%
  \BibitemOpen
  \bibfield  {author} {\bibinfo {author} {\bibfnamefont {L.~J.}\ \bibnamefont
  {Landau}},\ }\bibfield  {title} {\bibinfo {title} {Empirical two-point
  correlation functions},\ }\href {https://doi.org/10.1007/BF00732549}
  {\bibfield  {journal} {\bibinfo  {journal} {Foundations of Physics}\ }\textbf
  {\bibinfo {volume} {18}},\ \bibinfo {pages} {449} (\bibinfo {year}
  {1988})}\BibitemShut {NoStop}%
\bibitem [{\citenamefont {Avis}\ \emph {et~al.}(2006)\citenamefont {Avis},
  \citenamefont {Imai},\ and\ \citenamefont {Ito}}]{avis2006relationship}%
  \BibitemOpen
  \bibfield  {author} {\bibinfo {author} {\bibfnamefont {D.}~\bibnamefont
  {Avis}}, \bibinfo {author} {\bibfnamefont {H.}~\bibnamefont {Imai}},\ and\
  \bibinfo {author} {\bibfnamefont {T.}~\bibnamefont {Ito}},\ }\bibfield
  {title} {\bibinfo {title} {On the relationship between convex bodies related
  to correlation experiments with dichotomic observables},\ }\href
  {https://doi.org/10.1088/0305-4470/39/36/010} {\bibfield  {journal} {\bibinfo
   {journal} {Journal of Physics A: Mathematical and General}\ }\textbf
  {\bibinfo {volume} {39}},\ \bibinfo {pages} {11283} (\bibinfo {year}
  {2006})}\BibitemShut {NoStop}%
\bibitem [{\citenamefont {Navascu{\'e}s}\ \emph {et~al.}(2007)\citenamefont
  {Navascu{\'e}s}, \citenamefont {Pironio},\ and\ \citenamefont
  {Ac{\'\i}n}}]{navascues2007bounding}%
  \BibitemOpen
  \bibfield  {author} {\bibinfo {author} {\bibfnamefont {M.}~\bibnamefont
  {Navascu{\'e}s}}, \bibinfo {author} {\bibfnamefont {S.}~\bibnamefont
  {Pironio}},\ and\ \bibinfo {author} {\bibfnamefont {A.}~\bibnamefont
  {Ac{\'\i}n}},\ }\bibfield  {title} {\bibinfo {title} {Bounding the set of
  quantum correlations},\ }\href
  {https://doi.org/10.1103/PhysRevLett.98.010401} {\bibfield  {journal}
  {\bibinfo  {journal} {Physical Review Letters}\ }\textbf {\bibinfo {volume}
  {98}},\ \bibinfo {pages} {010401} (\bibinfo {year} {2007})}\BibitemShut
  {NoStop}%
\bibitem [{\citenamefont {Ishizaka}(2017)}]{Ishizaka2017cryptographic}%
  \BibitemOpen
  \bibfield  {author} {\bibinfo {author} {\bibfnamefont {S.}~\bibnamefont
  {Ishizaka}},\ }\bibfield  {title} {\bibinfo {title} {Cryptographic quantum
  bound on nonlocality},\ }\bibfield  {journal} {\bibinfo  {journal} {Physical
  Review A}\ }\textbf {\bibinfo {volume} {95}},\ \href
  {https://doi.org/10.1103/physreva.95.022108} {10.1103/physreva.95.022108}
  (\bibinfo {year} {2017})\BibitemShut {NoStop}%
\bibitem [{\citenamefont {Ishizaka}(2018)}]{Ishizaka2018necessary}%
  \BibitemOpen
  \bibfield  {author} {\bibinfo {author} {\bibfnamefont {S.}~\bibnamefont
  {Ishizaka}},\ }\bibfield  {title} {\bibinfo {title} {Necessary and sufficient
  criterion for extremal quantum correlations in the simplest bell scenario},\
  }\bibfield  {journal} {\bibinfo  {journal} {Physical Review A}\ }\textbf
  {\bibinfo {volume} {97}},\ \href {https://doi.org/10.1103/physreva.97.050102}
  {10.1103/physreva.97.050102} (\bibinfo {year} {2018})\BibitemShut {NoStop}%
\bibitem [{\citenamefont {Isobe}\ and\ \citenamefont
  {Tanimura}(2010)}]{isobe2010method}%
  \BibitemOpen
  \bibfield  {author} {\bibinfo {author} {\bibfnamefont {T.}~\bibnamefont
  {Isobe}}\ and\ \bibinfo {author} {\bibfnamefont {S.}~\bibnamefont
  {Tanimura}},\ }\bibfield  {title} {\bibinfo {title} {A method for systematic
  construction of {B}ell-like inequalities and a proposal of a new type of
  test},\ }\href {https://doi.org/10.1143/PTP.124.191} {\bibfield  {journal}
  {\bibinfo  {journal} {Progress of theoretical physics}\ }\textbf {\bibinfo
  {volume} {124}},\ \bibinfo {pages} {191} (\bibinfo {year}
  {2010})}\BibitemShut {NoStop}%
\bibitem [{\citenamefont {Tsirel'son}(1993)}]{tsirelson1993some}%
  \BibitemOpen
  \bibfield  {author} {\bibinfo {author} {\bibfnamefont {B.~S.}\ \bibnamefont
  {Tsirel'son}},\ }\bibfield  {title} {\bibinfo {title} {Some results and
  problems on quantum {B}ell-type inequalities},\ }\href@noop {} {\bibfield
  {journal} {\bibinfo  {journal} {Hadronic Journal Supplement}\ }\textbf
  {\bibinfo {volume} {8}},\ \bibinfo {pages} {329} (\bibinfo {year}
  {1993})}\BibitemShut {NoStop}%
\bibitem [{\citenamefont {Halliwell}(2014)}]{halliwell2014two}%
  \BibitemOpen
  \bibfield  {author} {\bibinfo {author} {\bibfnamefont {J.}~\bibnamefont
  {Halliwell}},\ }\bibfield  {title} {\bibinfo {title} {Two proofs of {F}ine's
  theorem},\ }\href {https://doi.org/10.1016/j.physleta.2014.08.012} {\bibfield
   {journal} {\bibinfo  {journal} {Physics Letters A}\ }\textbf {\bibinfo
  {volume} {378}},\ \bibinfo {pages} {2945} (\bibinfo {year}
  {2014})}\BibitemShut {NoStop}%
\bibitem [{\citenamefont {Lee}\ and\ \citenamefont
  {Tsutsui}(2017)}]{lee2017quasi}%
  \BibitemOpen
  \bibfield  {author} {\bibinfo {author} {\bibfnamefont {J.}~\bibnamefont
  {Lee}}\ and\ \bibinfo {author} {\bibfnamefont {I.}~\bibnamefont {Tsutsui}},\
  }\bibfield  {title} {\bibinfo {title} {Quasi-probabilities in conditioned
  quantum measurement and a geometric/statistical interpretation of
  {A}haronov's weak value},\ }\href {https://doi.org/10.1093/ptep/ptx024}
  {\bibfield  {journal} {\bibinfo  {journal} {Progress of Theoretical and
  Experimental Physics}\ }\textbf {\bibinfo {volume} {2017}},\ \bibinfo {pages}
  {052A01} (\bibinfo {year} {2017})}\BibitemShut {NoStop}%
\bibitem [{\citenamefont {Lee}\ and\ \citenamefont
  {Tsutsui}(2018)}]{lee2018general}%
  \BibitemOpen
  \bibfield  {author} {\bibinfo {author} {\bibfnamefont {J.}~\bibnamefont
  {Lee}}\ and\ \bibinfo {author} {\bibfnamefont {I.}~\bibnamefont {Tsutsui}},\
  }\bibfield  {title} {\bibinfo {title} {A general framework of
  quasi-probabilities and the statistical behaviour of non-commuting quantum
  observables},\ }in\ \href {https://doi.org/10.1007/978-981-13-2487-1_9}
  {\emph {\bibinfo {booktitle} {Reality and Measurement in Algebraic Quantum
  Theory: NWW 2015, Nagoya, Japan, March 9-13}}}\ (\bibinfo {organization}
  {Springer},\ \bibinfo {year} {2018})\ pp.\ \bibinfo {pages}
  {195--228}\BibitemShut {NoStop}%
\bibitem [{\citenamefont {Umekawa}\ \emph {et~al.}(2024)\citenamefont
  {Umekawa}, \citenamefont {Lee},\ and\ \citenamefont
  {Hatano}}]{umekawa2024advantages}%
  \BibitemOpen
  \bibfield  {author} {\bibinfo {author} {\bibfnamefont {S.}~\bibnamefont
  {Umekawa}}, \bibinfo {author} {\bibfnamefont {J.}~\bibnamefont {Lee}},\ and\
  \bibinfo {author} {\bibfnamefont {N.}~\bibnamefont {Hatano}},\ }\bibfield
  {title} {\bibinfo {title} {{Advantages of the Kirkwood--Dirac distribution
  among general quasi-probabilities on finite-state quantum systems}},\ }\href
  {https://doi.org/10.1093/ptep/ptae005} {\bibfield  {journal} {\bibinfo
  {journal} {Progress of Theoretical and Experimental Physics}\ }\textbf
  {\bibinfo {volume} {2024}},\ \bibinfo {pages} {023A02} (\bibinfo {year}
  {2024})}\BibitemShut {NoStop}%
\bibitem [{\citenamefont {Kimura}(2003)}]{kimura2003bloch}%
  \BibitemOpen
  \bibfield  {author} {\bibinfo {author} {\bibfnamefont {G.}~\bibnamefont
  {Kimura}},\ }\bibfield  {title} {\bibinfo {title} {The bloch vector for
  {$N$}-level systems},\ }\href {https://doi.org/10.1016/S0375-9601(03)00941-1}
  {\bibfield  {journal} {\bibinfo  {journal} {Physics Letters A}\ }\textbf
  {\bibinfo {volume} {314}},\ \bibinfo {pages} {339} (\bibinfo {year}
  {2003})}\BibitemShut {NoStop}%
\bibitem [{\citenamefont {Dangniam}\ and\ \citenamefont
  {Ferrie}(2015)}]{dangniam2015quantum}%
  \BibitemOpen
  \bibfield  {author} {\bibinfo {author} {\bibfnamefont {N.}~\bibnamefont
  {Dangniam}}\ and\ \bibinfo {author} {\bibfnamefont {C.}~\bibnamefont
  {Ferrie}},\ }\bibfield  {title} {\bibinfo {title} {Quantum {B}ochner's
  theorem for phase spaces built on projective representations},\ }\href
  {https://doi.org/10.1088/1751-8113/48/11/115305} {\bibfield  {journal}
  {\bibinfo  {journal} {Journal of Physics A: Mathematical and Theoretical}\
  }\textbf {\bibinfo {volume} {48}},\ \bibinfo {pages} {115305} (\bibinfo
  {year} {2015})}\BibitemShut {NoStop}%
\end{thebibliography}%

\end{document}